\documentclass[12pt,letterpaper]{article}
\usepackage{fullpage}
\usepackage[top=2cm, bottom=4.5cm, left=2.5cm, right=2.5cm]{geometry}
\usepackage{amsmath,amsthm,amsfonts,amssymb,amscd}
\usepackage{enumerate}
\usepackage{graphicx}
\usepackage{listings}
\usepackage{float}
\allowdisplaybreaks
\usepackage{natbib}
\usepackage{subfig}

\usepackage{lipsum}
\usepackage{epsfig}
\usepackage{bm}
\usepackage{multicol}
\usepackage{url}
\usepackage[utf8]{inputenc}
\usepackage{mathtools}
\usepackage{extarrows}

\usepackage{color}
\definecolor{dkgreen}{rgb}{0,0.6,0}
\definecolor{gray}{rgb}{0.5,0.5,0.5}
\definecolor{mauve}{rgb}{0.58,0,0.82}

\usepackage{booktabs}
\usepackage{longtable}
\usepackage{array}
\usepackage{multirow}
\usepackage{wrapfig}
\usepackage{colortbl}
\usepackage{pdflscape}
\usepackage{tabu}
\usepackage{threeparttable}
\usepackage{threeparttablex}
\usepackage[normalem]{ulem}
\usepackage{makecell}
\usepackage{xcolor}

\usepackage[section]{placeins}

\graphicspath{{figures_oct2021/}}

\usepackage[utf8]{inputenc}

\title{A dynamic spatial filtering approach to mitigate underestimation bias in field calibrated low-cost sensor air-pollution data}
\author{Claire Heffernan$^{1,a}$, Roger Peng$^{2}$, Drew R. Gentner$^3$, \\
	Kirsten Koehler$^4$, Abhirup Datta$^{1,b}$}
\date{\footnotesize $^1$Department of Biostatistics, Johns Hopkins University, $^a$cmheff@jhu.edu, $^b$abhidatta@jhu.edu\\
	$^2$Department of Statistics and Data Sciences, University of Texas, Austin, rdpeng@jhu.edu\\
	$^3$Department of Chemical \& Environmental Engineering, Yale University, drew.gentner@yale.edu\\
	$^4$Department of Environmental Health and Engineering, Johns Hopkins University, kkoehle1@jhu.edu}

\newcommand{\calW}{\mathcal{W}}
\newcommand{\calD}{\mathcal{D}}
\newcommand{\calF}{\mathcal{F}}

\newcommand{\bv}{\mathbf{v}}

\newcommand{\bu}{\mathbf{u}}
\newcommand{\bV}{\mathbf{V}}

\newcommand{\bx}{\mathbf{x}}
\newcommand{\by}{\mathbf{y}}

\newcommand{\bs}{\mathbf{s}}
\newcommand{\bS}{\mathbf{S}}
\newcommand{\bl}{\widehat{\beta_1}}
\newcommand{\bo}{\widehat{\beta_0}}

\newcommand{\bH}{\mathbf{H}}

\newcommand{\bI}{\mathbf{I}}
\newcommand{\bX}{\mathbf{X}}
\newcommand{\bA}{\mathbf{A}}

\newcommand{\bC}{\mathbf{C}}

\newcommand{\bz}{\mathbf{z}}

\newcommand{\bZ}{\mathbf{Z}}
\newcommand{\bones}{\mathbf{1}}

\newcommand{\st}{(\bs,t)}

\newcommand{\slt}{(\bS_B,t)}
\newcommand{\sot}{(\bS_{A\cup C},t)}

\newcommand{\sdt}{(\bS_D,t)}

\newcommand{\balpha}{ {\boldsymbol \phi} }

\newcommand{\bbeta}{ {\boldsymbol \beta} }

\newcommand{\bgamma}{ {\boldsymbol \gamma} }

\newcommand{\bmu}{ {\boldsymbol \mu} }

\newcommand{\bSigma}{ {\boldsymbol \Sigma} }

\newtheorem{theorem}{Proposition}
\newtheorem{alg}{Algorithm}

\begin{document}
\maketitle

\begin{abstract}
Low-cost air pollution sensors, offering hyper-local characterization of pollutant concentrations, are becoming increasingly prevalent in environmental and public health research. However, low-cost air pollution data can be noisy, biased by environmental conditions, and usually need to be field-calibrated by collocating low-cost sensors with reference-grade instruments. We show, theoretically and empirically, that the common procedure of regression-based calibration using collocated data systematically underestimates high air pollution concentrations, which are critical to diagnose from a health perspective. Current calibration practices also often fail to utilize the spatial correlation in pollutant concentrations. We propose a novel spatial filtering approach to collocation-based calibration of low-cost networks that mitigates the underestimation issue by using an inverse regression. The inverse-regression also allows for incorporating spatial correlations by a second-stage model for the true pollutant concentrations using a conditional Gaussian Process. Our approach works with one or more collocated sites in the network and is dynamic, leveraging spatial correlation with the latest available reference data. Through extensive simulations, we demonstrate how the spatial filtering substantially improves estimation of pollutant concentrations, and measures peak concentrations with greater accuracy. We apply the methodology for calibration of a low-cost PM$_{2.5}$ network in Baltimore, Maryland, and diagnose air pollution peaks that are missed by the regression-calibration. 
\end{abstract}


\section{Introduction}

Air pollution is regulated nationally in the United States using reference-grade instruments that conform to measurement standards like the Federal Reference Method (FRM) or
Federal Equivalent Method (FEM) set by the Environmental Protection Agency (EPA) \citep{EPA}. 
However, regulatory networks offer sparse geographical coverage  \citep{apte2017high} and hence the gold standard data cannot assess disparities in air-quality at fine scale spatial resolution. 

To fill the knowledge gap left by the regulatory monitoring, local networks of low-cost air pollution sensors are being increasingly deployed in many areas, including Los Angeles \citep{lu2021beyond}, Salt Lake City \citep{chadwick2021understanding}, Denver \citep{considine2021improving}, Berkeley \citep{kim2018berkeley} and the San Francisco Bay area \citep{apte2017high}. These sensors are orders of magnitude cheaper than the high-precision and high-accuracy regulatory devices. For example, the widely used Plantower low-cost sensors for fine particulate matter (PM$_{2.5}$) would cost around $100$ times less than 
reference-grade FRM monitors 
even when including the pricing of added components for the installing the low-cost sensors like housing, power, and data collection, storage, and transmission. Low-cost sensors can thus be deployed in larger numbers, creating dense monitoring networks that have high spatial resolution, which allows for neighborhood-level estimates of air pollution concentrations. 
The {\em hyper-local} 
characterization of exposures from low-cost sensors promises insights on air pollution and its health impacts at spatial scales beyond the scope of the sparse regulatory networks.

The data abundance of low-cost sensors comes at the expense of data quality. 
Sensor data quality depends on many variables including the manufacturer, sensor type, meteorological factors like relative humidity and temperature,
the chemical composition of particulates, time since installation, and cross-sensitivity to other pollutants, among others. Even sensors from the same manufacturer can perform differently under varying ambient conditions. 
Hence, raw data from these low-cost networks is not an accurate representation of the pollutant surface.


To enhance the data accuracy of low-cost sensor networks, two common types of calibration can be performed. Before the sensors are deployed in the field, laboratory calibration is often performed by exposing them to known pollutant concentrations under different regulated conditions (relative humidity, temperature) and deriving calibration equations from the measurements \citep{tryner2020laboratory,levy2018field}. These can be nonlinear parametric equations of these meteorological variables, but often do not capture the full range of possible ambient conditions and so further calibration is often needed. Field-calibration is often used to supplement laboratory calibration. This is done by collocating some of the sensors in the network with one or more high-quality reference instruments in the region \citep{zimmerman2018machine,Topalovic2019,datta2020statistical}. 
The paired time-series of collocated reference and low-cost measurements is used to train a regression model which subsequently calibrates data from other sensors in the network. 
The focus of this paper is on field calibration, which we will simply refer to as calibration from now on.

Different regression approaches to calibration  include multiple linear regression \citep{bigi2018performance,bi2020incorporating,ardon2020measurements,si2020evaluation,barkjohn2021development,datta2020statistical,romero2020development}, random effects models \citep{nordio2013estimating}, land-use regression  \citep{clougherty2013intra,larson2009mobile}, and machine learning methods like random forests \citep{lim2019mapping,zimmerman2018machine},  neural networks \citep{Topalovic2019} and boosting \citep{johnson2018using}. 
Field-calibration using some form of regression is one of the most widely used methods for calibrating low-cost air pollution data. For example, a United States wide regression-calibration equation has recently been recommended for calibration of the PM$_{2.5}$ sensors used in the PurpleAir network \citep{barkjohn2021development}. While regression-calibration reduce the bias of raw low-cost sensor data, this manuscript demonstrates 
two major limitations of this widely adopted approach.
\begin{enumerate}
    \item 
    High air pollution exposures disproportionately affect health, and we will show theoretically and empirically that regression-calibration systematically underestimates high levels of exposure. 
    \item Air pollution concentrations exhibit similarity across space, but regression-calibration is applied separately to each sensor in the network. This practice does not leverage this correlation -- neither among the low-cost data at different sites in the network, nor between the low-cost data and concurrent measurements from nearby reference devices. 
\end{enumerate}

In light of this issue of regression-calibration underestimating air pollution peaks, our goal is to develop a simple calibration approach for low-cost sensors that mitigates the peak underestimation issue, is spatially informed, and is dynamic in the sense that the calibration equation can be updated 
using the latest reference data without having to retrain the entire calibration model.
We propose a novel and simple approach to collocation-based dynamic field-calibration of low-cost sensor networks that mitigates the aforementioned shortcomings of regression-calibration. We first address the underestimation of  regression-calibration by switching to an inverse regression model, where the low-cost measurement is regressed on the true pollutant concentration and other covariates. We motivate this change  by making connections of the forward and inverse regression models to Berkson and classical measurement errors, respectively. Low-cost data is a noisy observation of the latent true pollutant concentration at the same location. Hence, the inverse regression, modeling classical measurement error, is more appropriate. We prove that the underestimation issue is not present in the inverse model. 


To leverage spatial correlation among pollutant concentrations and  concurrent reference data when making predictions, we then extend the inverse regression to a {\em spatial filtering} method 
for spatially informed and dynamic predictions of the true pollutant concentrations from the low-cost data. The inverse regression is the observation model part of the filter. As the true pollutant concentration is now the independent variable (covariate) in the inverse regression, we can seamlessly add a second-stage 
model for the true pollutant concentrations to capture the spatial correlation. Since the true pollutant surface is partially observed at the reference sites, we use a conditional Gaussian Process spatial model to incorporate this information on concurrent reference data. This corresponds to the state-transition part of the filter. Our method thus filters the low-cost data over space given the gold standard measurements at reference sites, which results in a smooth estimated pollutant surface. This spatial filtering is different to filtering approaches applied previously for modeling air pollution data where the filtering occurs in time. The advantage of spatial filtering is that it leverages the current data from all the reference devices for dictating the spatial state-transition model, resulting in a dynamic calibration. Many common calibration methods used for low-cost data only use the reference data from the collocated reference devices and only for fixed time-window for training the regression model thereby resulting in a static calibration equation that does not use current reference data.


We offer both a frequentist and a Bayesian implementation of the spatial filtering. 
The advantage of the Bayesian implementation is that the uncertainty of the model parameter estimation is propagated in the filtering, while in the faster frequentist implementation, only the parameter estimates from a preliminary step are plugged into the filtering step. 
Extensive numerical studies using simulated data were used to evaluate the method. We see that across a wide range of scenarios, compared to regression-calibration, spatial filtering offers consistently improved overall root mean squared error (RMSE) and better identification of high pollution events. 
We apply the filtering method to calibrate PM$_{2.5}$ data from a  low-cost sensor network data in Baltimore. 
The spatial filtering performs much better than regression-calibration in identifying high pollution days and is used to create maps of PM$_{2.5}$ concentrations in the city. 

\section{SEARCH low-cost PM$_{2.5}$ network in Baltimore}\label{sec:search}

\begin{figure}[t]
\centering
\includegraphics[width=2in]{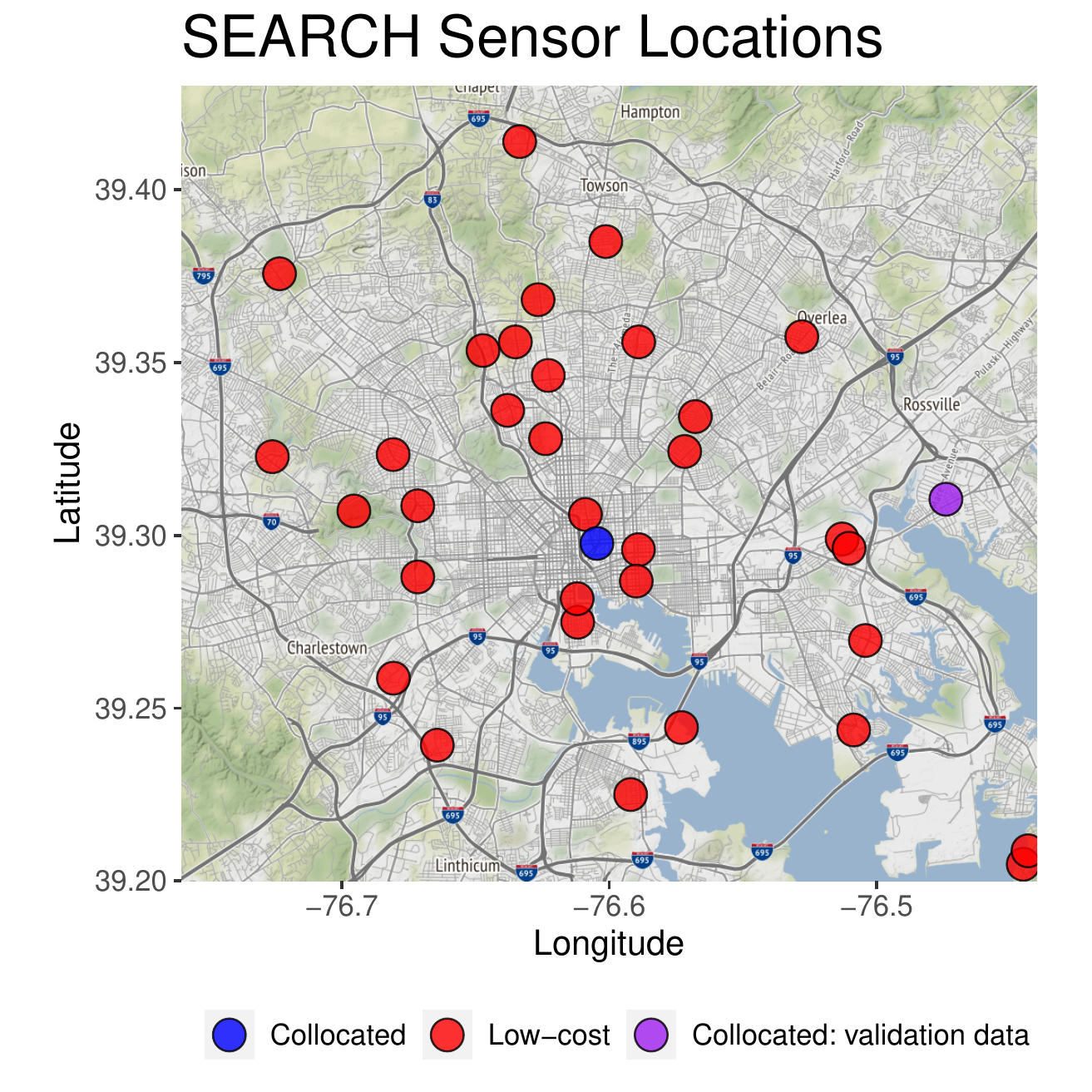}\includegraphics[width=3.5in]{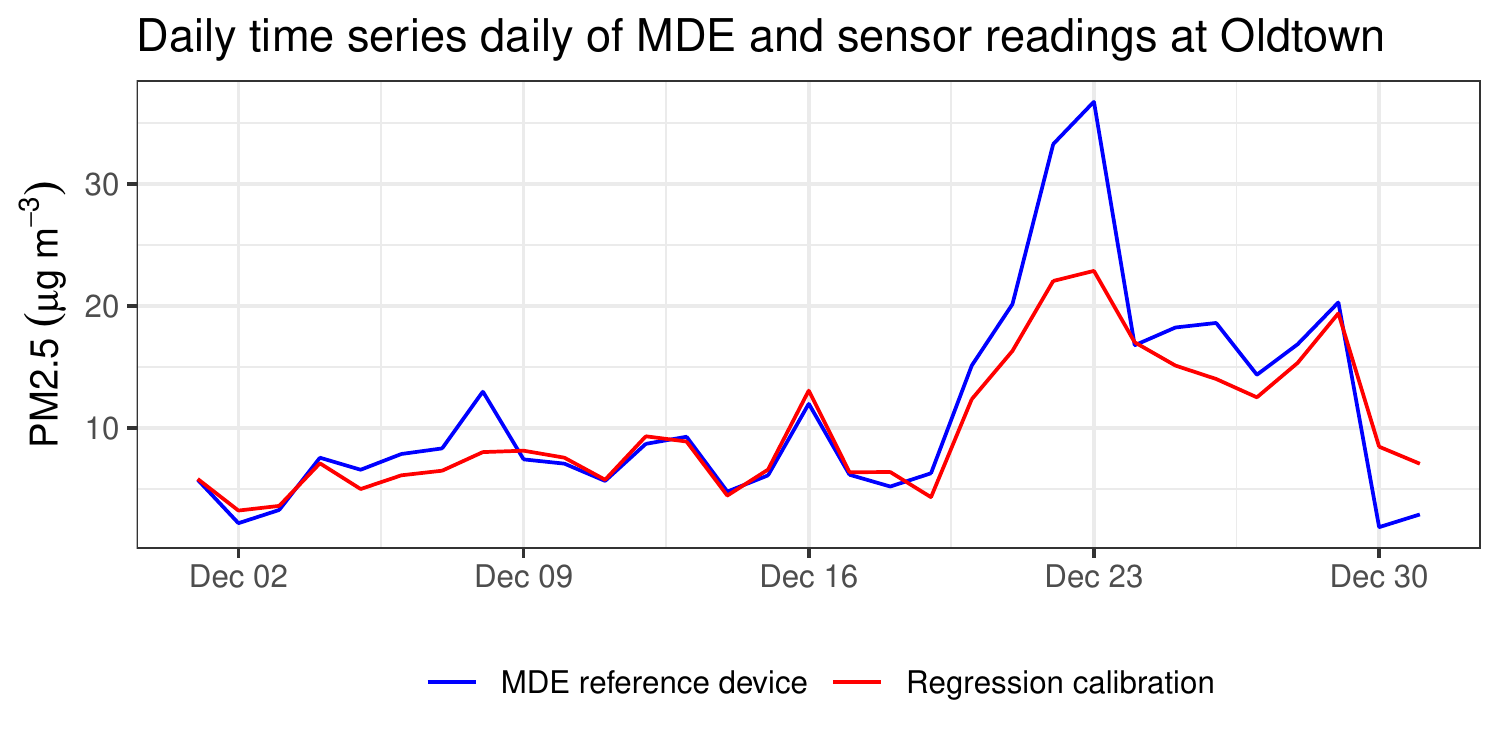}
\caption{SEARCH low-cost network: (Left) Map of the SEARCH network in Baltimore. The blue site is the collocated site at Oldtown which has a reference instrument for PM$_{2.5}$ from the Maryland Department of Energy (MDE) and two low-cost sensors. The red sites are non-collocated monitors. The purple site, at Essex outside of the city, has a low-cost sensor and a reference instrument that measures PM$_{2.5}$ every 6 days. This device will be used to validate calibration methods since its reference device has much lower temporal frequency. (Right) Daily PM$_{2.5}$ time series in December 2019 of data from Oldtown MDE reference instrument and regression-calibrated data from the collocated SEARCH low-cost sensor.}\label{ts_dec2019} 
\end{figure}

We first illustrate the underestimation of the regression-calibration model using PM$_{2.5}$ data from a network of low-cost air pollution sensors \citep{buehler2021stationary} in Baltimore, Maryland. Within Baltimore City limits, there is only one regulatory site (at Oldtown) managed by  the Maryland Department of Energy (MDE) that measures hourly PM$_{2.5}$ in the city using a reference monitor (an FEM Beta Attenuation Monitor (BAM)). Additionally, there is one reference device on the outskirts of the city at the Essex site, which measures PM$_{2.5}$ every 6 days. This regulatory PM$_{2.5}$ data from only two sites is not sufficient to provide insight about local fluctuations in air quality. Understanding such intra-urban variation is critical to study issues of environmental injustice and health within the city. 
To obtain spatially resolved data on air quality in Baltimore, the Solutions to Energy, Air, Climate, and Health (SEARCH) Center has been operating a low-cost air-pollution sensor network in Baltimore. The network had sensors at 36 locations between December 2019 and May 2020, which are shown in Figure \ref{ts_dec2019}. Each sensor measures multiple pollutants including PM$_{2.5}$ as well as relative humidity (RH) and temperature (T). Details on the design of the SEARCH network and the PM$_{2.5}$ sensors used are discussed in Section \ref{sec:network} of the Supplement.

Given the lack of spatially resolved regulatory PM$_{2.5}$ data in Baltimore, data from the SEARCH low-cost network is of importance as it will help study the spatial variation in air quality within the city and its association with health, socioeconomic and other variables.
However, the PM$_{2.5}$ sensors in the SEARCH network, Plantower A003, a common brand of sensor, tend to overestimate reference measurements \citep{ardon2020measurements}.
 A recent study \citep{barkjohn2021development}
found that the overestimation for these samples is about a factor of two, and higher in humid environments. \cite{datta2020statistical} found biases of similar magnitude in the SEARCH raw PM$_{2.5}$ data which demonstrated the need for calibration before any use of the data.

There are two sensors collocated with a reference device at the Oldtown location. \cite{datta2020statistical} 
used this field-collocation data from Oldtown to estimate a {\em gain-offset model} \citep{Balzano2007} for regression-calibration of the low-cost PM$_{2.5}$ data: \begin{equation}\label{eq:gainoffset}
\begin{array}{cc}
E(x(\bs,t))=o(\bs,t) + g(\bs,t) y(\bs,t),\\
\mbox{ where } o(\bs,t) = \balpha'\bz(\bs,t) \mbox{ and } g(\bs,t) = \bgamma'\bz(\bs,t),\;
\end{array}
\end{equation} where 
$x(\bs,t)$ denotes the reference PM$_{2.5}$ data 
at location $\bs$ and time $t$ and $y(\bs,t)$ is the low-cost data. The gain $g$ (multiplicative bias) and offset $o$ (additive bias) were modeled as linear functions of the covariates $\bz(\bs,t)$ (RH, T, a weekend indicator, and a daylight indicator). The model parameters were estimated on a training window using least squares. 

The calibrated low-cost PM$_{2.5}$ data from this regression-calibration model were substantially more accurate  than the raw or lab-corrected low-cost data. However, the baseline PM$_{2.5}$ concentrations in Baltimore is generally around $8 \mu g/m^3$ for the study period of \cite{datta2020statistical}, so the improvement in accuracy after calibration primarily reflected mitigation of biases in the low-cost data at low concentrations.
The performance of this model specifically during windows of high pollutant concentrations was not studied. Figure \ref{ts_dec2019} (right) presents comparisons in December 2019  between the daily predictions from the model of \cite{datta2020statistical} for the SEARCH low-cost sensor at Oldtown and the reference instrument at that site. The regression-calibration model clearly underestimates when the true PM$_{2.5}$ reaches unhealthy levels on December 23, 2019. The predicted daily concentrations from the regression-calibration model ($\sim 20 \mu g/m^3$) is nearly half of the true concentration ($\sim 35 \mu g/m^3$). 

The World Health Organization (WHO) recently reduced their recommendation for annual average PM$_{2.5}$ concentrations from 10 $\mu g/m^3$ to 5 $\mu g/m^3$ \citep{WHO}. The 24-hour average PM$_{2.5}$ concentration standard by the WHO is 15 $\mu g/m^3$, which is a 99th percentile standard that should only be exceeded 3-4 times per year. The EPA's standards are 12 $\mu g/m^3$ for the annual average and 35 $\mu g/m^3$ for the 24-hour average 98th percentile \citep{EPA_NAAQS}. 
For reporting daily air quality, the EPA's {\em Air Quality Index} (AQI) threshold to classify concentrations as ``moderate" is 12 $\mu g/m^3$, reflecting the need for maintaining concentrations well below the 98$^{th}$ percentile daily standard  of 35 $\mu g/m^3$ (which is the threshold for the ``unhealthy" classification). 

Table \ref{fpr_dec2019} presents the proportion of times 
hourly moderate or unhealthy observations (according to the aforementioned cutoffs) are misclassified by the regression-calibration 
in Baltimore. Although, the AQI cut-offs are for daily level, it is important to also properly calibrate hourly measurements as the daily concentration is obtained by averaging them.
 Across the period from December 2019 through May 2020, 23\% of the moderate or unhealthy instances, as measured by the reference instruments, are incorrectly predicted by the regression calibrated low-cost data as being good. 
This example shows the misclassification of PM$_{2.5}$ concentrations by the regression-calibration model for high values of true PM$_{2.5}$ concentrations. As high levels of exposure affect health adversely, it is critical for calibration techniques for low-cost sensors to be aware of this asymmetry in risks of exposure misclassification and be able to accurately identify days of high air pollution events. 

\begin{table}
\caption{Misclassification rates according to AQI classification of the regression-calibration model as currently used to calibrate the SEARCH network, for hourly data in December 2019 - May 2020. US AQI Classifications are: Good PM$_{2.5}$ is less than $12 \mu g/m^3$, Moderate/Unhealthy is $12.1 \mu g/m^3$ or more.}
\begin{tabular}[t]{>{\centering\arraybackslash}p{3em}ccccc}
\toprule
\multicolumn{2}{c}{ } & \multicolumn{2}{c}{Prediction Classification} & \multicolumn{1}{c}{ } \\
\cmidrule(l{3pt}r{3pt}){3-4}
 &  & Good (\%) & Moderate/Unhealthy (\%) & Sample Size\\
\midrule
 & Good & 96 & 4 & 3672\\
\cmidrule{2-5}
\multirow{-2}{3em}{\centering\arraybackslash \textbf{True PM2.5}} & Moderate/Unhealthy & 23 & 77 & 428\\
\bottomrule
\end{tabular}
\label{fpr_dec2019}
\end{table}

In the next Section we present a novel approach that mitigates the underestimation issue. The proposed method will be very useful in cities having poor air quality with frequent peaks in concentrations. 
However, the approach will also be applicable to calibrate low-cost sensor air-pollution networks in cities like Baltimore with fewer peaks and lower baseline levels. In the US, only around 5\% of the population live in ambient concentrations over the annual standard of 12 $\mu g/m^3$, while around 60\% of the population had annual exposures below 8 $\mu g/m^3$ \citep{jbaily2022air}. In this regard, Baltimore is very representative of the PM$_{2.5}$ concentrations experiences by the majority of Americans. 
The WHO annual standard of 5 $\mu g/m^3$ shows that concentrations well below the ``moderate'' threshold of 12 $\mu g/m^3$ are of importance to health. Although this is an annual standard, it is important to properly calibrate hourly or daily measurements at all concentrations to assess compliance to the annual standard. 
 Therefore, a calibration method needs to be accurate across the whole gamut of pollutant concentrations. 
 In Section \ref{sec:search_analysis}, we will demonstrate how our proposed method successfully calibrates low-cost sensor PM$_{2.5}$ data in Baltimore capturing both the occasional peaks and the baseline lower concentrations. 


\section{Methods}

\subsection{Low-cost air pollution networks}\label{sec:low_cost_networks}

\begin{figure}[h]
\centering
\includegraphics[scale=0.7]{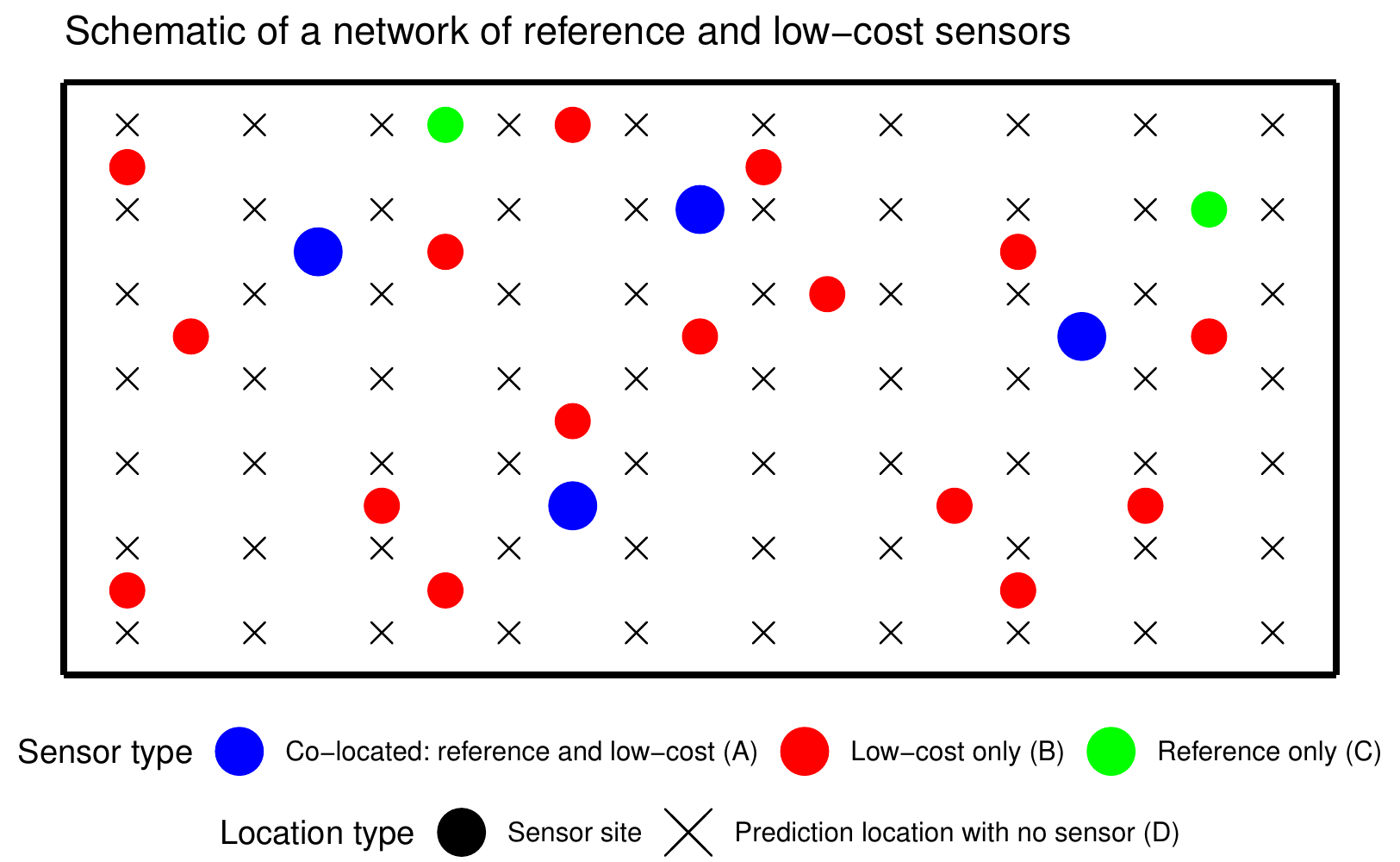}
\caption{Schematic of a low-cost air pollution network in an area with 6 reference devices (blue and green sites) and 20 low-cost sensors (blue and red sites), where the blue sites are collocated sites. The grid of crosses indicates locations at which a PM$_{2.5}$ prediction is desired but there are no instruments of either type. }\label{schematic}
\end{figure}

A schematic of a general low-cost air pollution network is shown in Figure \ref{schematic}. The locations in the schematic can be split into four sets. The blue sites have a reference device and a collocated low-cost sensor at that location, and will be referred to as Set A. In general, this set will have a very few sites as regulatory monitoring networks using reference devices are quite sparse. We assume that there is at least one reference device in the area, i.e., at least one site in Set A, enabling collocation of a low-cost sensor to learn the biases in the low-cost data. The case where Set A is empty will be mentioned in Section \ref{sec:nocol}. The red sites, Set B, only have low-cost sensors. This set is typically numerous. Set C is the green sites, where there is a reference device and no low-cost sensor. Like Set A, this set will also typically have very a few sites as the reference network is sparse, and it can even be empty if there is a low-cost sensor placed at every reference site in the area. Lastly, in addition to calibrating the low-cost data at the network sites, another goal is to predict pollutant concentrations at a dense grid of locations which are then interpolated to create maps. Set D represents such a grid of prediction locations, denoted by the black crosses. 
Reference samplers also have some measurement error but data from these are widely used as the gold standard \citep{frm}.  Throughout this paper, we assume that the reference devices have negligible measurement error, while the low-cost devices have measurement errors that we wish to account for.

\subsection{Regression-calibration}\label{sec:reg-cal}
The gain-offset model (\ref{eq:gainoffset}) subsumes a large class of regression-calibration models. For example, if no covariates $z$ are considered, it reduces to the most basic calibration model 
\begin{align}\label{reg-cal-no-cov}
    E(x\st)=\beta_0+\beta_1y\st
\end{align}
with constant gain $\beta_1$ and offset $\beta_0$. In many sensor calibration problems, (\ref{reg-cal-no-cov}) is popularly used \citep{Miskell2018,Balzano2007,zheng2019gaussian}. 
A popular choice is calibration using the multiple linear regression (MLR) model 
\begin{align*}
E(x\st)&=\beta_0+\beta_1 y\st+\boldsymbol{\beta_2}'\bz\st
\end{align*}
where $\bz$ is the set of covariates. MLR is widely used for calibrating low-cost air pollution sensors \citep{bigi2018performance,bi2020incorporating,ardon2020measurements,si2020evaluation,barkjohn2021development,romero2020development}. The choice of covariates depends on the network design and type of pollutant and will typically include meteorological variables, daily, weekly or seasonal periodicity variables, time since installation, land-use variables, etc. MLR is also a special case of the gain-offset model (\ref{eq:gainoffset}) with a constant gain $\beta_1$ and the offset $\beta_0+\boldsymbol{\beta_2}'\bz\st$ being a linear model of the covariates. 

We consider the gain-offset model for regression-calibration in its most general form (\ref{eq:gainoffset}) as it subsumes all the aforementioned special cases and allows to also model both the gain and the offset as functions of covariates. Rewriting (\ref{eq:gainoffset}) as 
\begin{align}
E(x\st)&=\beta_0+\beta_1 y\st+\boldsymbol{\beta_2}'\bz\st+\boldsymbol{\beta_3}'\bz\st y\st
\label{reg-cal}
\end{align}
we note that it corresponds to a regression model which allows for interaction of the low-cost data $y$ with each covariate in $\bz$. The model coefficients $\beta_0,\beta_1,\bbeta_2,\bbeta_3$ 
can be fit using least squares on the data from the collocated sites (Set $A$ in Figure \ref{schematic}) and predictions can be made across the entire network (Set $A \cup B$). 

This regression-calibration model suffers from two major limitations: (a) the predictions $\hat{x}$ underestimate $x$ when $x$ is large, and (b) the model does not consider the spatial structure of air pollution. We illustrate the underestimation issue here and discuss (b) in Section \ref{sec:kf}.

\begin{theorem}\label{th:under_berkson}
Consider a data generation process relating low-cost pollutant measurements and the true pollutant values given by Equation (\ref{reg-cal}) and with i.i.d. errors $\epsilon$. Assume that the covariates and low-cost measurements are bounded  and that $Var(\epsilon)<\infty$. Then the bias $\hat{x}-x$ of the predictions $\hat{x}$ from regression-calibration is asymptotically negatively correlated with the true pollutant concentration $x$. 
\end{theorem}

The proof is given in Appendix \ref{proof1}. Note that the regression-calibration predictions $\hat x_i$ of the reference concentrations $x_i$ are ordinary least squares estimates. Hence, irrespective of the data generation mechanism, these predictions $\hat x_i$ satisfy the identity $\sum_{i=1}^n (\hat x_i - x_i) =0$. The residuals $e_i=\hat x_i - x_i$ are thus centered around zero implying that there are both positive and negative residuals. This combined with the negative correlation between $e_i$ and $x_i$, as guaranteed from Proposition \ref{th:under_berkson}, allows us to conclude that the $e_i$'s are predominantly negative for large $x_i$'s. 
Thus, for higher concentrations of the true pollutant $x$, the bias $\hat x - x$ will be negative and thus the regression-calibration  estimate $\hat x$ tends to underestimate the pollution concentration $x$, as observed in Figure \ref{ts_dec2019} and Table \ref{fpr_dec2019}. 
Proposition \ref{th:under_berkson} is a general result for linear models and shows that, even under the assumption of that model (\ref{eq:gainoffset}) is  correctly specified, the regression-calibration is inherently flawed for accurately capturing peak concentrations. The linear assumption is on the parameters and not on the functional form of the covariates, and hence the result can be valid even for some  non-linear regression functions (e.g., the gain-offset model (\ref{eq:gainoffset}) includes interactions between the low-cost data and the covariates).

\subsection{Inverse regression model}\label{sec:inverse}
To address the underestimation issue, we consider an inverse model, where the true pollutant concentration is used as an independent variable and the response is the low-cost sensor pollutant reading: 
\begin{align}
\label{obsmodel}
\begin{split}
y\st&=\beta_0+\beta_1x\st+\boldsymbol{\beta_2}'\bz\st+\boldsymbol{\beta_3}'\bz\st x\st +\epsilon\st
\end{split}
\end{align} 
This is equivalent to the inverse gain-offset model 
\begin{equation}\label{eq:obsmodel}
E(y(\bs,t)) =o(\bs,t) + g(\bs,t) x(\bs,t).
\end{equation}
with linear offset $\beta_0 + \bbeta_2'\bz\st$ and gain $\beta_1 + \bbeta_3'\bz\st$.

We argue that the inverse model is more organic and interpretable. The gains and offsets can be viewed as biases of the low-cost data from the true pollutant concentrations. The forward (\ref{eq:gainoffset}) and inverse (\ref{eq:obsmodel}) gain-offset models correspond to, respectively, the Berkson and classical measurement errors \citep{Fuller87} for the low-cost data .
In air pollution modeling, the Berkson error model is suitable when the observed values are spatial aggregations over geographical areas \citep{zeger2000exposure}. However, in the case of collocated calibration, the low-cost data and reference data are at the same sites, so there is no such geographical aggregation involved. The classical error model (\ref{eq:obsmodel}), which assumes that the observed low-cost measurements are more noisy than the underlying true pollutant concentrations at the same location, is a more appropriate representation. 

We first show that the underestimation issue of regression-calibration persists even under the classical error model for the data generation process. 

\begin{theorem}\label{th:under_classical}
Consider the classical error model  (\ref{obsmodel}) for the low-cost data, with i.i.d. errors and no other covariates. Then the bias $\hat{x}-x$ of the predictions from a regression-calibration model (\ref{reg-cal-no-cov}) fitted to this data is asymptotically negatively correlated with the true pollutant concentration $x$. 
\end{theorem}

The proof is included in Appendix \ref{proof2}. Propositions \ref{th:under_berkson} and \ref{th:under_classical} prove that under both models of measurement error for the low-cost data, the regression-calibration residuals will be negatively correlated with the true pollutant concentration, leading to underestimation when the true concentration levels are high. 

The classical measurement error is a more natural model for the low-cost data, as argued above. Hence, we propose using this inverse regression model (\ref{obsmodel}) for fitting the low-cost data. Once the model is fit, for a given value $y$ of the low-cost data, we can predict the true pollutant concentrations by simply inverting the regression equation: 
\begin{align}\label{eq:pred}
	\begin{split}
	\hat{x}&=\frac{y-\bo-\widehat{\boldsymbol{\beta_2}}'\bz}{\bl+\widehat{\boldsymbol{\beta_3}}'\bz}
	\end{split}
\end{align}
We show that the inverse regression model does not suffer from the underestimation issue. 



\begin{theorem}\label{th:inverse}
Consider the classical error data generation process (\ref{obsmodel}) for the low-cost pollutant measurements given the true pollutant values, and assume i.i.d. errors. Also, assume that the variables $x$ and $\bz$ are bounded,  $\epsilon\sim N(0,\tau^2)$, $\beta_1 + \bbeta_3' z$ are bounded away from zero (i.e., there exists some $a >0$ such that $P(|\beta_1 + \bbeta_3' \bz| > a) =1$), and that $\bX'\bX/n$ converges in probability to a positive definite matrix, where $\bX$ is the matrix of independent variables. Then the bias of the predictions (\ref{eq:pred}) from the inverse regression model 
is asymptotically uncorrelated with the true pollutant concentrations. 
\end{theorem}

The proof is in Appendix \ref{proof3}. Thus, if the inverse model is fit, the residuals are uncorrelated with the value of the true pollutant concentration and predictions will not suffer from the underestimation issue when the concentration is high. 
The assumption of $\beta_1 + \bbeta_3' z$ lying away from zero is necessary since for covariates values lying on or near this hyperplane,  (\ref{eq:pred}) involves division by a near-zero quantity that can result in some predictions of unrealistically high magnitude. 
In practice, however, this assumption may be nearly violated, and in finite samples there can be instability in the predictions from this inverse model. In the next Section we discuss how to mitigate this issue in practice via augmenting the inverse-regression model with a second-stage spatial model for the true pollutant surface to complete a spatial filtering algorithm. The spatial model will essentially enforce shrinkage over space towards concurrent reference data, and stabilize the predictions.  

\subsection{Gaussian Process Filtering}\label{sec:kf}
The second major limitation of regression-calibration is that the approach does not leverage spatial correlation in the air pollutant concentrations. 
Subsequent to training the regression model on the collocated data, the model calibrates the low-cost data at each site in the network independently. Leveraging the spatial structure in the low-cost data across the network sites can potentially improve quality of the calibration 

A related issue is the static nature of the calibration equation owing to not using concurrent reference data available. To elaborate, for estimating the regression-calibration model, of the reference instruments in Sets A and C (Figure \ref{schematic}), only the data from Set A (the collocated sites) is used for a fixed training window $\calW$. When calibrating the low-cost network data for a subsequent time $t$, data from the reference sites in set A or C will usually be available for that time. This data is not utilized in regression-calibration, despite carrying valuable information. 
The low-cost data at collocated or nearby network sites are likely to be correlated with this concurrent reference data and a dynamic calibration approach leveraging this information will better capture true pollutant concentrations. 

To address the two issues, we extend the inverse regression model to a novel spatial filtering approach that accommodates both types of spatial correlation --- among the low-cost data at different locations (sites A and B), and between the low-cost data and the reference data  (sites A and C). 
We consider a two stage model. The first stage is the inverse regression model (\ref{obsmodel}).
Unlike the regression-calibration, the inverse regression has the true pollutant concentration $x$ as the independent variable. This allows a 
second-stage geospatial model for $x$ to capture the spatial correlation in true pollution concentrations. 

We propose a second-stage Gaussian process (GP) model for the pollutant concentrations $\bx_t(\cdot) \sim GP(\mu_t,C_t)$ where $\bx_t=\{x(\bs,t) : s \in \calD\}$ is the pollutant surface over the spatial domain $\calD$ at time $t$, $\mu_t$ is the surface mean, and $C_t$ is the GP covariance function such that $C_t(\bs_i,\bs_j) = Cov(x(\bs_i,t),x(\bs_j,t))$. 
GPs are widely used to model smooth spatial surfaces owing to the convenient representation of finite GP realizations as multivariate normal distributions which facilitates predictions at new locations (kriging) via simple conditional normal distributions.
The mean function $\mu_t$ can be modeled using covariates if there are sufficient reference sites (sites A and C). Otherwise, it can simply be modeled as a time-specific constant, as we do here. Any valid family of covariance function can be used for $C_t$, including but not limited to the exponential, Matérn, and squared exponential families. 
We assume temporarily that all parameters are known. This includes the inverse regression model coefficients $\boldsymbol{\beta}$ and error variance $\tau^2$, as well as the mean $\mu_t$ and the parameters of the covariance function $C_t$ for the GP. We will discuss estimation of these parameters in Section \ref{sec:implement}.

Let $\bS_B$ be the coordinates of the $n$ non-collocated low-cost sites in Set B for which $y\st$ is observed but $x\st$ is not, and $\bS_{A\cup C}$ be the coordinates of the $p$ sites in Sets A and C where $x\st$ is known. Our goal is to infer on the true pollutant concentrations $\bx\slt$ in $\bS_B$ based on all available knowledge at time $t$, i.e., the true pollutant concentrations $\bx\sot$ at $\bS_{A\cup C}$ and the low-cost data $\by\slt$ at $\bS_B$. 
At time point $t$, the GP model implies the following conditional distribution for $\bx\slt$ 
\begin{align}
\label{kriging}
\begin{split}
\bx\slt|\bx\sot&\sim N(\tilde\bmu_t,\bSigma_t)\\
\tilde\bmu_t&=\mu_{t}\bones+\bC_{t,B,{A\cup C}}\bC_{t,{A\cup C},{A\cup C}}^{-1}(\bx\sot-\mu_{t}\bones)\\
\bSigma_t&=\bC_{t,B,B}-\bC_{t,B,{A\cup C}}\bC_{t,{A\cup C},{A\cup C}}^{-1}\bC_{t,{A\cup C},B}
\end{split} 
\end{align}
where $\bC_{t,i,j}=Cov(x(\bs_i,t),x(\bs_j,t))$. 

Equation (\ref{kriging}) differs from the common geospatial models where the entire $x(\bs,t)$ surface is latent and an  unconditional GP prior is used. In the setting of low-cost networks, the latent surface of true pollutant concentrations is partially observed at the reference sites $A \cup C$. Hence, (\ref{kriging}) is a conditional GP prior for the unobserved part of the surface given the available knowledge of the surface from  realizations at $A\cup C$. 

Equations (\ref{obsmodel}) and (\ref{kriging}) complete the specification of our spatial filter to obtain predictions of true pollutant concentrations based on all available low-cost and reference data. Our two-stage model can be perceived as a spatial analog 
of Kalman-filtering \citep{kalman1960new}. In Kalman-filters or other filtering approaches over time, a stochastic process is observed at one or few time points which dictates the evolution at a future time. This temporal evolution, based on partial realization of the stochastic process, is used to filter noisy observations at future time points. In low-cost networks, at each time point $t$, the stochastic process (pollutant surface) is over space. The low-cost data at the network sites $B$ are the noisy observations, and the reference data at the small set of locations $A \cup C$  are the partial realizations of the spatial process that informs about the true pollutant concentrations at the other locations owing to the spatial correlation in pollutant concentrations. Thus, equation (\ref{kriging}) is the {\em state-transition model} dictating the spatial evolution of the partially observed surface $x(\bs,t)$, while Equation (\ref{obsmodel}) is the  {\em observation model} for the noisy low-cost data $y(s,t)$. Together, these equations form a filtering setup for calibration and smoothing of low-cost networks, where the quantity of interest is $\bx\slt|\by\slt,\bx\sot$. 

To predict $\bx\slt$, we first write the observation model (\ref{obsmodel}) for a vector $\by\slt$ of $n$ low-cost observations. Let $\bZ\slt$ be the $n\times n_{cov}$ matrix of covariates at time $t$ and $\tau^2$ be the variance of the normally distributed errors $\epsilon_i$. We then have the observation model
\begin{align}\label{eq:obsmodelvec}
\begin{split}
    \by\slt&\sim N\left(\beta_0\bones+\bZ\slt\boldsymbol{\beta_2}+\left(\beta_1\bI+diag(\bZ\slt\boldsymbol{\beta_3})\right) \bx\slt,\tau^2\bI\right).
\end{split}
\end{align}

Note that during estimation of the observation model based on collocated data for a fixed time window $\calW$, the true pollution-level $\bx$ is known at the collocation sites A and the unknown quantities in (\ref{eq:obsmodelvec}) are the parameters $\beta_i$'s and $\tau^2$. However, at the filtering stage at a later time $t$, the pre-estimated parameters $\beta_i$'s and $\tau^2$ are known and the unknowns in  (\ref{eq:obsmodelvec}) are the true pollutant concentrations $\bx(\bS_B,t)$ at the current time $t$. 

Letting $\bH\slt=\beta_1\bI+diag(\bZ\slt\boldsymbol{\beta_3})$, we can rewrite the observation model as 
\begin{align*}
    \by\slt&\sim N\left(\beta_0\bones+\bZ\slt\boldsymbol{\beta_2}+\bH\slt \bx\slt,\tau^2\bI\right)
\end{align*}
    
Thus the observation model is now a linear model in the unknowns $\bx(\bS_B,t)$. We transform the observations $y\st$ to $u\st=y\st-\beta_0-\boldsymbol{\beta_2}'\bz\st$ and get 
\begin{align}\label{transformed obs}
    \bu\slt&=\by\slt-\beta_0\bones-\bZ\slt\boldsymbol{\beta_2}\sim N\left(\bH\slt \bx\slt,\tau^2\bI\right)
\end{align}

\begin{figure}
    \centering
    \includegraphics[scale=0.2]{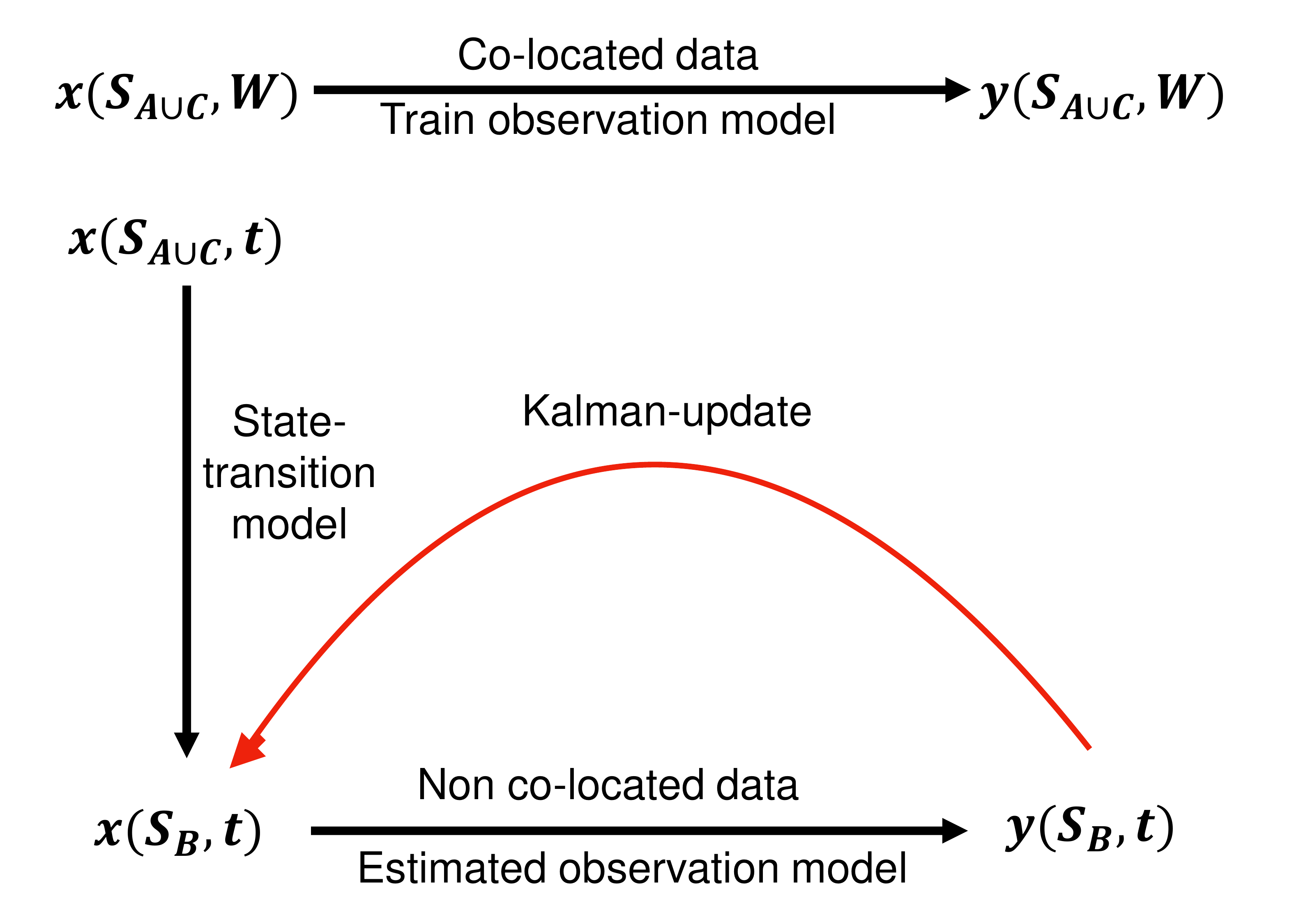}
    \caption{Schematic of the spatial filtering approach}
    \label{fig:kalman}
\end{figure}

The transformed observations $\bu$ can be considered as the measurements in a Kalman filter model, with the observation model defined by $\bH\slt$. The Kalman filter equations can be fit to the two stage model given by (\ref{kriging}) and (\ref{transformed obs}) to get
\begin{align}\label{kalman}
\begin{split}
&\hat{\bx}\slt=\tilde\mu_t\quad \text{(Predict)}\\
&\bx_{update}\slt=\\
&\quad\left(\bSigma_t^{-1}+\frac{1}{\tau^2}\bH\slt^2\right)^{-1}\left(\bSigma_t^{-1}\hat{\bx}\slt+\frac{1}{\tau^2}\bH\slt \bu\slt\right)\quad \text{(Update)}
\end{split} 
\end{align}

where $\tilde\bmu_t$ and $\bSigma_t$ are defined as in (\ref{kriging}). 
The schematic in Figure \ref{fig:kalman} summarizes the entire process. Our approach jointly predicts the pollutant concentration $\bx\slt$ at all the non-collocated sites given the known true pollutant values $\bx\sot$ at $\bS_{A\cup C}$ and the observed low-cost data $\by\slt$ at $\bS_B$.  
An  initial update (`predict step') of $\bx\slt \;|\; \bx\sot$ is dictated by the state-transition model, i.e., the conditional GP distribution in (\ref{kriging}). The  final update of $\bx\slt$ given the low-cost data $\by\slt$ is analogous to the Kalman update step, and gives the network-wide calibrated and smoothed estimate of the pollutant surface. The filtering relies on pre-estimation of the observation model and the spatial parameters which will be discussed in Section \ref{sec:implement}. 

Unlike regression-calibration, the calibration equation (\ref{kalman}) from spatial filtering is dynamic in nature. 
This is evident from the Kalman-update in  (\ref{kalman}) which becomes a weighted sum of the kriging prediction (\ref{kriging}) based on current reference data and the measurement from the inverse regression model (\ref{obsmodel}), with the weights $\bSigma_t$ and $\bH_t$ being time-specific and estimated from the data. Thus the calibration is informed by the current concentrations of true pollutants at the reference sites. 

The conditional GP model, using all available reference data, simultaneously incorporates the spatial correlation between the low-cost data and the reference data (via the kriging prediction $\tilde \bmu_t$) and the correlation among the low-cost sites (via the kriging covariance $\bSigma_t$). This effectuates a spatially smooth estimate of $\bx\slt$ unlike regression-calibration which treats data from each site independently. Also, leveraging of the spatial information is essentially a spatial shrinkage method that mitigates the instability issue of the naive predictions from the inverse regression (\ref{eq:pred}). Unstable predictions correspond to low-precision (near-zero diagonal entries in the $\bH$ matrix) and will be naturally down-weighted in (\ref{kalman}). 



We note that in our approach, the filtering is over space at each time point, unlike most Kalman filter applications for spatio-temporal air pollution data that filter over time. We discuss this difference in more details in Discussion. 

\subsubsection{Predicting on a grid of locations}\label{sec:grid}
The spatial filtering approach offers a coherent way to obtain joint predictions of the true pollutant concentrations at any arbitrary set of locations with neither reference nor low-cost sensors (Set D in Figure \ref{schematic})  to create smooth maps of the pollutant concentrations in the area. 

The joint posterior likelihood of the unknown true pollutant concentrations conditional on the observed data can be rewritten as: 
\begin{align}\label{eq:pred_grid_posterior}
\begin{split}
&p\left(\bx(\bS_B),\bx(\bS_D) | \bx(\bS_A),\bx(\bS_C), \by(\bS_B) \right) \\
&= p\left(\bx(\bS_D) | \bx(\bS_A),\bx(\bS_B),\bx(\bS_C), \by(\bS_B) \right) \times p\left(\bx(\bS_B) | \bx(\bS_A),\bx(\bS_C), \by(\bS_B) \right) \\
&= p\left(\bx(\bS_D) | \bx(\bS_A),\bx(\bS_B),\bx(\bS_C) \right) \times p\left(\bx(\bS_B) | \bx(\bS_A),\bx(\bS_C), \by(\bS_B) \right) 
\end{split}
\end{align}
where the time-index $t$ is omitted and the final equality comes from the fact that $\by(\bS_B)$ and $\bx(\bS_D)$ are independent conditional on $\bx(\bS_B)$ as there are no observations in Set D. In this expression, the first term is a conditional normal distribution that can be obtained from kriging, since $x_t(\cdot)$ is a Gaussian Process. The second term is the posterior normal distribution already available from the update step of the previous section (Equation (\ref{kalman})). This facilitates straightforward prediction of $\bx(\bS_D)$ conditional on the observed data. We provide the details in the Supplement Section \ref{sec:supp_grid}. 

\subsection{Implementation}\label{sec:implement}

The filtering update in Equation (\ref{kalman}) assumes that the parameters of the observation model and the state-transition model are known. In practice, these parameters are unknown and will need to be estimated in addition to inferring the true values $\bx\slt$. The parameters $\boldsymbol{\beta_i}$'s and $\tau^2$ of the observation model  can be estimated over a training period $\calW$ where both $\bx(\bS_A,t)$ and $\by(\bS_A,t)$ are measured at the collocated sites $\bS_{A}$. Standard least squares optimization can be used to estimate the coefficients and the  observation model variance. Since there is typically abundant collocated data consisting of hourly collocated time-series for several weeks to months, these parameters can be estimated with high precision and the estimates can be plugged into the filtering updates. 

The parameters of the GP model, $\mu_t$ 
and all parameters in the covariance function $C_t$ 
are allowed to be time-varying to capture dynamic spatial correlation in the air pollution surface. They need to be estimated at each time point $t$ using all available data $\bx(\bS_{A\cup C},t)$ and $\by(\bS_B,t)$ for the time point. As the total number of sites with either a reference or a low-cost sensor will be relatively small or moderate, these estimates may have non-negligible variability and the Kalman-updates may be sensitive on the decision to propagate or not propagate this parameter uncertainty. We explore the impact of this choice by offering both a frequentist and a Bayesian implementation of the filtering method. 




\subsubsection{Frequentist}\label{sec:freq}

Under the frequentist implementation of the filtering approach, the observation model can be used to predict an initial value of the true $x$ from the observed $y$ using Equation (\ref{eq:pred}). 
We then have $\hat \bx_{init}\approx GP(\mu_t,C_t)$, 
where at the reference sites (A and C) $\hat x_{init}$ denotes the observed true $x$ and at the low-cost network sites (B) $\hat x_{init}$ denotes the initial predicted value of $x$ from Equation (\ref{eq:pred}). The likelihood for $\hat \bx_{init}\approx GP(\mu_t,C_t)$ Maximum likelihood Estimates for the spatial parameters can then be obtained by maximizing the GP likelihood with $\hat x_{init}$. 
We use the SpatialTools package \citep{spatialtools} for this optimization. 
Once the estimates 
for the mean $\mu_t$ and any spatial parameters in $C_t$
are obtained, Equation (\ref{kalman}) can be used to estimate $\hat \bx\slt$ on Set B with the estimated covariance matrix and observation model, and Equation (\ref{eq:predgrid}) of the Supplement can be used for predictions on the grid (Set D). 
We summarize the method in the following algorithm:

	\begin{alg}\label{alg:freq}\textup{
		\textbf{Frequentist GP Filter}
		\begin{enumerate}
			\item {\em Training observation model (\ref{obsmodel}) in the training window $\calW$ at collocated sites $A$:} 
			\begin{enumerate}
			    \item Run linear regression \texttt{lm$[y(A,\calW) \sim \bx(A,\calW) + \bz(A,\calW) + \bx(A,\calW)*\bz(A,\calW)]$} to obtain estimates of the coefficients $\boldsymbol{\beta_i}$'s for $i=0,\ldots,3$ and the variance $\tau^2$.
			\end{enumerate} 
			\item {\em Filtering at any time $t$:}
			\begin{enumerate}
				\item Make an initial prediction $\hat{\bx}_{init}\slt$ from (\ref{eq:pred}). Let $\hat \bx_t = (\bx(\bS_{A},t), \hat \bx(\bS_B,t),\bx(\bS_{C},t))$. 
				\item Obtain the GP maximum likelihood estimates of $\mu_t$ and the parameters $\theta_t$ of $C_t = C(\theta_t)$ based on the $\hat \bx_t$ from step (a) at locations $\bar D= A \cup B \cup C$:%
				$$ (\hat \mu_t, \hat \theta_t) = {\arg \max}_{( \mu_t, \theta_t)} N(\bx_t | \mu_t \bones, \bC( \theta_t)_{\bar D,\bar D}).$$ 
				\item Calculate the kriging mean $\tilde \mu_t$ and variance $\Sigma_t$ from (\ref{kriging}) using $(\hat \mu_t, \hat \theta_t)$
				\item Calculate $\bu\slt$ from $\by\slt$ using the mean in (\ref{transformed obs}). 
				\item Predict step: Calculate $\hat\bx\slt = \tilde \mu_t$  (the first equation in (\ref{kalman})).
				\item Update step: Calculate $\bx_{update}\slt$ using the second equation in (\ref{kalman}). 
				\item Grid predictions: Use Equation (\ref{eq:predgrid}) with $\hat \bx_B =\bx_{update}\slt$ to predict on Set D. 
			\end{enumerate}
		\end{enumerate}
	}
	\end{alg}
	
This method does not propagate the uncertainty of the spatial parameters into the Kalman updates. \cite{zimmerman1992mean} quantified the impact of not propagating the estimation uncertainty and established that this leads to overly confident predictions. They also offered a way to correct for this. However, the correction relies on strong assumptions on the underlying spatial covariance function and is based on asymptotics which are unlikely to be relevant for low-cost networks with  small to moderate number of locations.

\subsubsection{Bayesian}
Our spatial filtering, like  Kalman-filter or other filtering approaches, has an inherent Bayesian flavor as the update step in Figure \ref{fig:kalman} can be viewed as the posterior mean of $\bx\slt \;|\; \bx\sot,\by\slt$ given the conditional GP prior (\ref{kriging}) and the low-cost observations modeled as (\ref{eq:obsmodelvec}). Augmenting these two equations with additional priors for the remaining spatial hyper-parameters, we can use a Bayesian model to jointly estimate $\hat x$ and the GP model parameters, and propagate uncertainty in the spatial parameter estimation into the estimate of $x$. We still estimate the observation model and plug in its parameters before the Bayesian estimation. As mentioned before, these parameters will be estimated with high-precision given abundant collocated data, hence the associated uncertainty is negligible. However, if desired, these parameters can also be estimated jointly. 



The main advantage of the Bayesian formulation is that the division in (\ref{eq:pred}) is not explicitly performed unlike the frequentist implementation where it is used to obtain the initial estimate of $\bx\slt$. This means that the predictions that have highly inflated variances will not be directly used to estimate the spatial parameters, and are naturally  down-weighted in the Bayesian framework owing to incorporation of the prediction uncertainty. The main disadvantage of the Bayesian framework is that it is more computationally expensive than the frequentist implementation. We summarize the Bayesian estimation and prediction process in the following algorithm using an off-the-shelf sampler like RStan \citep{carpenter2017stan}:


	\begin{alg}\label{alg:Bayes}\textup{
			\textbf{Bayesian GP Filter}
		\begin{enumerate}
			\item {\em Training observation model (\ref{obsmodel}) in the training window $\calW$ at collocated sites $A$:} Same as Step 1 of Algorithm \ref{alg:freq}. 
			\item {\em Bayesian Filtering at any time $t$:}
				\begin{enumerate}
				\item Specify low-cost data likelihood from (\ref{eq:obsmodelvec}): 
				$$ \ell(\by) = N\left(\by\slt \,\big|\, \hat\beta_0\bones+\bZ\slt\boldsymbol{\hat\beta_2}+\left(\hat\beta_1\bI+diag(\bZ\slt\boldsymbol{\hat\beta_3})\right) \bx\slt,\tau^2\bI\right)$$
				\item Specify GP likelihood for the true pollutant surface at locations $\bar D= A \cup B \cup C$
				$$\ell(\bx) = N(\bx(\bS_{\bar D},t) \,\big|\, \mu_t \bones, \bC(\theta_t)_{\bar D,\bar D} )$$
					\item Assign priors $\pi(\mu_t,\theta_t)$ to $\mu_t$ and the parameters of $C_t$.
					\item MCMC: Using any sampler, draw  $N$ MCMC samples of $\bx\slt$, $\mu_t$, and $\theta_t$ from their joint posterior proportional to $\ell(\by) \times \ell(\bx) \times \pi(\mu_t,\theta_t)$ 
					\item Grid predictions: Draw from the posterior given in Equation (\ref{eq:jointkrig}) to predict on Set D. 
				\end{enumerate}
			\end{enumerate}}
	\end{alg}
	
\vskip -25pt \subsection{No collocation}\label{sec:nocol} Most low-cost sensor networks use some form of collocation with reference devices to estimate the biases in the low-cost data. Hence we have assumed throughout the methods development that Set A, the set of collocated sites, has at least one location. This enables training of the observation model for our GP filter and of the regression-calibration model of Section \ref{sec:reg-cal}. However, there can be exceptions to this where a low-cost air pollution network does not have exact collocation. We address the possibility of applying our method in the case where there are no collocated sites in Supplement \ref{sec:supplement_no_colocation}.

\subsection{Extension to modeling time}\label{sec:spacetime}

For the state-space model, a novelty of our approach is the filtering  in space which conditions the analysis on the concurrent information available on the true pollutant concentrations at the reference sites. We do not filter or smooth over time and, instead, estimate time-specific spatial state-space models. This is because the high-frequency low-cost data offer the opportunity to characterize ultra-short-term fluctuations of the pollutant concentrations. Filtering in time using typical temporal models can smooth out localized (in-time) peaks in concentrations, such as fireworks displays that increases concentrations drastically for just a few hours. Such localized peaks would be contrary to the expected evolution of concentrations over time and filtering in time would smooth it out.

Hence, if modeling in time is of interest, one needs to consider a sufficiently rich class of temporal models that can capture short-term peaks. Our spatial filtering framework can be easily extended to a spatio-temporal filter by considering both space and time dynamics in the state-space model. We present one such approach in Supplemental Section \ref{sec:supst}, based on suggestions from one reviewer. We also discuss there how our GP filter can also easily accommodate time structure in the observation model. 


\section{Simulation studies}\label{sec:sim}

We conduct several simulation studies to evaluate the performance of the proposed method compared to the original regression-calibration model. We consider varying degrees of underlying spatial correlation for air pollution concentrations, different low-cost and reference network designs and sample sizes, as well as various forms of model misspecification. 
For each simulation experiment, the 
$p$ collocated and $n$ non-collocated sites are sampled from the unit square, 1,000 time points of pollutant concentrations are generated to estimate the observation/regression models, and the methods are evaluated on 100 future time points. We first consider a correctly specified state-space model where at each time point, a vector $\bx_t$ of true pollution concentrations at a set of locations is generated from a GP model with an exponential covariance function $C_t(\bs_i,\bs_j)=\sigma_t^2\exp(-\phi_t||\bs_i-\bs_j||)$, whose specifications differ across simulations. Subsequently, we will also consider a misspecified state-space model, not using any GP but just a deterministic smooth surface to generate the true pollutants. The true concentrations are simulated independently across time. Then covariates are sampled from the following distributions, independently at each time point:
\begin{align*}
RH\sim U\left(24,76\right), 
T\sim U\left(17,45\right), 
weekend\sim Bern\left(\frac 27\right), 
daylight\sim Bern\left(\frac 23\right)
\end{align*}
where $U$ denotes the uniform distribution and $Bern(r)$ denotes the Bernoulli distribution. The ranges for temperature and relative humidity correspond to the respective ranges of them as measured by the SEARCH network in Baltimore in August and September of 2019. This time period corresponds to the testing period for evaluating the regression-calibration model for the Baltimore SEARCH network in \cite{datta2020statistical}. This choice of simulating the covariates is one way to provide some realistic marginal ranges for RH and T so that we can interpret them as relative humidity and temperature covariates. Note that since neither our proposed method (GP filter) nor the regression-calibration method make use of any temporal dependence between datapoints, we can simulate $\bx_t$ and the covariates independently across time. Any reordering of time points in the training data produces the same result with both methods. Given the true concentrations and covariates, the low-cost data are generated from an observation model. The coefficients for the observation model are obtained from fitting model (\ref{obsmodel}) on the Baltimore SEARCH network PM$_{2.5}$ data in August and September 2019. The datasets simulated in this way provides realistic coefficient values, which allows us to think of the simulated true concentrations and low-cost values as real world PM$_{2.5}$ measurements from reference and low-cost sensors. The observation model variance is chosen to be $\tau^2=2$. We use an observation model with covariates except when investigating the effect of covariate misspecification, i.e., of including extraneous covariates and of missing covariates.

We compare the regression-calibration model (RegCal) and the Bayesian implementation of the Gaussian Process filtering method (GP Filter). We also initially consider the standalone inverse regression model from Section \ref{sec:inverse} (Inverse). Lastly, we also consider a threshold exceedances model, the generalized Pareto model (Pareto) \citep{pickands1975statistical,holmes1999application} for calibration. This model makes predictions on data where the true concentrations is above some threshold, but do not calibrate any low-cost measurements that are 
below the threshold. We fit the Pareto model using the threshold of 12 $\mu g m^{-3}$ corresponding to the limit of an AQI classification of Good. Details of our implementation of the Pareto model are in Supplement Section \ref{sec:supplement_pareto}. To evaluate the performance of the method we use root mean square error (RMSE) and false negative rate (FNR) of AQI classification for PM$_{2.5}$ based on the latent true air pollution surface. FNR is defined as the proportion of observations where the true pollutant concentration at a site is a Moderate or Unhealthy AQI (PM$_{2.5} \geq 12$) but the prediction from the low-cost sensor data is a Good classification (PM$_{2.5}<12$). The Moderate and Unhealthy classifications are combined for the purpose of the FNR because our simulation setups, based on  the Baltimore PM$_{2.5}$ data,  have very few Unhealthy time points. $50$ replicate datasets are simulated for each simulation setting, and the results are averaged over these replicate datasets. 

\subsection{Simulation 1a: Correctly specified model}\label{sec:sim_1a}

We begin by assuming a correctly specified model with 1 collocated site and 50 non-collocated sites (similar design as the Baltimore SEARCH network). We let $\phi=\frac{3}{\sqrt{2}}$ and $\mu=7$, and $\sigma^2\in\{5,10,15,20\}$. Different values of $\sigma^2$ are used to evaluate the performance of the model with different spatial noise to random noise ratios ($\sigma^2/\tau^2$) in the low-cost data. 

\begin{figure}[!t]
	\centering
	\includegraphics[height=1.8in,trim={0 0 80 0},clip]{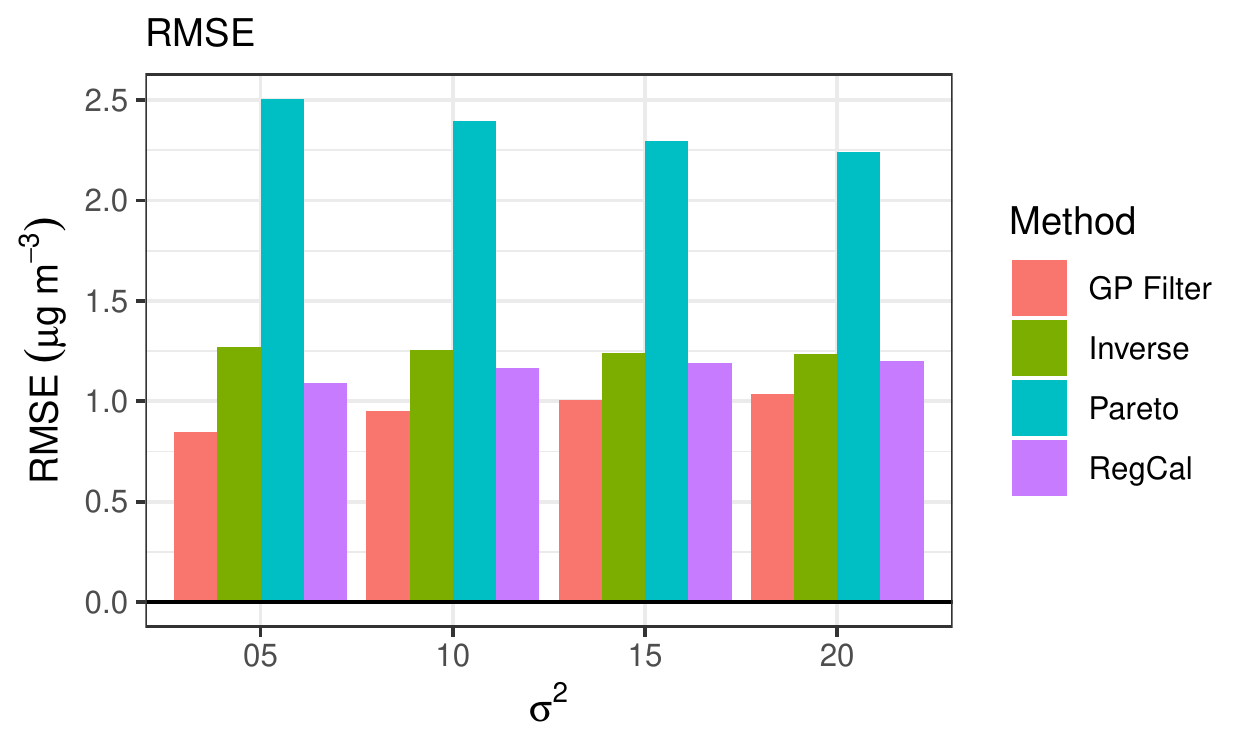}\includegraphics[height=1.8in,trim={0 0 0 0},clip]{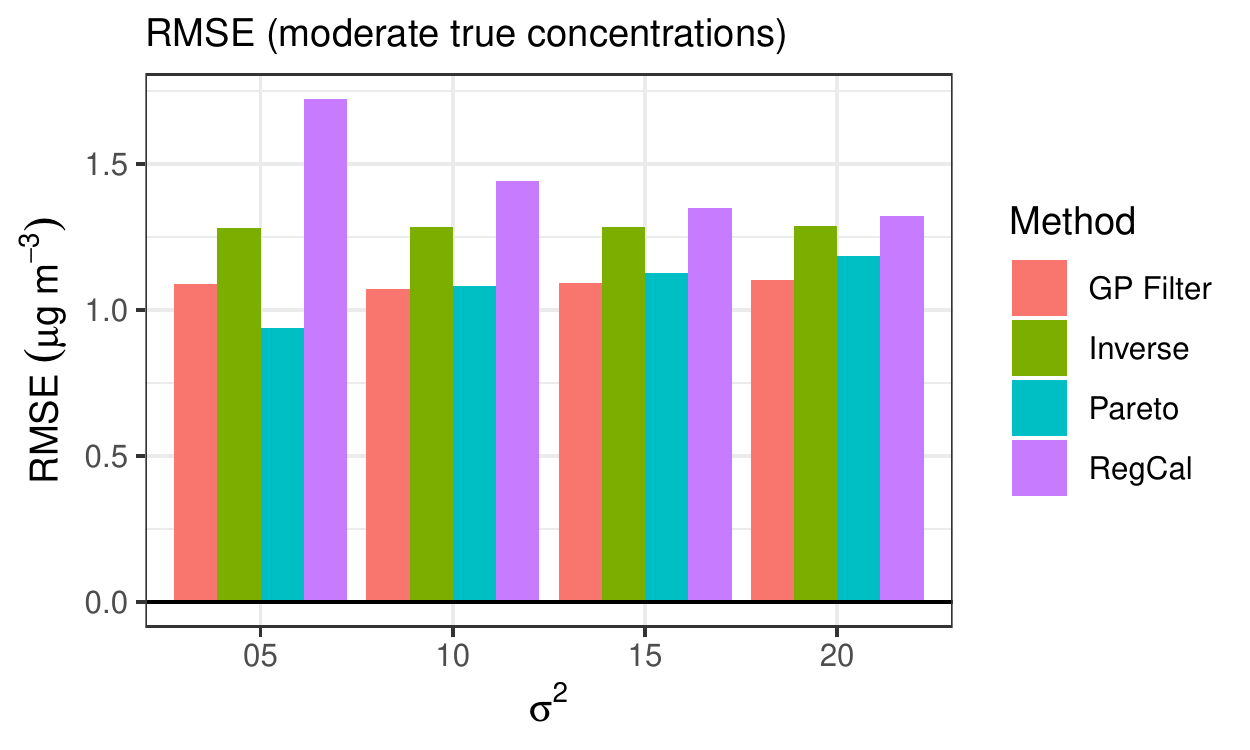}
	\caption{RMSE for setting 1a, a correctly specified model, averaged over 50 datasets. Results for four $\sigma^2$ values are shown. (Left) RMSE over all time points. (Right) RMSE over time points with moderate/unhealthy true concentrations ($X>12$). 
	}
	\label{sim_1a_rmse}
\end{figure}

Figure \ref{sim_1a_rmse} shows the RMSE (averaged over 50 datasets) of the methods compared for this setting, as well as the RMSE for only true moderate or unhealthy concentrations. For all values of $\sigma^2$, the GP Filter has $20\%$ or more lower overall RMSE than RegCal. The inverse regression model performs slightly worse than RegCal overall, while the Pareto has by far the worst overall RMSE. This is because it leaves much all of the low-cost data below the threshold unchanged which compares poorly against the reference data. As the spatial variance increases, the RMSE of both the GP Filter and RegCal methods increases. This is expected as with higher spatial variability, the predictions at each site have higher uncertainty. 
 The RMSE for Pareto regression decreases as $\sigma^2$ increases, which is because as spatial variability increases, there are more high low-cost values ($\geq 12 \mu gm^{-3}$), so a greater proportion of data are calibrated by the Pareto model. 

When the RMSE is restricted to only moderate/unhealthy observations, some new patterns emerge. The Pareto has very good RMSE on moderate observations, comparable or slightly lower than the GP Filter  and performing better than the other two methods. 
RegCal consistently has the worst RMSE on this set, illustrating the consequence of the underestimation of high concentrations, with the inverse model performing  better than RegCal. 
It is important to accurately calibrate data at both low and high levels of exposure. 
The GP filter consistently performs well in both ends of the spectrum. Since both the Pareto and the inverse regression perform very poorly overall, 
 we focus on only the RegCal and GP Filter when presenting the rest of the results. 

\begin{figure}
\raggedright
	\includegraphics[height=1.8in,trim={0 0 80 0},clip]{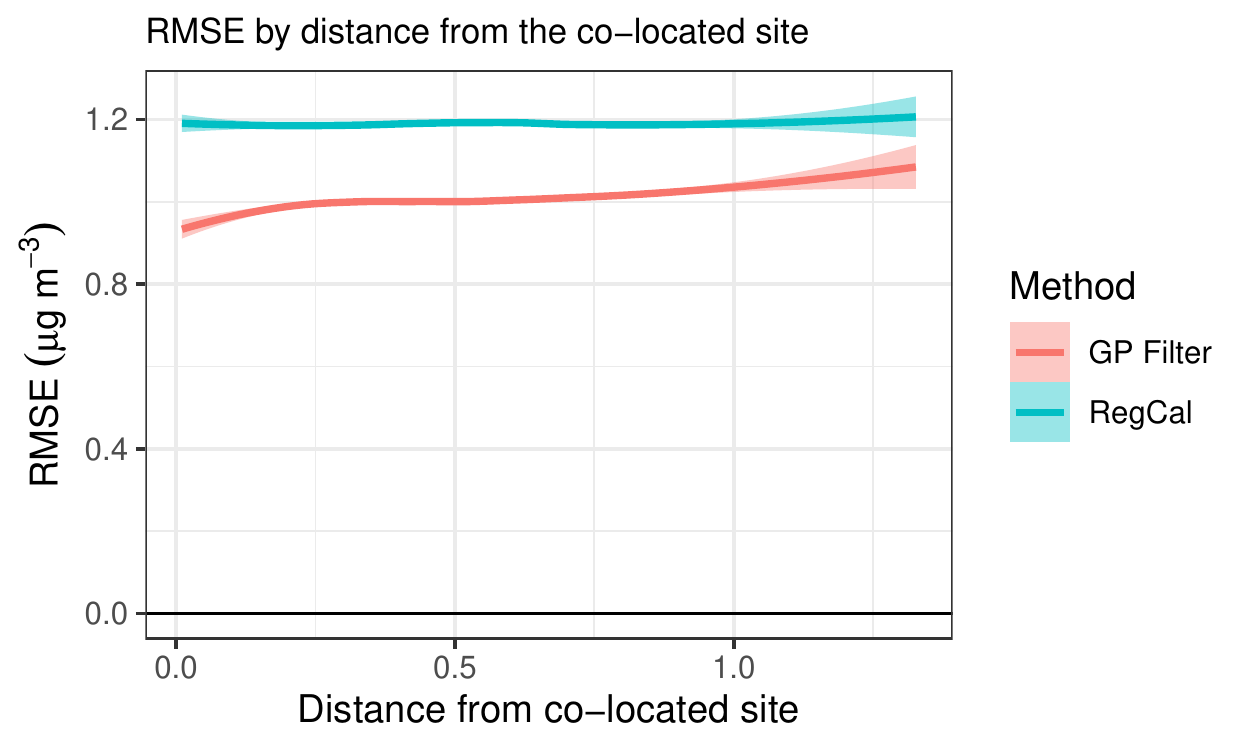}\includegraphics[height=1.8in,trim={0 0 0 0},clip]{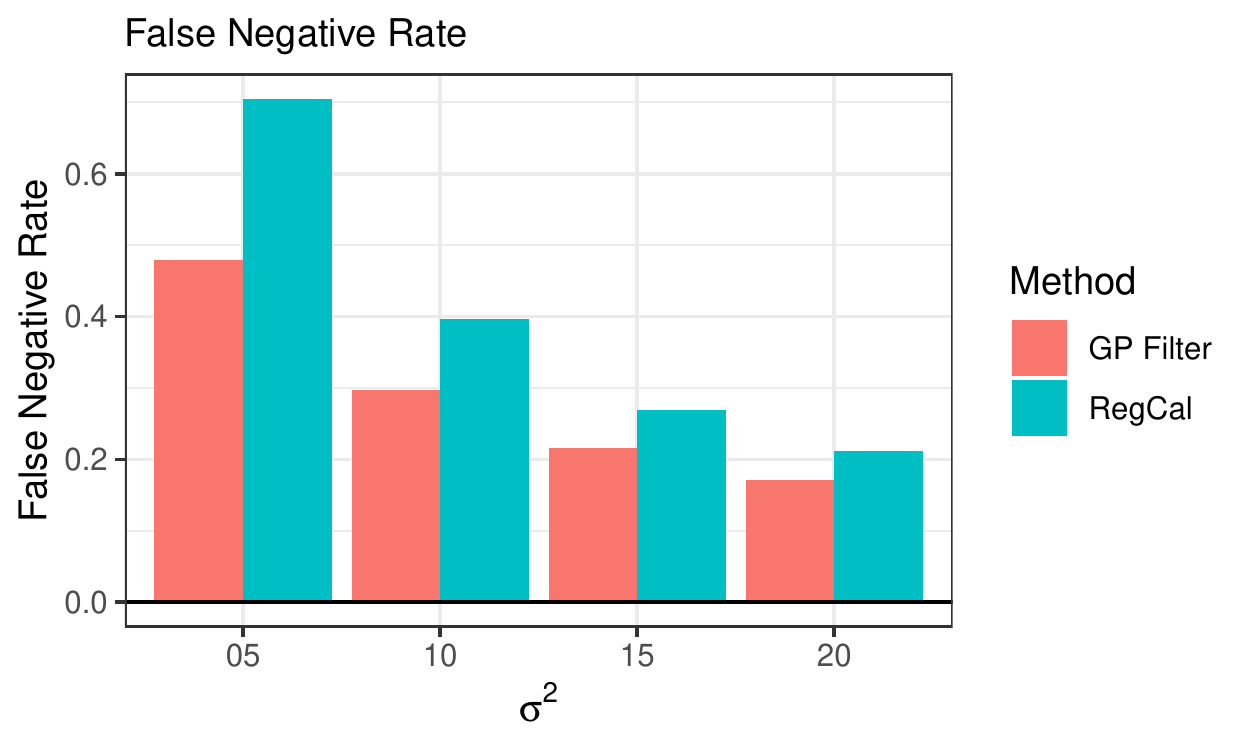}
	\includegraphics[height=1.8in,trim={0 0 80 0},clip]{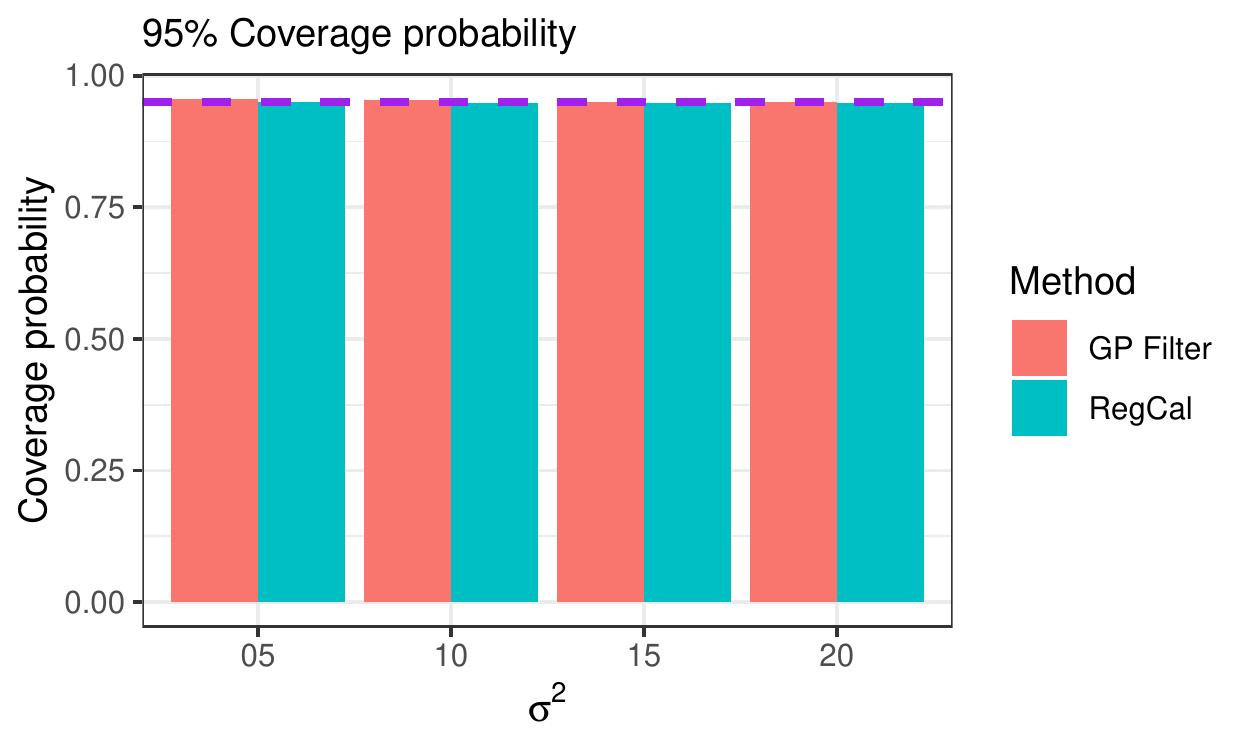}\includegraphics[height=1.8in,trim={0 0 80 0},clip]{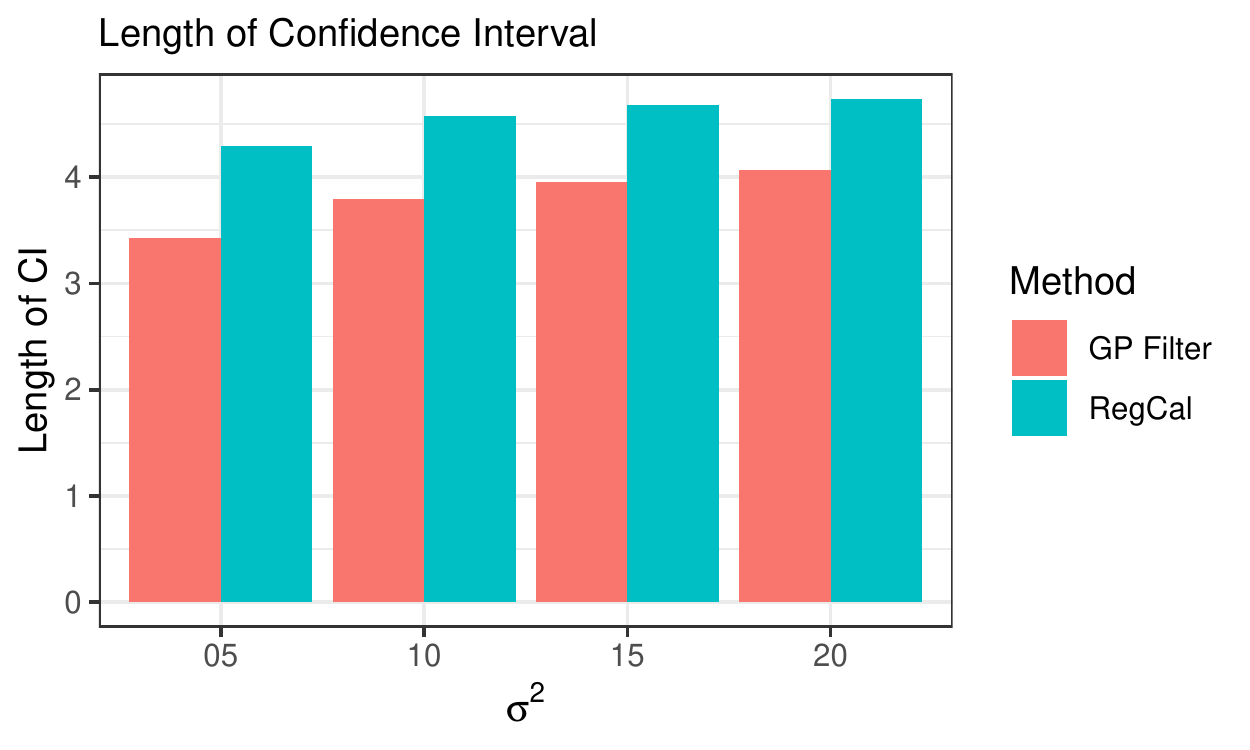}
	\caption{Results for setting 1a, averaged over 50 datasets: (Top left) RMSE by distance of site from the collocated reference site for $\sigma^2=15$. (Top Right) False negative rate. (Bottom left) Coverage probability, with the 95\% probability denoted by the dashed purple line. (Bottom right) Length of 95\% confidence interval. 
	}
	\label{sim_1a}
\end{figure}

Figure \ref{sim_1a} shows further results of the correctly specified model. The top right panel shows how the RMSE changes as a function of the distance from the reference instrument across all the datasets, for one choice of $\sigma^2(=15)$. We see that for the filtering approach the RMSE decreases for sensors closer to the reference site. This is because the filtering model, via use of the conditional GP (\ref{kriging}), accounts for the underlying spatial correlation in the true pollution surface. So the closer sites will have predicted pollutant concentrations closer to the reference value. Even at further distances, the RMSE for the filtering method is substantially lower than that for the regression-calibration. The RMSE by distance plots for all other $\sigma^2$ values are included in the supplemental materials and reveal similar trends (Figure \ref{1a_rmse_dist_all}).

The top right panel plots the FNR of inaccurately classifying moderate or unhealthy AQI days as good. Once again, the GP Filter has consistently smaller FNR than RegCal. 
The bottom left panel shows that the prediction interval estimates for both methods have coverage probabilities close to 95\%. However, the GP Filter has narrower interval widths (bottom right) than RegCal, showing that it offers improved precision for the predictions by leveraging the latest reference data.  

The GP Filter was also be implemented in a frequentist way, 
as discussed in Section \ref{sec:implement}. The performance of the two implementations are very similar, the frequentist approach is much faster (Table \ref{sim_1a_runtime}) and might be the pragmatic choice for large datasets. However, many low-cost networks typically do not have very large sample sizes so the computation time for the Bayesian model will generally be manageable. Also, for larger networks, computations for the Bayesian implementation can be expedited by replacing the GP priors by the scalable Nearest Neighbor Gaussian Processes \citep{nngp,dnngp,finley2019efficient}.


\subsection{Simulation 1b: Impact of network design}

\begin{figure}
	\centering
	\includegraphics[height=1.8in,trim={0 0 73 0},clip]{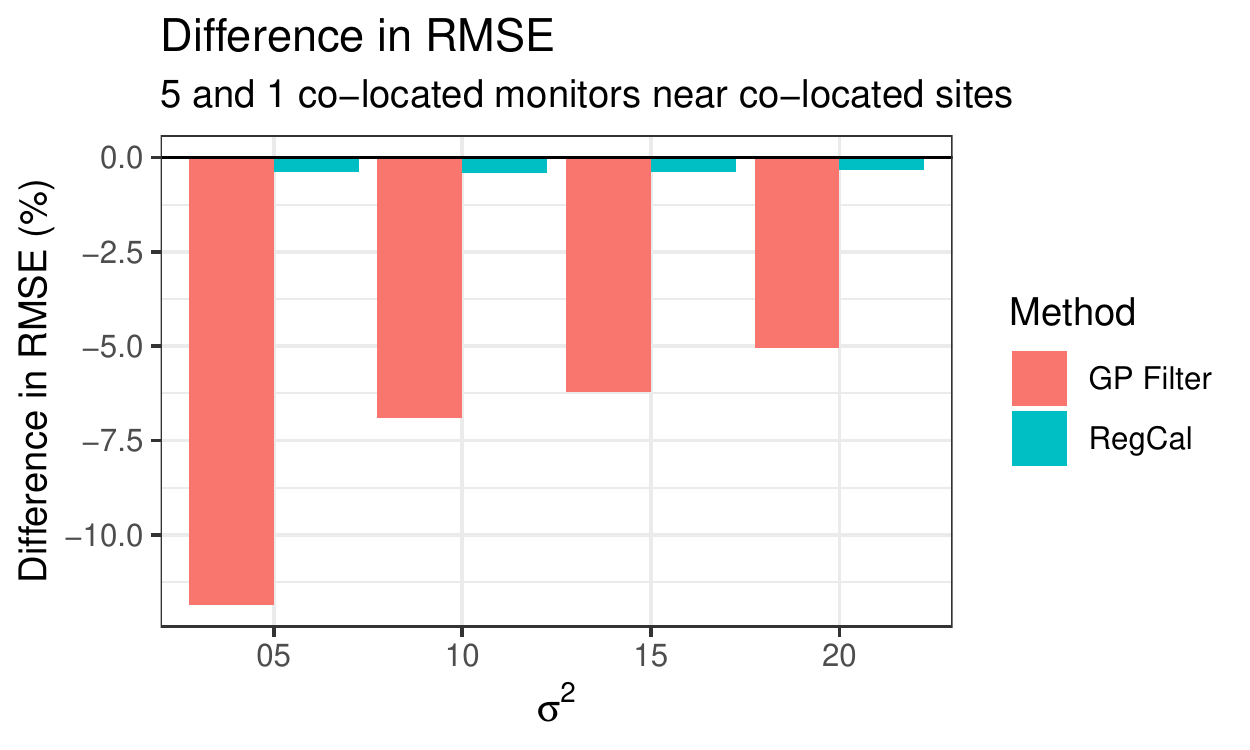}\includegraphics[height=1.8in]{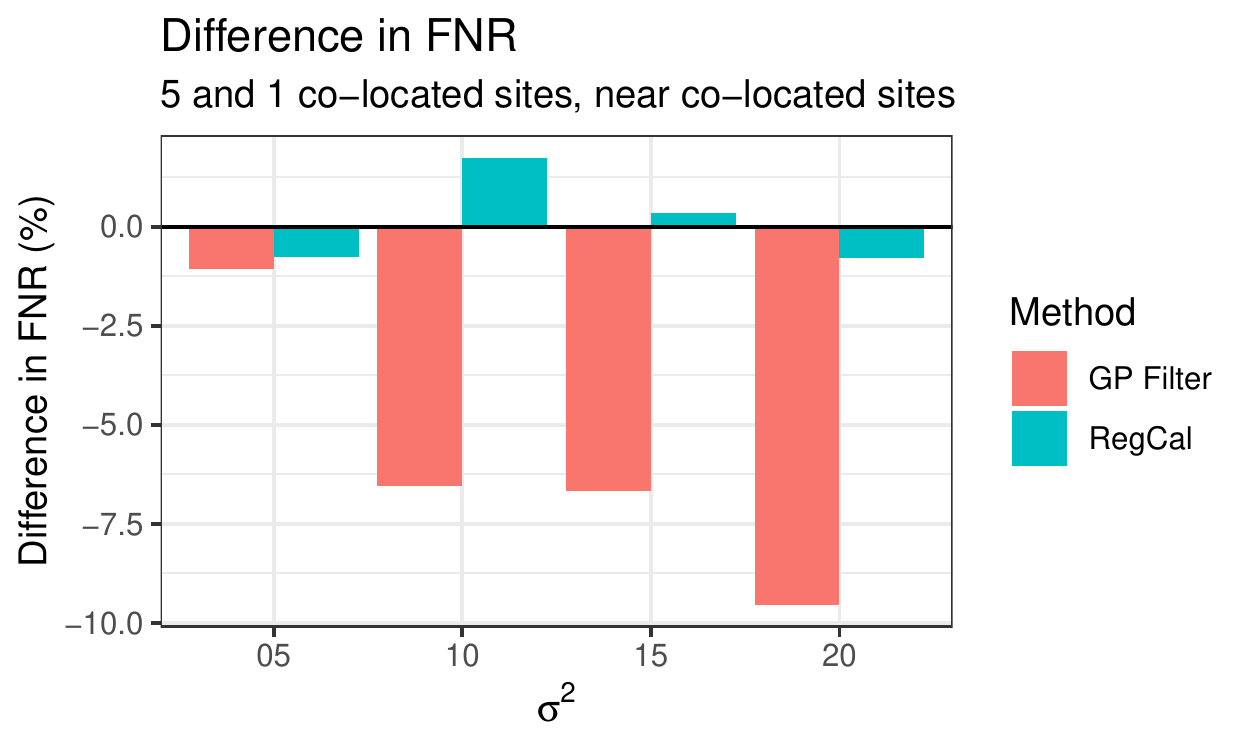}
	\includegraphics[height=1.8in,trim={0 0 73 0},clip]{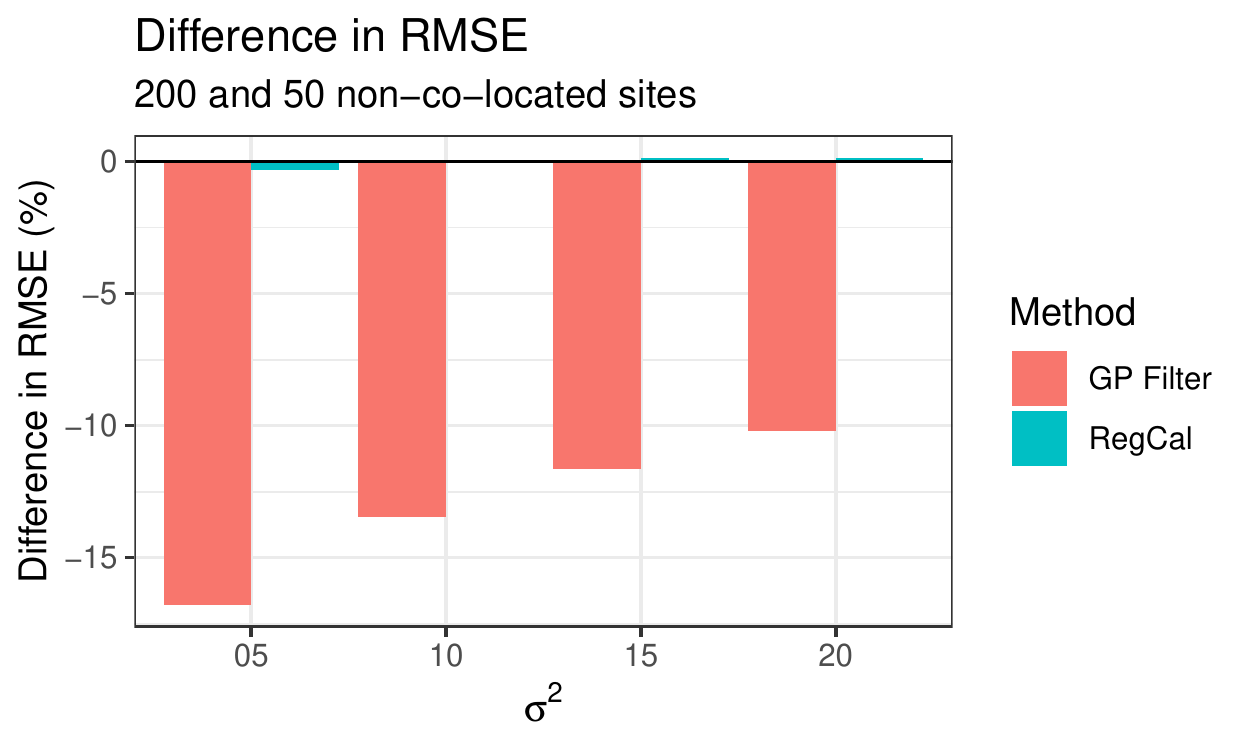}\includegraphics[height=1.8in]{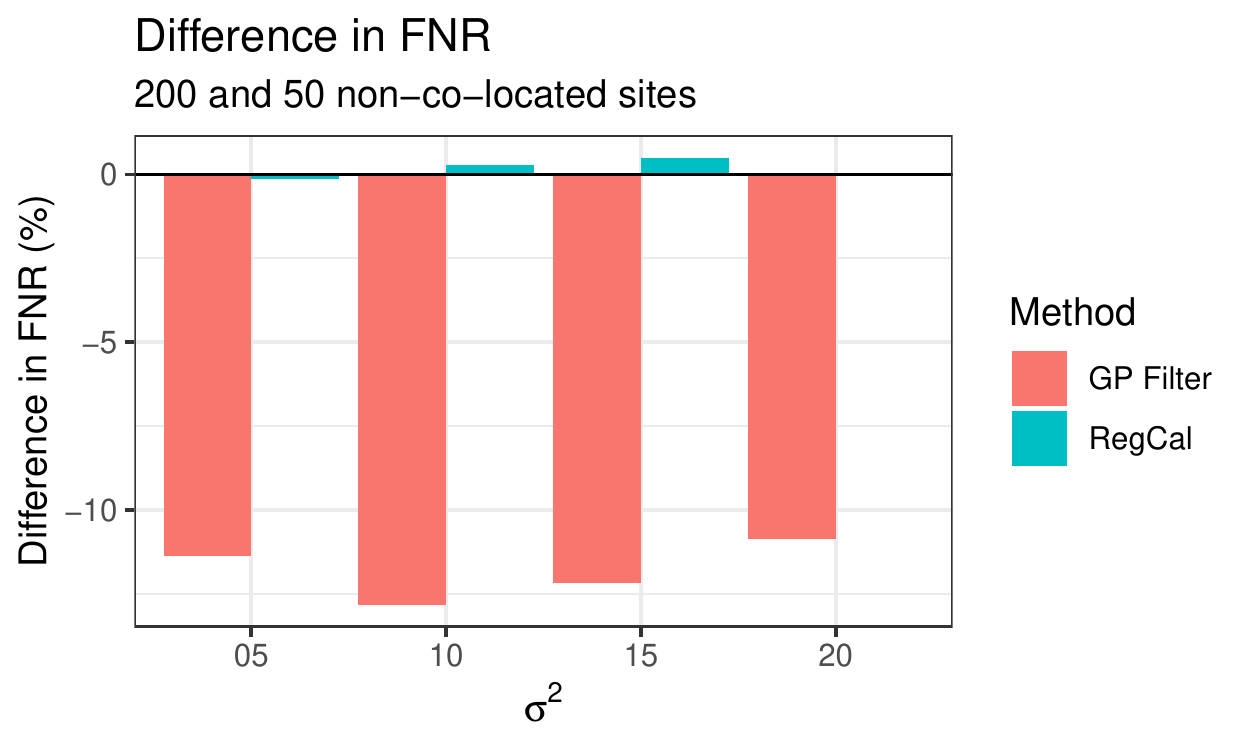}
	\caption{Results of setting 1b. (Top) Comparison of the change in performance of the methods when increasing the number of collocated sites from 1 to 5, with percent difference in RMSE in top-left and percent difference in FNR in top-right. (Bottom) Comparison  of the change in performance of the methods when increasing the number of low-cost sensors from 50 to  200 non-collocated sites with percent difference in RMSE in bottom-left and percent difference in FNR in bottom-right. 
	}
	\label{sim_1b}
\end{figure}

The previous simulation setup emulated the Baltimore SEARCH low-cost network design with around $50$ low-cost sites and only one reference site.  We conduct additional simulations with other network designs where we increase the numbers of both the low-cost sites and the reference sites. We first consider $p=5$ reference sites with collocated low-cost sensors, and keep all other settings and parameter choices the same from setting 1a. Figure \ref{sim_1b} shows the differences in model comparison metrics when increasing the number of reference sites from $1$ to $5$. We focus on the  performance on sites within 0.1 units of any additional reference site and the results are averages over 50 replicate datasets. We see that the GP filter benefits much from more additional reference data, as it uses them dynamically update the calibration via filtering. As the number of reference sites increases from $1$ to $5$, for sites near the reference sites, the RMSE of the filtering method decreases by $5-12\%$ for all setups as shown in the top left panel. The FNR of the filtering method decreases by up to 10\% in the sites around the reference sites, as shown in the top right panel. The RMSE or FNR of the regression-calibration remains roughly unchanged with increase in number of reference sites, as the method does not leverage any of the increased spatial information available from more reference data. 

We then consider a network with only 1 reference site for collocation but increase the number of non-collocated low-cost network sites from $n=50$ to $200$. 
We see that with an increase in density of deployment of the low-cost network, the RMSE of the filtering method consistently decreases by $10-18\%$ (bottom left panel). 
This is because with more total sensors, the spatial parameters are better estimated (Figure \ref{1b_parameters}). 
The FNR decreases by 10-13\% when 200 non-collocated sensors are used (bottom right panel). Once again the regression-calibration does not benefit from increased density of low-cost sensors. 

\subsection{Simulation 2: Misspecified observation model}\label{sec:misspec}

\begin{figure}[t]
	\centering
	\includegraphics[width=4.5in]{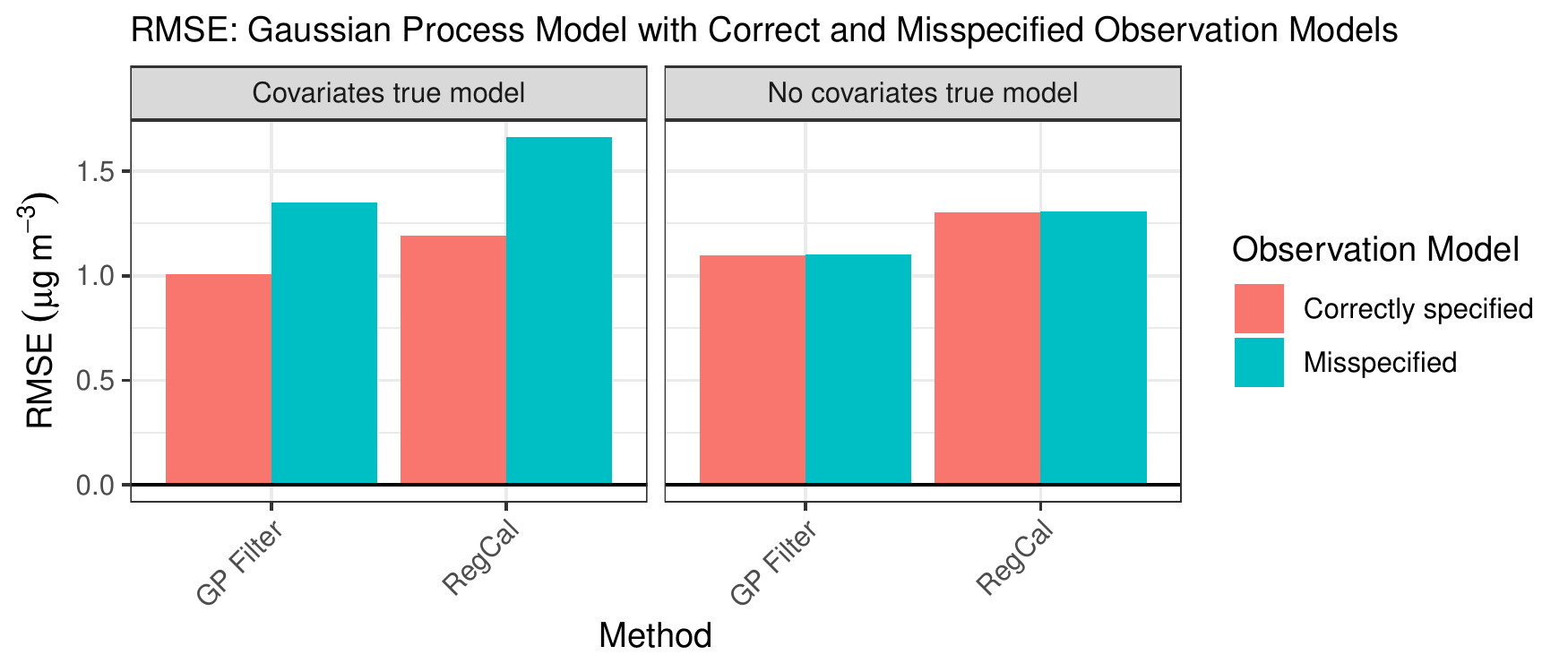}
	\caption{Comparison of RMSE of a correctly specified and a misspecified covariate set when $\sigma^2=15$, averaged over 50 datasets. True observation models with and without covariates are used.}
	\label{2_rmse_15}
\end{figure}

We now consider situations where the covariates used in 
the observation model are incorrectly specified. We consider both the case where the set of covariates is under or over specified. 
The results when $\sigma^2=15$ are shown in Figure \ref{2_rmse_15}. We see that for the setting of under-parametrization, i.e., when covariates used in the true data generation mechanism are not used in fitting, the RMSE is expectedly higher for the misspecified model than the correctly specified one. 
However, the RMSE is still lower for the misspecified GP Filter method compared to the corresponding misspecified regression-calibration. 
With over-parametrization, i.e., when there are no covariates in the true model but they are included in the model fitted, adding covariates results in only a tiny increase in RMSE. The RMSE and FNR for other $\sigma^2$ values are included in the supplemental materials (Figures \ref{2_rmse}, \ref{2_fpr}). The FNR also increases when true covariates are omitted from the observation model.

\subsection{Simulation 3: Misspecified state-transition model}\label{sec:sim_ps}

\begin{figure}[h]
	\centering
	\includegraphics[width=4.5in]{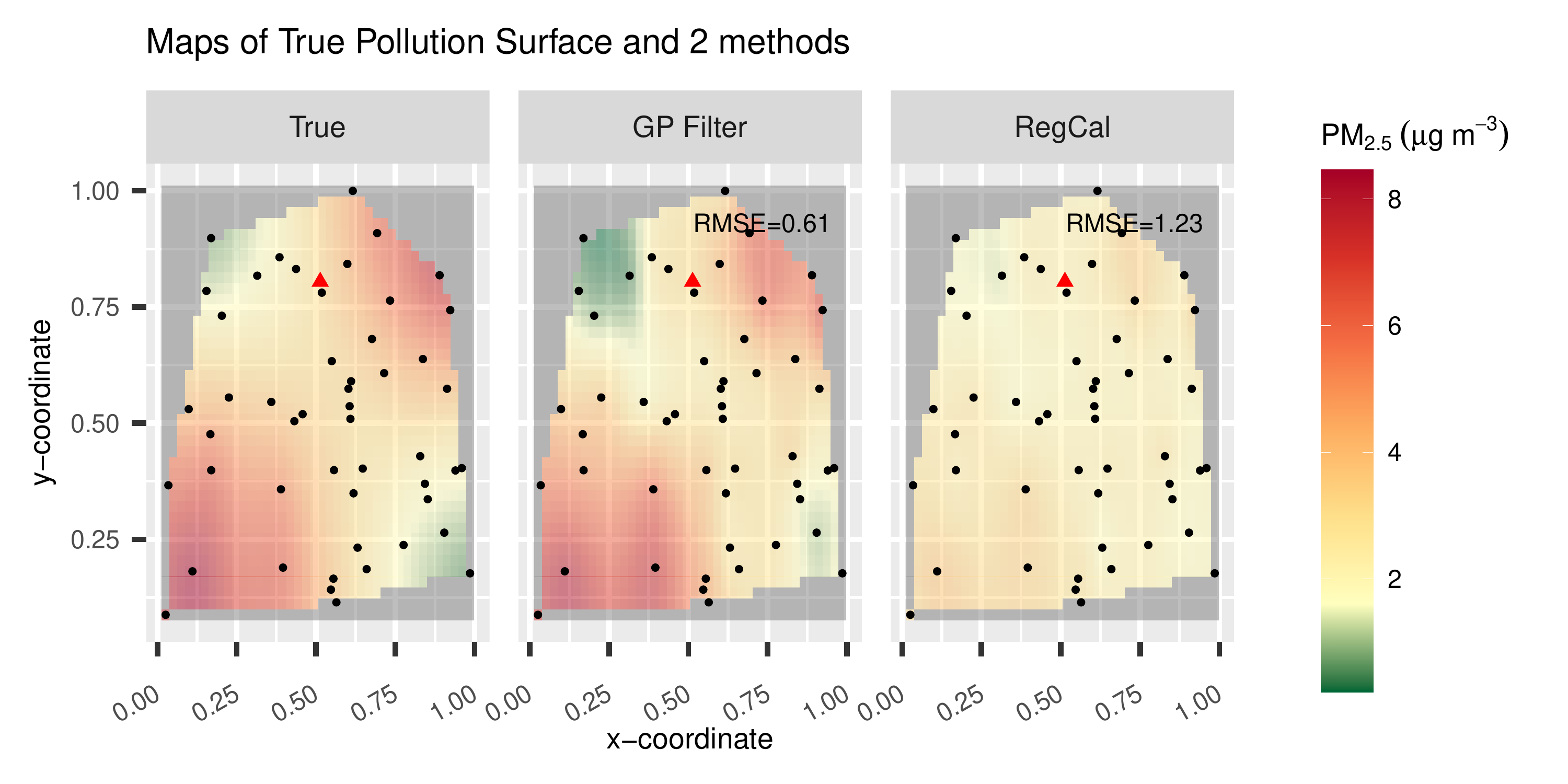}
	\caption{Comparison of the two methods for a misspecified spatial surface. The left panel is the simulated true PM$_{2.5}$ surface at one time point. The red triangle is the collocated site, and the black circles are the 50 non-collocated sites. The middle and right panels plot the predicted surfaces for that time point from the GP filter and RegCal respectively. The RMSE of each method over all the test time points is included in the respective panels.}
	\label{3_map}
\end{figure}

We now simulate data from a misspecified latent air pollution model. Instead of generating the true pollution surface from a Gaussian Process, we generate a fixed smooth spatial surface representing concentrations in an area with two sources of air pollution. The two point sources are randomly selected in the unit square at locations $\bs^*_{1}$,$\bs^*_{2}$. At each time point, the emission from each source ($p_i(t)$, $i=1,2$) are sampled from a $U(2,9)$ distribution. We then let the PM$_{2.5}$ surface be defined as a distance-based kernel weighted sum of the two emissions, i.e., $x(\bs,t)=\sum_{i=1}^2p_i(t)\exp\{||\bs-\bs_{i}^*||^2/(2\gamma)\}$, where $\gamma$ is a scale parameter for how slow the concentrations decay around the source. We use $\gamma\in\{0.1,0.4,0.7,1\}$. One example simulated pollutant surface is shown in Figure \ref{3_map} (left).  


The RMSE and FNR of these simulations are similar to the correctly specified model and are included in the Supplement (Figures \ref{3_rmse}, \ref{3_fpr}). The GP Filter performs substantially better than the regression-calibration, even when the true data is not generated using a GP. This is not surprising as GP are widely used as a non-parametric technique to estimate smooth functions or surfaces and have established theoretical guarantees about accurate surface estimation \citep{choi2005posterior,van2008rates}. 
Figure \ref{3_map} shows maps of the true pollution surface and the predictions from the two methods in one dataset at a single time point. The GP Filter captures the peaks of the true pollution surface, whereas the regression-calibration method underestimates most of the higher pollutant concentrations. We also see that the RMSE over all 100 validation time points in that dataset is significantly lower for the GP Filter $(0.61)$ than the regression-calibration method $(1.23)$. Therefore, even in the case of a misspecified model, the GP Filter is a considerable improvement over regression-calibration both in overall RMSE and for capturing peaks in air pollution. This implies that  the improvement afforded by the GP filter over regression-calibration is robust to misspecification of the GP covariance function or even a fully misspecified state-transition model. 

\section{Analysis of Baltimore SEARCH low-cost PM$_{2.5}$ network data}\label{sec:search_analysis}

We now apply the GP Filter to calibrate PM$_{2.5}$ data from the SEARCH network in Baltimore and compare it to the regression-calibration (RegCal) model previously developed for this network \citep{datta2020statistical}. As discussed in Section \ref{sec:search}, within Baltimore city, regulatory PM$_{2.5}$ data is only available at one site (Oldtown). Therefore, to understand intra-urban spatial variation in PM$_{2.5}$ and how they correlate with different socio-economic, health, and demographic variables, it is critical to have spatially resolved PM$_{2.5}$ data as offered by the SEARCH network. 

\begin{figure}
    \centering
    \includegraphics[height=2.4in,trim={0 0 200 0},clip]{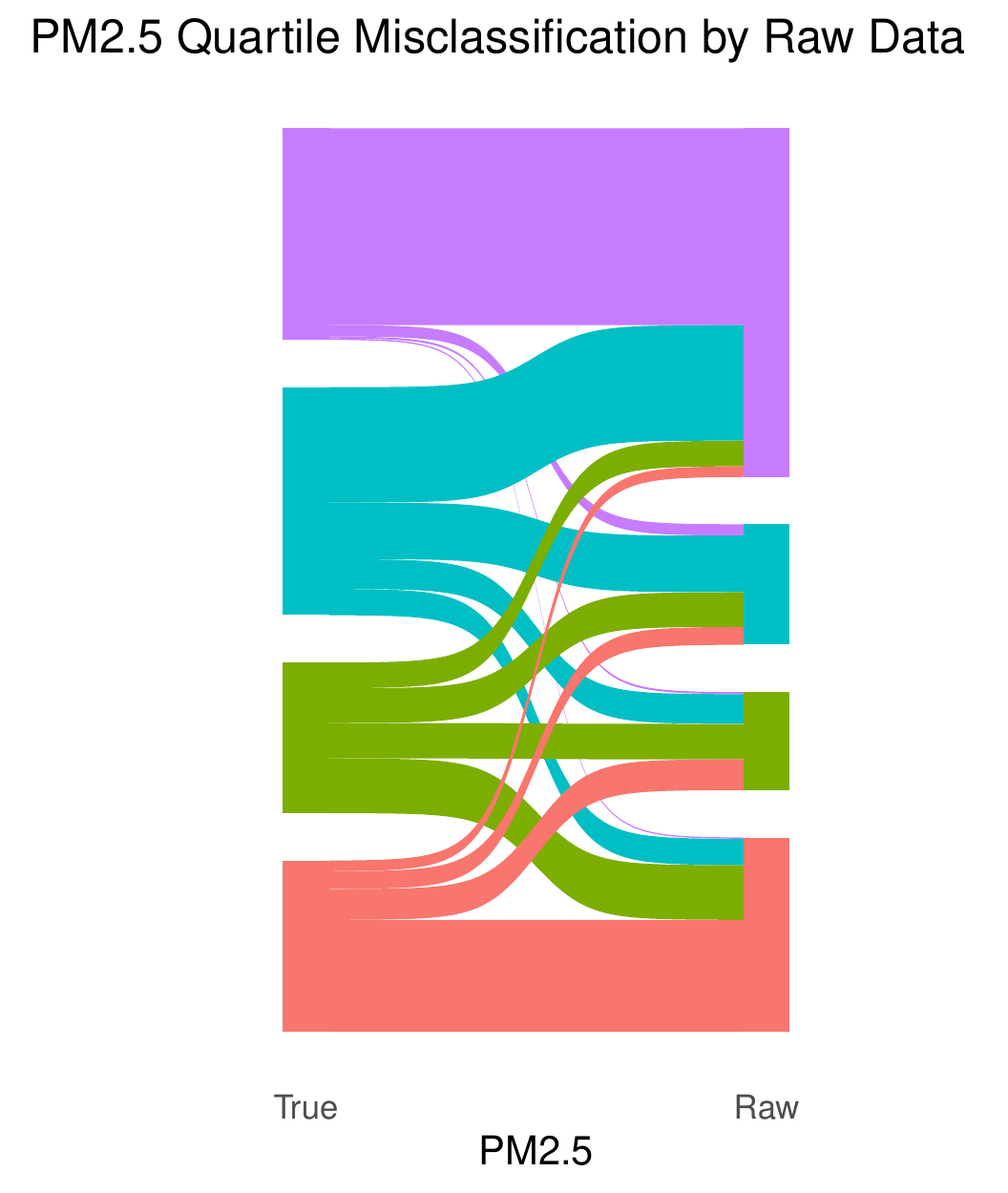}
    \includegraphics[height=2.4in]{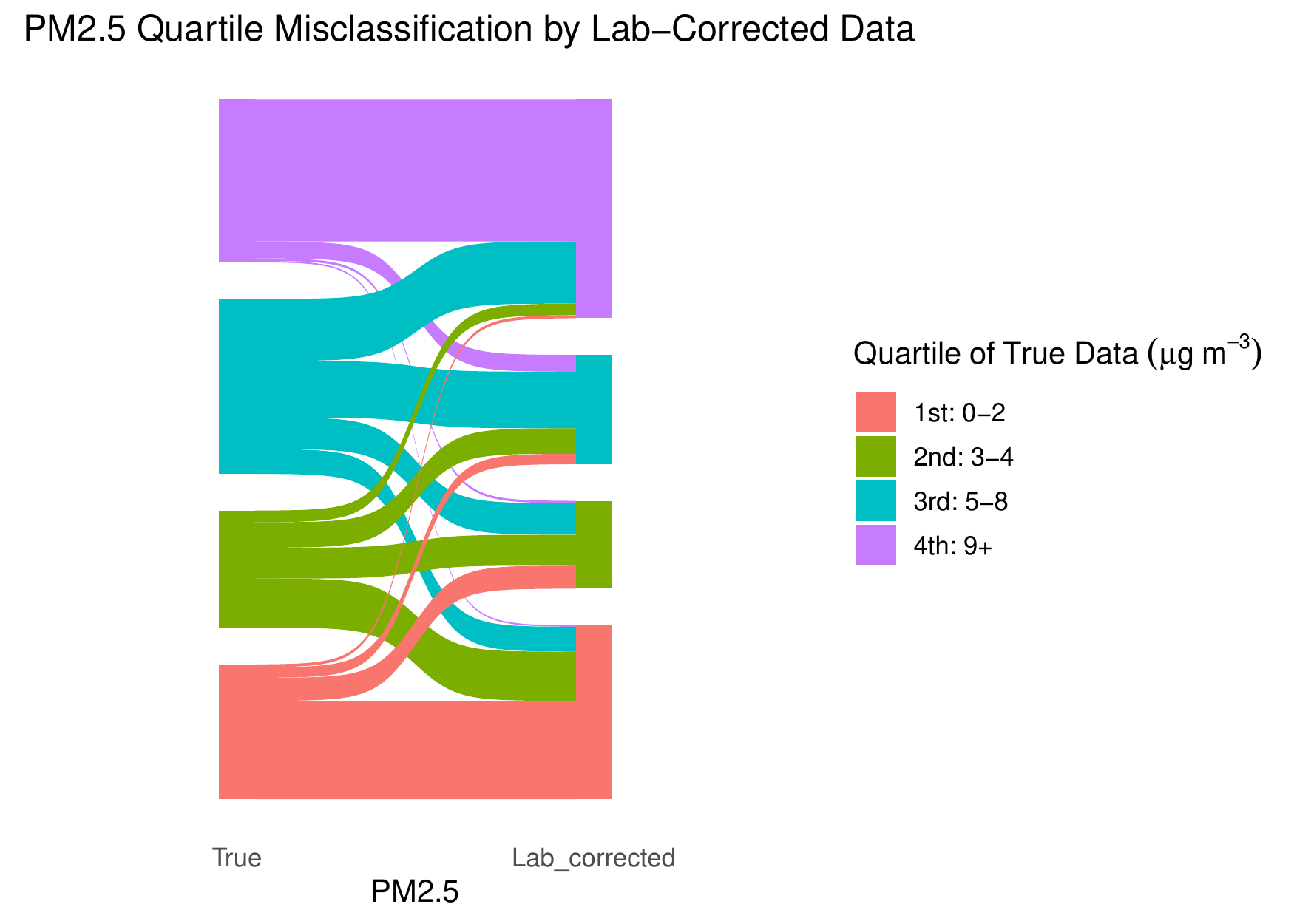}
    \caption{Sankey plots for Baltimore PM$_{2.5}$ data at Oldtown which classify both the true PM$_{2.5}$ concentrations and the collocated low-cost data into the quartiles of the distribution of the true PM$_{2.5}$ data. Left plot uses the raw low-cost data, right plot uses the low-cost data calibrated only using laboratory-based correction equation without any statistical calibration.}
    \label{fig:sankey}
\end{figure}

The baseline concentrations in Baltimore are generally lower than the moderate AQI threshold of $12 \mu g m^{-3}$. However, there is considerable bias in the SEARCH low-cost data even at these  low-concentrations. To illustrate this, Figure \ref{fig:sankey} presents the quartiles of the true PM$_{2.5}$ concentrations at Oldtown and how they compare with the collocated raw (or just laboratory corrected) low-cost PM$_{2.5}$ concentrations. These Sankey plots visualize how data at same time points in the two data sources (true and low-cost PM$_{2.5}$ concentrations) are different. 
We see that within every quartile of the true concentrations, 
a large percentage of the collocated and concurrent low-cost sensor measurements are in a range that corresponds to a different quartile of the true concentrations. 
The combined range of the first three quartiles of true PM$_{2.5}$ concentrations is less than 9 $\mu g/m^3$ (well below the ``moderate'' threshold). This heavy bias of low-cost data even at lower concentrations is consistent with the findings of \cite{datta2020statistical}. If using the low-cost data without any statistical calibration, it would lead to rampant exposure misclassification at all levels. Thus it is critical to properly calibrate the SEARCH low-cost data at both low and high concentrations before any use of the data in spatially-resolved analysis of air-quality and other variables in Baltimore.

\begin{figure}
	\centering
	\subfloat[][Daily time series at site near Oldtown]{
	\includegraphics[width=3.2in]{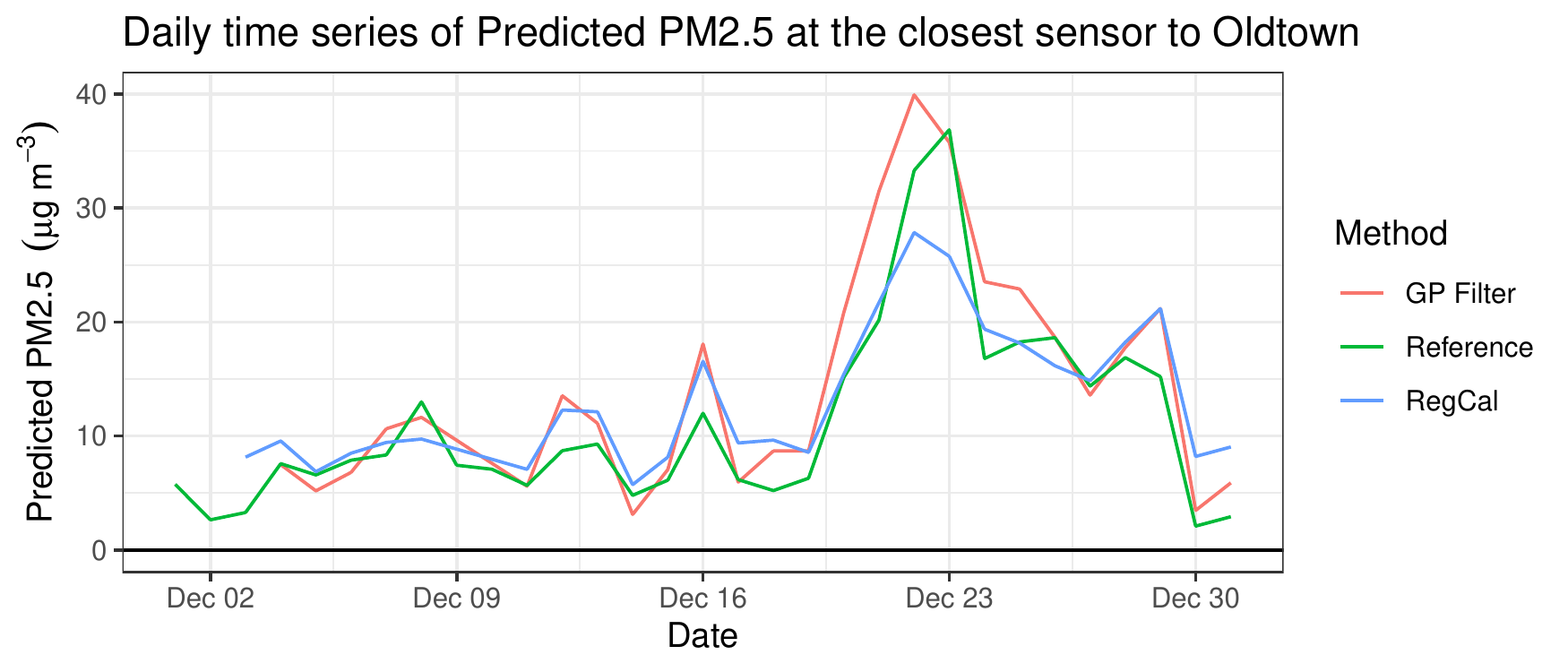}
\label{search_ts_19}}
\subfloat[][Hourly residuals]{
	\includegraphics[width=2.3in]{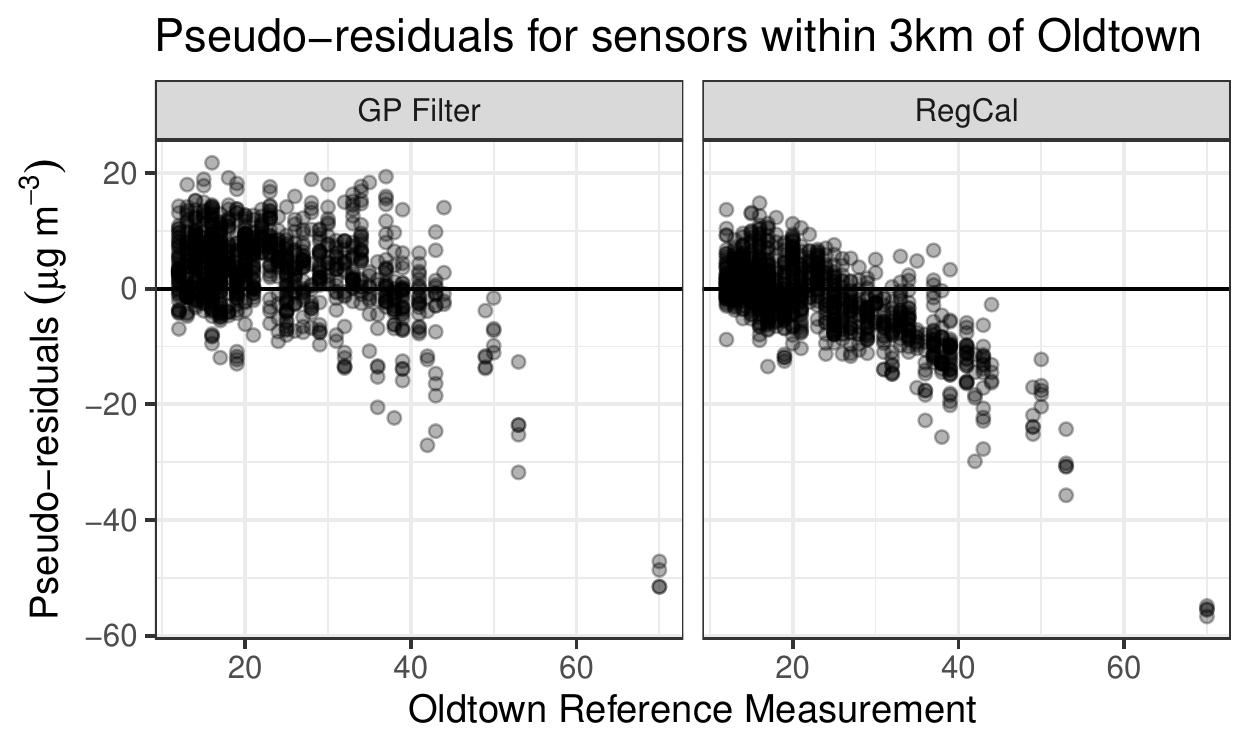}
\label{search_resid}}
	\caption{(a) Daily time series in December 2019. The GP Filter and RegCal predictions are at the closest sensor to the reference site at Oldtown. The reference time series is the PM$_{2.5}$ recorded by the reference-grade BAM at the Oldtown site. (b) Hourly pseudo-residuals ($prediction-Oldtown$) for SEARCH sensors within 3km of the Oldtown MDE device at time points where the Oldtown PM$_{2.5}$ AQI level is moderate or unhealthy ($>12 \mu \text{g/m}^3$).}
	\label{search_oldtown}
\end{figure}

For the calibration models, the same set of covariates are used as considered previously, i.e., RH, T, and a weekend indicator. 
We perform an analysis on six months of data, from December 2019 through May 2020. We present the result for  December 2019 here because this month had a period with moderate/unhealthy concentrations. The results of the full analysis, as well as detailed sensitivity and model validation analysis, are in the Supplement Section \ref{sec:supplement_search}. For testing in December 2019, the regression-calibration model and the observation model of the filtering method are trained on 749 hourly observations from November 2019. 
Subsequent to training the observation model, most of our analysis was performed at the daily level. The GP Filter can be applied at any temporal resolution, but we choose daily observations since our validation data is at the daily scale. However, the analysis were also repeated at the hourly level for producing three results that need larger sample sizes. 
The misclassification rates of Table \ref{fpr_dec2019} is based on hourly data to increase sample size of moderate/unhealthy time points. The residual plot in Figure \ref{search_resid} is also showing hourly residuals, again to increase the number of data points within a 3km radius of the Oldtown site. Supplemental Figure \ref{supp_search_acf} shows the partial autocorrelation plot of hourly residuals so that the autocorrelation at a finer timescale can be observed. 
All these results using the hourly analysis are explicitly specified.

For the GP filter, we consider four different choices of covariance function. The results here are for the GP Filter  applied using the best choice -- an exponential covariance function with a nugget effect and with the value of the spatial decay parameter fixed to its maximum likelihood estimator. However, all four choices of the covariance function yielded similar results (see Figures \ref{supp_search_ts_4covariance} and \ref{supp_search_metrics}).
Since filtering uses the latest available reference data from the collocation site (Oldtown), the Oldtown data cannot be used for evaluation. Instead, we test the models using data from the reference instrument at Essex, the purple site in Figure \ref{ts_dec2019}, which is not used for training any model. The Essex site collects PM$_{2.5}$ measurements every 6 days, so 5 observations are available for testing in December. At both the Oldtown and Essex sites, there are two low-cost sensors at the same location, data from which are averaged to create a single low-cost time-series for each location. 

We use the daily observations from the Essex site to assess the calibration methods. Since the validation is on the daily level, for training, we average the hourly observations and omit monitors that had data for less than 16 hours per day. We also omit the daylight terms from the observation model of the GP filter and the RegCal model as the time-scale for the analysis was daily. To provide a fair comparison to the GP Filter, we retrain the RegCal model on the same training data as the GP filter, rather than use the legacy model coefficients from \cite{datta2020statistical}.  

\begin{figure}
\centering
\subfloat[][RMSE.]{
	\includegraphics[height=1.7in,trim={0 0 70 0},clip]{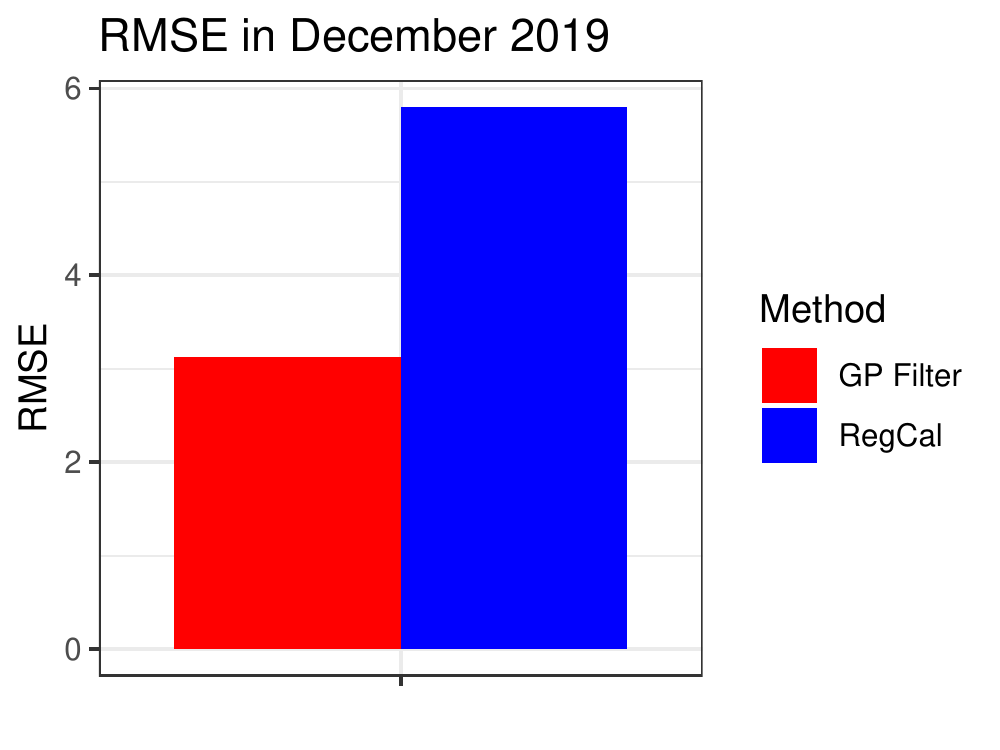}
\label{search_rmse_essex}}
\subfloat[][Daily predictions and 95\% confidence intervals.]{
	\includegraphics[height=1.7in,trim={0 0 0 0},clip]{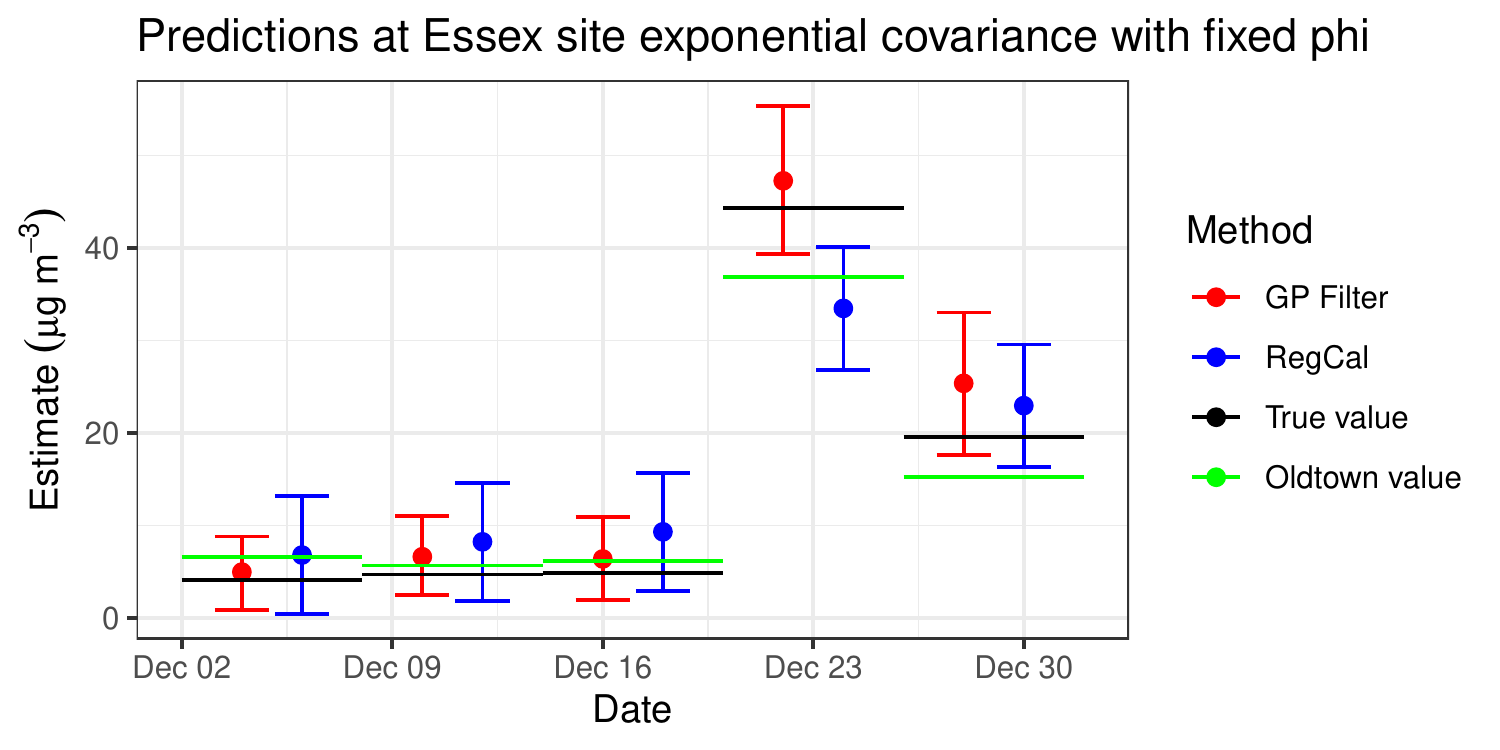}
\label{search_cov}}
\caption{Out-of-sample comparison of GP filtering and regression-calibration for the SEARCH low-cost network in December 2019: (a) RMSE at the Essex site. (b) Predictions and 95\% confidence intervals at Essex. The horizontal black lines are the true measurements each day at Essex, and the horizontal green lines are the true daily concentrations at Oldtown. }
\end{figure}

The RMSE of the two methods are shown in Figure \ref{search_rmse_essex}. The RMSE of the GP Filter is much lower than the regression-calibration. For a more in-depth understanding of the performance, we also look at the predictions from the methods at each individual date in Figure \ref{search_cov}. The figure shows that the GP Filter prediction is always closer to the true Essex value, and the prediction intervals always capture the true values. The baseline PM$_{2.5}$ concentrations are better estimated by the GP Filter. The regression-calibration point estimate considerably underestimates true PM$_{2.5}$ concentration on December 23, 2019 which was in the unhealthy AQI range, and the prediction interval does not capture the true value for that day. The filtering method does not suffer from such substantial under-prediction. For days with lower true concentrations, the GP Filter has narrower prediction intervals than the regression-calibration model, showing  more certainty about the true PM$_{2.5}$ concentration. The intervals for the GP Filter get wider on the days with higher concentrations, and they always capture the true concentration at Essex.

\begin{figure}
	\centering
	\subfloat[][Mean predictions]{
	\includegraphics[width=4.5in]{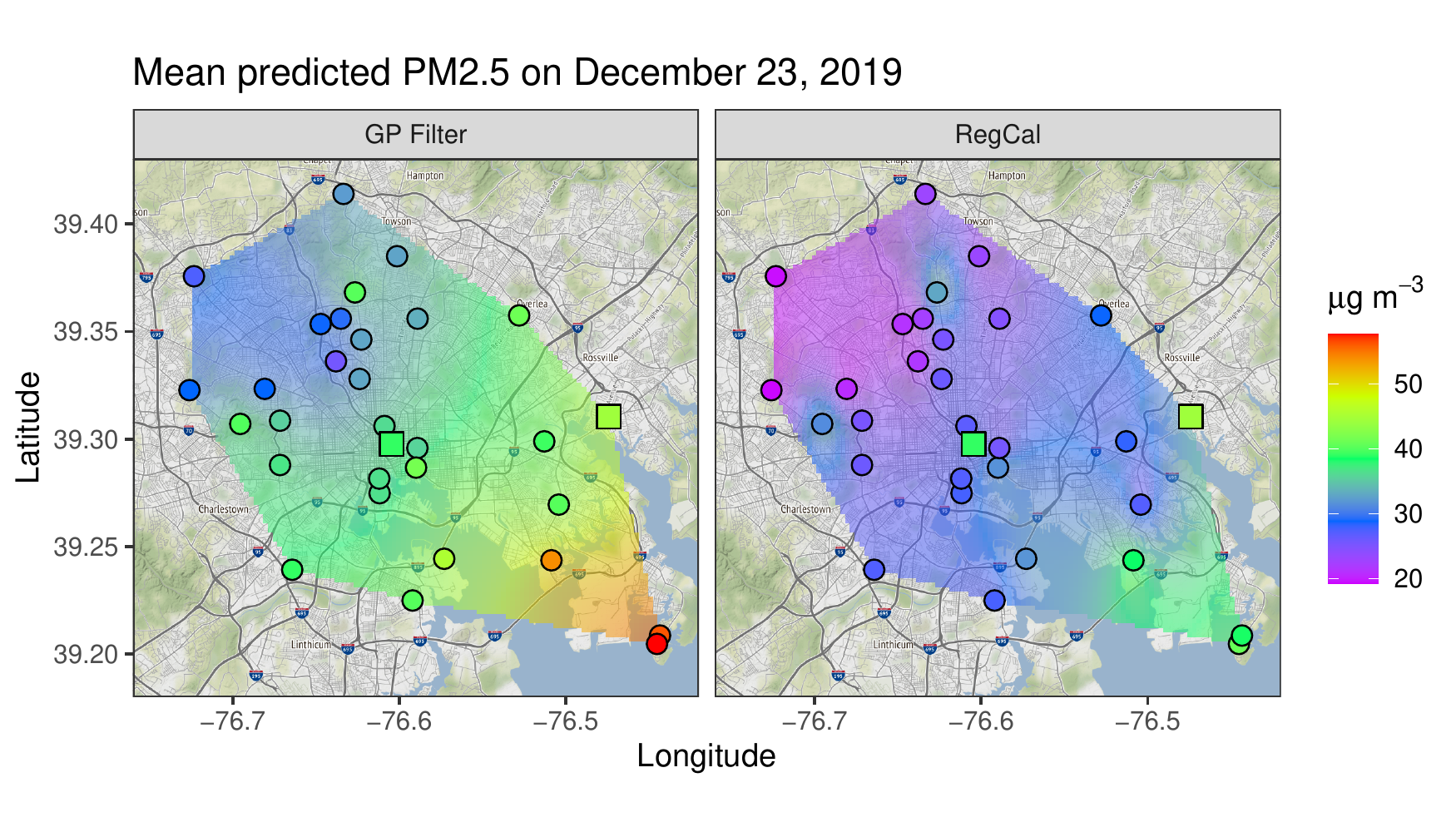}}

\subfloat[][Length of confidence intervals]{
	\includegraphics[width=4.5in]{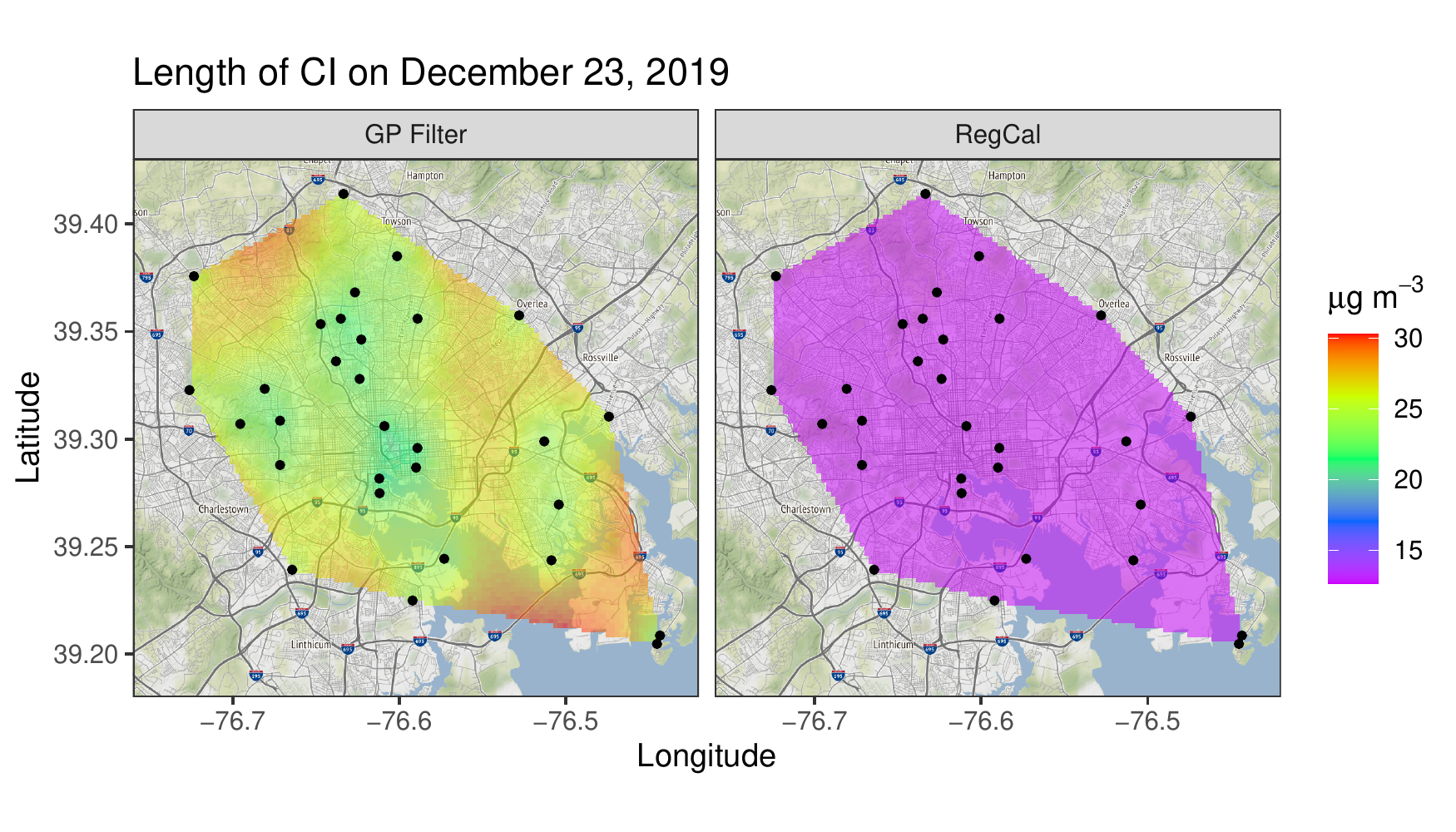}}
	\caption{(a) Interpolated maps of predictions across SEARCH network on December 23. The left panel shows the predictions from the GP Filter, using kriging according to the fitted spatial surface. The second panel shows an interpolation of the predictions from RegCal. The two squares denote the reference measurements at the Oldtown (center) and Essex (east) sites. The circles denote the calibrated low-cost PM$_{2.5}$ data from each method at the sites in the SEARCH network. (b) Length of the confidence interval around the predictions across the city. For the GP Filter, the kriging procedure can construct intervals. For RegCal, the lower bound of the length of the confidence interval is plotted.}
	\label{search_dec23}
\end{figure}

Figure \ref{search_dec23} shows maps of the predicted PM$_{2.5}$ surfaces on December 23, 2019, as well as the uncertainty in predictions across the city. For the GP Filter, samples can be drawn from the kriging distribution (Equation (\ref{eq:jointkrig})) given the low-cost and reference data, which enables the mean and a prediction interval to be calculated directly from our method, taking the spatial structure of the network into account. RegCal does not provide a way to make predictions at sites where there are no low-cost sensors, as it uses the sensor RH and T measurements for specifying the model for the true PM$_{2.5}$. So we use the MBA package in R to interpolate the means across the network, and for plotting the uncertainty we plot the lower bound of the length of the interval using only the error variance $\tau^2$. This is a lower bound on the uncertainty since a prediction interval would account for the uncertainty in the parameter estimates, but without having low-cost observations and covariate information across the entire city, it is not possible to truly construct prediction intervals. On this day, the Oldtown and Essex MDE PM$_{2.5}$ concentrations both correspond to an unhealthy AQI, and the regression-calibration has much lower estimates than the GP Filter methods over the entire city. The map produced by GP Filter matches the observed PM$_{2.5}$ at Oldtown and Essex more closely than the RegCal. Additionally, the uncertainty in the estimates by the GP Filter demonstrates another novelty of the method. The spatial model allows for the predictions on a grid of locations across the city, and the uncertainty is lower around the locations of sensors, as is expected, and higher far from the sensors. Meanwhile, it is not possible to quantify the uncertainty in the interpolation for RegCal, which is why only the lower bound is plotted.

\begin{figure}
\centering
\subfloat[][Average PM$_{2.5}$]{
	\includegraphics[height=2.5in,trim={0 0 0 0},clip]{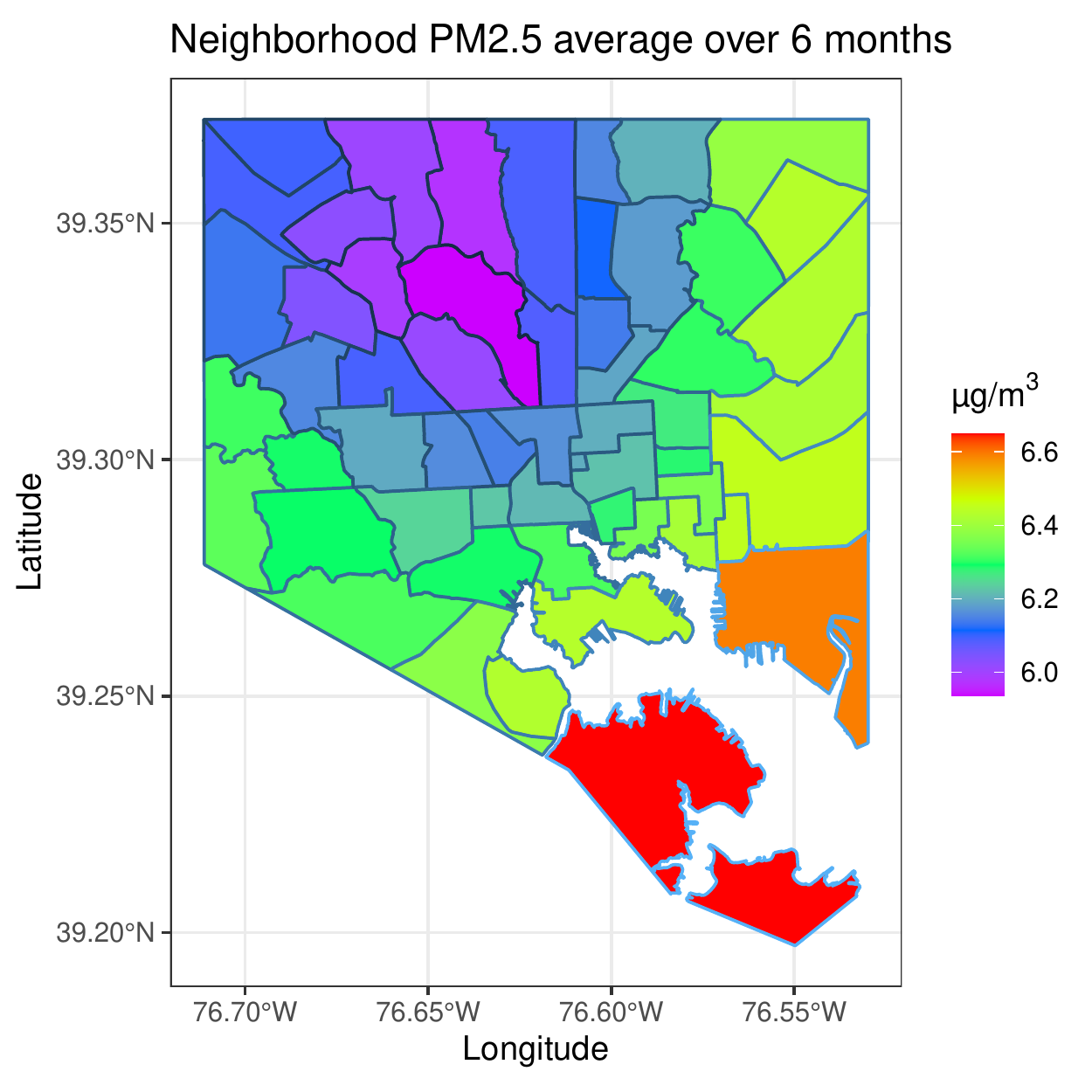}}
\subfloat[][95th percentile of PM$_{2.5}$]{
	\includegraphics[height=2.5in,trim={0 0 0 0},clip]{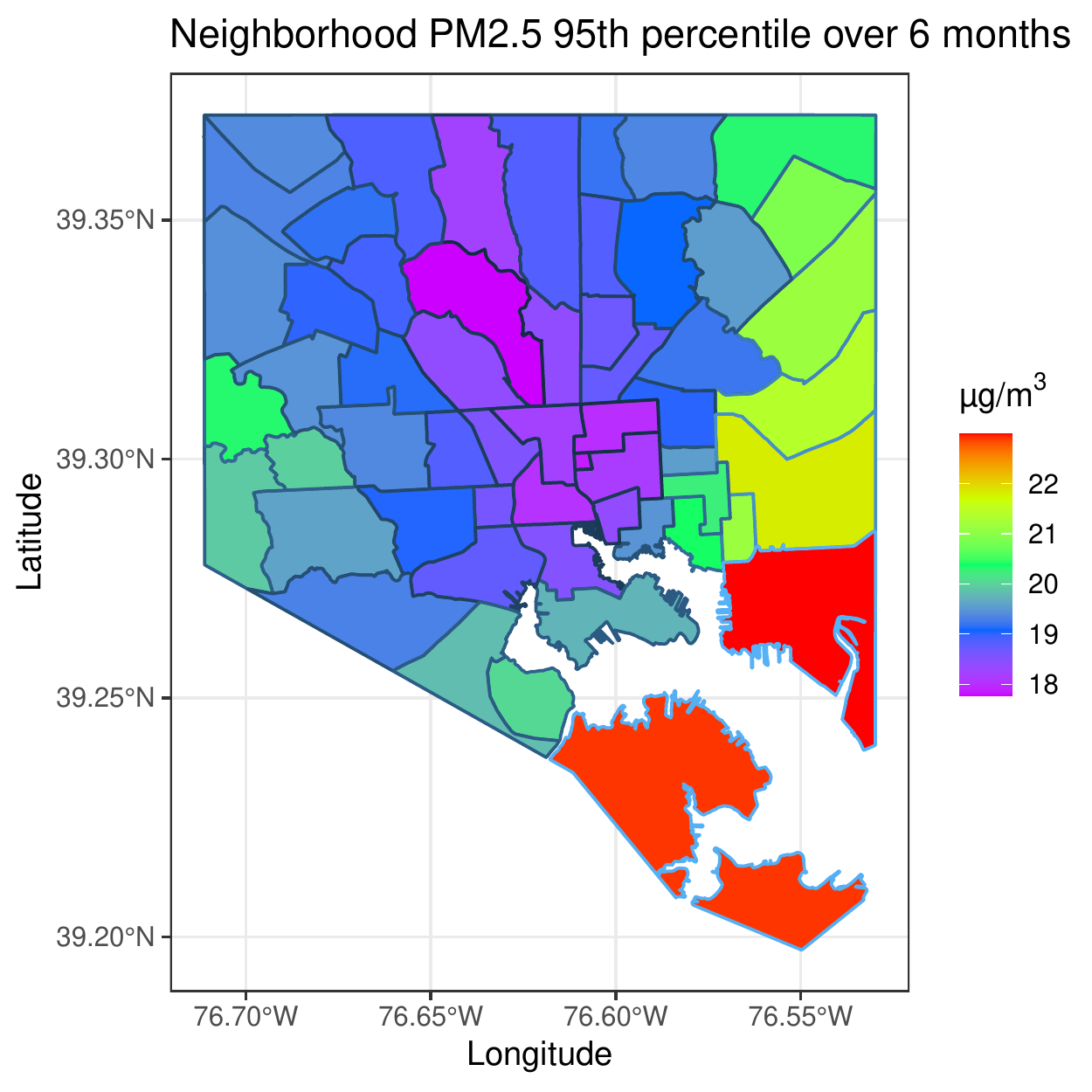}}
\caption{Maps by neighborhood in Baltimore of (a) average PM$_{2.5}$ across 6 months (b) 95th percentile of PM$_{2.5}$ across 6 months}
\label{search_map_neighborhood}
\end{figure}

We also present a map of the PM$_{2.5}$ concentrations summarized across the 6 month period we considered in the full analysis, December 2019 - May 2020 (Figure \ref{search_map_neighborhood}). We see that the neighborhoods with the lowest PM$_{2.5}$ concentration are in the northwest of the city and the highest averages are in the south and east. All neighborhoods have similar averages which are around $6-6.6 \mu g/m^3$. However, more variability appears when looking at the map of the the upper 95th percentile of PM$_{2.5}$ in the city which varies between $18-23 \mu g m^{-3}$. 
The north and center of the city had the lowest 95th percentile, and the south, southwest and east again have high $95\%$ quantiles. These maps show where people are most exposed to high concentrations of air pollution. 

We also perform some validation near Oldtown to check certain properties of the model. Since we have hourly reference data at Oldtown, we now retrain the model at the hourly level, and the observation model and RegCal model both now include the daylight indicator. To further investigate the under-prediction issue, we look at the hourly predictions from the two methods at the SEARCH site closest to the Oldtown reference instrument, so that the reference time series from Oldtown can be used to approximate the true PM$_{2.5}$ time series at that location. We use this site instead of the reference site of Oldtown itself for comparison since the prediction from the GP Filter at Oldtown agrees exactly with the known reference PM$_{2.5}$ concentration measured at that location, due to the exact interpolation property of kriging. Figure \ref{search_ts_19} shows that the filtering method results in predictions that are closer to the (nearby) reference, especially when the reference concentrations spike around December 23.

A pseudo-residual plot is used for model diagnostics in Figure \ref{search_resid} (right), where the true PM$_{2.5}$ concentration used for all sensors is the Oldtown concentration since that is the only site where the true PM$_{2.5}$ concentration is observed. Only sensors within 3km of Oldtown are considered so that the Oldtown reference measurement is a reasonable substitute for the true value. 
The residuals from the GP filter generally does not exhibit strong correlation with the true pollutant concentrations except when the true concentrations are very high ($> 50 \mu g /m^3$) where there is some underestimation. The RegCal residuals exhibit strong negative correlation with the true pollution concentrations throughout its range. The residuals are negative for moderate or unhealthy PM$_{2.5}$ concentrations, thereby considerably underestimating them. This behavior of systematic  underestimation is consistent with the theoretical results of Propositions \ref{th:under_classical} and \ref{th:under_berkson}.  

\section{Discussion}\label{sec:discussion}
The promise of low-cost air pollution networks is indisputable, owing to their cost-efficacy and spatio-temporal richness of their output. However, low-cost air pollution data can be highly biased and variable and any responsible use of the data in scientific studies mandates thorough quality control and evaluation of the data. While field-calibration using regression based on collocated reference devices has become the state-of-the-art for low-cost data correction, we show that this practice a) underestimates peaks in air pollution, and b) does not leverage the spatial correlation in pollution concentrations across an area. 

We present a simple but novel dynamic calibration approach via a spatial filtering that uses inverse-regression to mitigate the under-estimation issue and a conditional Gaussian Process model to leverage the spatial correlation. 
Our filtering approach works with as few as one collocation site. 
Our simulations showed that even a network of 50 non-collocated sites and 1 collocated site can provide enough spatial information to make better joint calibration and predictions than if each sensor is calibrated individually. 
In fact, across all the simulation settings and in the data analysis on the SEARCH network, it is evident that the Gaussian process filtering method significantly outperforms the regression-calibration method. The RMSE and the FNR for identifying high pollutant concentrations are both lower in the GP Filter, and the method is robust to various forms of misspecification. 
The GP Filter thus positions itself as an extremely useful method to calibrate data in cities with high average concentrations or with many peaks. Even in cities with lower average concentrations, such as Baltimore, the GP Filter provides an accurate assessment of baseline concentrations and the occasional peaks, which is also valuable for a health association study.

Another highlight of our method is the ability to predict on any grid of locations unlike regression-calibration that can only predict at sites with low-cost sensors. This allows us to predict average pollutant concentrations in different neighborhoods in a city, which informs on the disparities in pollutant exposure along various socio-economic and demographic gradients. 
Recently, the EPA formed an Office of Environmental Justice and External Civil Rights, showing the agency’s commitment to addressing environmental injustice in the United States. Understanding disproportionate impacts of air pollution on communities requires quantifying the within-city differences in air pollution concentrations, so the ability to predict concentrations across the entire area is an advantage of the GP Filter compared to RegCal or to just using regulatory data. 

In the SEARCH network in Baltimore the sensor locations were chosen from diverse ambient settings using a weighted design (see Section \ref{sec:search}). Hence, preferential sampling is less of a concern here than in other commercial networks where data is available from any household which buys a sensor and hence can have biased sampling in more affluent areas. \cite{zidek2014reducing} proposed an adaptive design for sensor locations to address such preferential sampling bias. However, in practice, finding suitable locations and hosts for long-term deployment of a sensor is challenging and executing such a dynamic strategy is also very resource-intensive. Hence, an adaptive design is often not feasible. 
The uncertainty estimates from our spatial model does allow us to determine where future sensors should be installed -- in the locations with the highest uncertainty. For example, in the uncertainty map of Figure \ref{search_dec23}, we see that some of the areas with the highest uncertainty are in the northwest and southeast of the city, as well as portions of the east of the city. 
These uncertainty maps provided by GP filter can be used to strategize future sensor placement. 

\subsection{Related methodological literature}\label{sec:related_methods}

Our method combines elements of Kalman filtering, GP models, general multivariate spatio-temporal regression, and extreme value calibration. We briefly discuss how it is similar to relevant methods in each of these fields.

There is a large literature on co-kriging or Bayesian melding approaches that jointly model multi-network data on the same variable but with different magnitudes of measurement error and different spatial coverage \citep[see for example][]{zimmerman2005complementary,fuentes2005model}. These are general purpose methods that has been used in many different spatial or spatio-temporal applications \citep{cowles2002combining,cowles2003bayesian}. Our proposed two-stage model can be viewed broadly as a specialized version of this framework, one that includes data from one (reference) network in modeling the bias of the other (low-cost) network, along with adding meteorological covariates, and interaction terms. However, there are notable differences arising from the specific application of calibrating low-cost sensors. Our theoretical results on peak underestimation by regression-calibration and mitigation of the issue by the inverse regression, are valid even for just a pair of collocated devices (one low-cost and one reference) without consideration of any network.  These results highlight an important deficiency of the common calibration technique and are of independent importance even without the subsequent model development for the entire network. To our knowledge, this underestimation of high pollutant levels by regression-calibration has not been studied previously. These theoretical insights, in turn, prescribe a natural directionality in the subsequent two stage model --- modeling the low-cost data $y$ conditional on the reference data $x$ and adding a marginal spatial model for $x$. A model specified in the opposite direction, i.e., the regression-calibration model augmented with a marginal spatial model for the low-cost data, would also constitute a complementary co-kriging or Bayesian melding approach but would fail to address the underestimation issue. Thus if  low-cost sensor network calibration is viewed as a co-kriging problem, this manuscript shows that is important to model the directionality properly. 
The proposed implementation of our model using Kalman updates is also different from those adapted in co-kriging models where estimation and prediction proceed simultaneously. Due to the high-frequency (hourly) nature of the data, such joint modeling of the entire data would be both computationally challenging and require re-estimation of the entire model with every new data-point. In our approach, the observation model is estimated apriori using abundant training data and a Kalman filtering approach is used with this `known' observation model at each new time point to offer scalable predictions.

Gaussian Process methods have also been abundant in the co-kriging literature and recently \cite{zheng2019gaussian} used GPs to smooth and calibrate data from a low-cost network. However, their calibration equation still used a forward regression with the low-cost data as the independent variable and the true concentrations as the response. Hence this does not mitigate the underestimation issue and implicitly assumes that the true pollutant surface is noisier than the low-cost surface. Also, the approach required the presence of many ($\sim 20$) reference instruments in the area to capture the spatial structure to estimate device-specific calibration equations. This is unrealistic in many applications (e.g., Baltimore has only one continuous reference PM$_{2.5}$ measurement). Our spatial filtering approach is more parsimonious in terms of resource needs and can be applied with as few as one reference site in the region which facilitates both training of the observation model as well as positing the conditional GP model for spatial smoothing.

Our spatial filtering method filters in space and is notably different from existing filtering approaches for spatio-temporal data like the spatio-temporal filter or kriged Kalman-filter  \citep{mardia1998kriged,sahu2005bayesian}, and related methods adapted for air pollution modeling  \citep{van2000analysis,wu2020high,tang2013inversion,june2021operational}. all of which filter in time. 
The high-frequency low-cost sensor data offer the opportunity to characterize ultra-short-term fluctuations of the pollutant concentrations. Filtering in time using lower frequency reference measurements will smooth these out by placing too much weight on lower baseline concentrations and treating a true peak in concentrations as a random measurement error to be smoothed out. The unique setting of collocated calibration mandates filtering in space instead of time.  Filtering across space is necessary to smoothly interpolate the data beyond the network locations to create continuous pollutant maps. The available reference data 
dictates the state-transition model (\ref{kriging}) across space by modeling the spatial correlation. 
However, if desired time dependence can be  accommodated in our framework  both in the observation model and the state-space model (see Section \ref{sec:spacetime}).




The simulation studies revealed that, if only the moderate/unhealthy concentrations were of interest, a threshold exceedance type calibration \citep{davison1990models} like the Pareto model considered in Section \ref{sec:sim_1a} could be applied to model the exceedances in the peak values using generalized Pareto distribution \citep{pickands1975statistical} (as we see in Figure \ref{sim_1a_rmse} (right)). However, the biggest drawback of threshold-based approaches is that they do not calibrate the noisy low-cost data at low concentrations, leading to very poor overall performance (Figure \ref{sim_1a_rmse} (left)). Thus, these approaches cannot be applied for calibrating air pollution data in a city like Baltimore with concentrations predominantly below threshold (see Figure \ref{supp_search_ts_4covariance}). Baltimore is very representative of many other US cities in terms of air quality, and these baseline concentrations represent the air pollution that someone is exposed to a majority of the time. There is now overwhelming evidence that all concentrations of pollution can be detrimental to health. Accurate assessments of pollution concentrations below the current regulatory limits are needed to continue to build evidence in support of revisions of the air pollution standards. Multiple recent studies had concluded that PM$_{2.5}$ is associated with increased risk of mortality even at concentrations below current national air quality standards \citep{EPA_ISA,di2017association,di2017air,wei2020causal,shi2021national,ward2021long}. The EPA itself, in its most recent Policy Assessment for particulate matter noted, in regards to possible thresholds in the concentration response curve, that studies ``consistently demonstrate a linear relationship with no evidence of a threshold" \citep{EPA_policy_assessment}. 
Additionally, the World Health Organization (WHO) recently reduced their recommendation for annual average PM$_{2.5}$ concentrations from 10 $\mu g/m^3$ to 5 $\mu g/m^3$. 
So it is important to adequately measure typical low exposures in a city or area. Low-cost sensor data is very biased even at these low concentrations (Figure \ref{fig:sankey}) and a calibration approach needs to calibrate data across a wide range of concentrations above and below the threshold for the calibrated data to be realistic and useful in health association studies, as well as other applications, such as climate research.

Some other shortcomings of threshold-based approaches include low sample size for training, the threshold can be only created using the observed low-cost data, which leads to exposure misclassification both above and below the threshold in opposite directions, and that the current extensions of Pareto regression to accommodate spatial correlation will not be able to estimate the spatial random effects with only one or very few collocated sites with reference data. These issues are expanded on in Supplement \ref{sec:supplement_pareto}.
 
Our proposed method does not rely on any threshold, calibrating the low-cost data at all concentrations capturing both baseline low-levels and occasional peaks. It also models spatial correlation in PM$_{2.5}$ concentrations with as few as one regulatory site in the region. This results in a dynamic calibration of low-cost network data, leveraging spatial correlation with latest measurements from nearby regulatory sites.


\subsection{Future work:} In the future, we plan to study in details the Baltimore PM$_{2.5}$ maps created from application of the method as well as conduct association studies with these predicted PM$_{2.5}$ levels across the city with various neighborhood-level socio-economic, demographic or health indicators. It is important to note that both the regression-calibration and our model assume that the relationship between the true pollutant concentration and the low-cost measurement is the same across all sensors, so the same observation model trained at one site can be used across the network. This assumption may not be valid in all gas sensor models, where there is a large amount of unit-to-unit variability. This unit-specific effect is not included in our model, so it may not be directly applicable to gas sensor networks. 
A calibration approach tailored to gas measurements has been proposed \citep{kim2018berkeley}. This method uses gas cross-sensitivities, regional concentrations measured by reference monitors (which need not be collocated), co-emitted gases, ozone uniformity over space, and chemical conservation equations to calibrate a low-cost network of gas 
sensors. 
This approach takes the chemistry of pollutants into account and calibrates each one according to its particular behavior, but does not impose an explicit spatial model for air pollutant as our method does or investigate under-prediction of high concentrations. One future direction is to assess feasibility of our method for calibrating networks with device-specific biases. 

We also identify multiple possible extensions of this method that can be considered for future research. Currently, a Gaussian error term is assumed. Generalizations of the Kalman filter that allow for non-Gaussian distributions of the error \citep{wuthrich2016robust} can be incorporated into the method. Also, the gains and offsets used in the observation model are modeled as linear functions of the covariates. Non-linear calibration models have also been considered for calibration of low-cost networks \citep{Topalovic2019,lim2019mapping,zimmerman2018machine,johnson2018using}. More flexible non-linear observation models can be developed for the filtering to potentially improve fitting complex variable relationships.

\section{Appendix: Proofs of Propositions}
In the Appendix, Propositions 1 and 3 are only proved in the case without covariates. When covariates are present, the proofs of these propositions are more technical and are provided in the Supplement Section \ref{sec:supp_proofs}. 

\subsection{Proposition 1 Proof}\label{proof1}

\begin{proof}
We first prove the result for the case without covariates (model (\ref{reg-cal-no-cov})). Consider data $(y_i,x_i)$, $i=1,\dots n$. If the true model is $x_i=\beta_0+\beta_1y_i +\epsilon_i$, and linear regression is used to estimate coefficients $\bo,\bl$, then  
\begin{align*}
\widehat{Cov}(\hat{\bx},\bx)&=\frac 1n\sum_{i=1}^{n}\hat{x}_ix_i-\frac 1n\sum_{i=1}^{n}\hat{x}_i\frac 1n\sum_{i=1}^nx_i\\
&=\frac 1n\sum_{i=1}^n\left(\bo+\bl y_i\right)\left(\beta_0+\beta_1y_i+\epsilon_i\right)-\left(\bo+\bl \bar{y}\right)\left(\beta_0+\beta_1\bar{y}+\bar{\epsilon}\right)\\
&=\beta_1\bl\left(\frac 1n\sum_{i=1}^ny_i^2-\bar{y}^2\right)+\bl\left(\frac 1n\sum_{i=1}^ny_i\epsilon_i-\bar{y}\bar{\epsilon}\right)\\
\widehat{Cov}(\hat{\bx}-\bx,\bx)&=\widehat{Cov}(\hat{\bx},\bx) - \widehat{Cov}(\bx,\bx) \\
&=\beta_1\bl\left(\frac 1n\sum_{i=1}^ny_i^2-\bar{y}^2\right)+\bl\left(\frac 1n\sum_{i=1}^ny_i\epsilon_i-\bar{y}\bar{\epsilon}\right)-\left(\frac 1n\sum_{i=1}^nx_i^2-\bar{x}^2\right)\\
&= \beta_1\bl s_y^2+\bl s_{y,\epsilon}-s_x^2
\end{align*} 
where $s_x^2,s_y^2$ are the sample variance of $\bx$ and and $\by$ and $s_{y,\epsilon}$ is the sample covariance of $\by$ and $\boldsymbol{\epsilon}$. 
This quantity asymptotes to $\beta_1^2Var(Y)+\beta_1Cov(Y,\epsilon)-Var(X)$ by convergence of sample variances and covariances to their population analogs, consistency of $\hat \beta_1$ to $\beta_1$, and Slutsky's theorem. Finally, noting that $\beta_1 = Cov(X,Y)/Var(Y)$ and ${Cov}(Y,\epsilon)=0$ we have 
$$ \lim \widehat{Cov}(\hat{\bx}-\bx,\bx) = \frac{Cov(X,Y)^2}{Var(Y)}-Var(X)<0$$ by the Cauchy-Schwartz inequality. 
Therefore, the bias $\hat{\bx}-\bx$ is negatively correlated with the true response $\bx$.

The general case with covariates in the model is proved in the supplement.

\end{proof}

\subsection{Proposition 2 Proof}\label{proof2}

\begin{proof}
Let $y_i=\beta_0+\beta_1x_i +\epsilon_i$ be the true classical error model and we fit the regression-calibration model $x_i=\hat\alpha_0+\hat\alpha_1y_i$ using least squares. Then
\begin{align*}
\widehat{Cov}(\hat{\bx}-\bx,\bx)
&=\frac 1n\sum_{i=1}^{n}\hat{x}_ix_i-\frac 1n\sum_{i=1}^{n}\hat{x}_i\frac 1n\sum_{i=1}^nx_i-\left(\frac 1n\sum_{i=1}^nx_i^2-\bar{x}^2\right)\\
&=\frac 1n\sum_{i=1}^n\left(\hat\alpha_0+\hat\alpha_1 y_i\right)x_i-\left(\hat\alpha_0+\hat\alpha_1 \bar{y}\right)\bar{x}-s_x^2\\
&=\frac 1n\sum_{i=1}^n\left(\hat\alpha_0+\hat\alpha_1 (\beta_0+\beta_1x_i +\epsilon_i)\right)x_i-\left(\hat\alpha_0+\hat\alpha_1 (\beta_0+\beta_1\bar{x} +\bar{\epsilon})\right)\bar{x}-s_x^2\\
&=\hat\alpha_1\beta_1\frac 1n\sum_{i=1}^nx_i^2+\hat\alpha_1\frac 1n\sum_{i=1}^nx_i\epsilon_i-(\hat\alpha_1\beta_1\bar{x}^2+\hat\alpha_1\bar{x}\bar{\epsilon})-s_x^2\\
&=\frac{s_{xy}}{s_y^2}\beta_1\left(\frac 1n\sum_{i=1}^nx_i^2-\bar{x}^2\right)+\frac{s_{xy}}{s_y^2}\left(\frac 1n\sum_{i=1}^nx_i\epsilon_i-\bar{x}\bar{\epsilon}\right)-s_x^2\text{ since } \hat\alpha_1=\frac{s_{xy}}{s_y^2}\\
&=\frac{s_{xy}}{s_y^2}\beta_1s_x^2+\frac{s_{xy}}{s_y^2}s_{x,\epsilon}-s_x^2\\
&\overset{P}{\to}\frac{Cov(X,Y)}{Var(Y)}\beta_1Var(X)+\frac{Cov(X,Y)}{Var(Y)}Cov(X,\epsilon)-Var(X)\\
&\phantom{\overset{P}{\to}}\text{by consistency and Slutsky's}\\
&=\frac{Cov(X,Y)^2}{Var(Y)}-Var(X)\quad\text{since } \beta_1=\frac{Cov(X,Y)}{Var(X)}\text{ and } Cov(X,\epsilon)=0\\
&<0\quad\text{by the Cauchy-Schwartz inequality.}
\end{align*}
\end{proof}

\subsection{Proposition 3 Proof}\label{proof3}

\begin{proof}
We first consider a true model with no covariates, $y_i=\beta_0+\beta_1x_i +\epsilon_i$ where we only need the assumption $\beta_1\neq 0$. If the estimated coefficients from least squares are $\bo,\bl$, then 
\begin{align*}
&\widehat{Cov}(\hat{\bx}-\bx,\bx)\\
&=\frac 1n\sum_{i=1}^{n}(\hat{x}_i-x_i)x_i-\frac 1n\sum_{i=1}^{n}(\hat{x}_i-x_i)\frac 1n\sum_{i=1}^nx_i\\
&=\frac 1n\sum_{i=1}^{n}\left(\frac{(\beta_0+\beta_1x_i+\epsilon_i)-\bo}{\bl}-x_i\right)x_i-\frac 1n\sum_{i=1}^{n}\left(\frac{(\beta_0+\beta_1x_i+\epsilon_i)-\bo}{\bl}-x_i\right)\frac 1n\sum_{i=1}^nx_i\\
&=\frac{1}{\bl}\left((\beta_1-\bl)\frac 1n\sum_{i=1}^nx_i^2-(\beta_1-\bl)\bar{x}^2+\frac 1n\sum_{i=1}^{n}x_i\epsilon_i-\bar{\epsilon}\bar{x}\right)\\
&\to 0.
\end{align*}
The limiting result holds by by consistency of the least squares estimates and since the errors $\epsilon$ have $E[\epsilon]=0$  and $X \perp \epsilon$.

The general case of a model with covariates is proved in the supplement.
\end{proof}

\subsection*{Acknowledgments}
 The authors would like to thank Colby Buehler (Yale) and Misti Levy Zamora (U. Conn.) for their contributions to the SEARCH network deployment.

\subsection*{Funding}
CH was partially supported by the Fonds de recherche du Québec - Nature et Technologies bourse de maîtrise B1X, and partially supported by the National Science Foundation Graduate Research Fellowship Program under Grant No. DGE2139757. AD, RP, and KK were partially supported by National Institute of Environmental Health Sciences (NIEHS) grant R01 ES033739. AD was partially supported by National Science Foundation (NSF) Division of Mathematical Sciences grant DMS-1915803. KK, DRG, RP, AD and CH acknowledge support from the assistance agreement no. RD835871 awarded by the U.S. Environmental Protection Agency to Yale University. It has not been formally reviewed by the EPA. The views expressed in this document are solely those of the authors and do not necessarily reflect those of the agency. The EPA does not endorse any products or commercial services mentioned in this publication. DRG acknowledges HKF Technology (a Kindwell Company) for also supporting the sensor development.

\medskip

\bibliographystyle{apalike}
\bibliography{mybib}

\clearpage
\appendix

\renewcommand\thesection{S\arabic{section}}
\renewcommand\theequation{S\arabic{equation}}
\renewcommand\thefigure{S\arabic{figure}}
\renewcommand\thetable{S\arabic{table}}
\setcounter{figure}{0}
\section*{Supplement to ``A dynamic spatial filtering approach to mitigate underestimation bias in field calibrated low-cost sensor air-pollution data''}

\section{Supplemental proofs}\label{sec:supp_proofs}
\subsection{Proposition 1 Proof with covariates}

We proved proposition 1 in the case where the regression-calibration model does not contain covariates in the appendix. Now, we consider the case with covariates in the model. 

\begin{proof}
For the general case, i.e., when the regression-calibration model contains covariates, as in Equation (\ref{reg-cal}), the true model can be written as $x_i=\beta_0+\beta_1y_i+\boldsymbol{\beta_2}'\bz_i+\boldsymbol{\beta_3}'\bz_i y_i+\epsilon_i=\bv_i'\boldsymbol{\beta}_0+\bv_i'\boldsymbol{\beta}_1y_i+\epsilon$ for $i=1\dots n$, where $\bv_i=(1,\bz_i')$ and $\boldsymbol{\beta}_0=(\beta_0,\boldsymbol{\beta_2}')'$, $\boldsymbol{\beta}_1=(\beta_1,\boldsymbol{\beta_3}')'$. 

Assume that the covariates $\bv_i$ and the measured pollutant concentrations $y_i$ are bounded above. Also assume $Var(\epsilon)<\infty$. 
Then
\begin{align*}
	&\widehat{Cov}(\hat{\bx}-\bx,\bx)\\
	&=\frac 1n\sum_{i=1}^n(\hat{x_i}-x_i)x_i-\frac 1n\sum_{i=1}^n(\hat{x_i}-x_i)\frac 1n\sum_{i=1}^nx_i\\
	&=\frac 1n\sum_{i=1}^n(\bv_i'\widehat{\boldsymbol{\beta}_0}+\bv_i'\widehat{\boldsymbol{\beta}_1}y_i-\bv_i'\boldsymbol{\beta}_0-\bv_i'\boldsymbol{\beta}_1y_i-\epsilon_i)(\bv_i'\boldsymbol{\beta}_0+\bv_i'\boldsymbol{\beta}_1y_i+\epsilon_i)\\
	&\phantom{=}
	-\frac 1n\sum_{i=1}^n(\hat{x_i}-x_i)\frac 1n\sum_{i=1}^nx_i\\
	(\ast)&=\frac 1n\sum_{i=1}^n\Big(\bv_i'(\widehat{\boldsymbol{\beta}_0}-\boldsymbol{\beta}_0)\boldsymbol{\beta}_0'\bv_i+\bv_i'(\widehat{\boldsymbol{\beta}_0}-\boldsymbol{\beta}_0)\boldsymbol{\beta}_1'\bv_iy_i+\bv_i'(\widehat{\boldsymbol{\beta}_0}-\boldsymbol{\beta}_0)\epsilon_i\\
	&\phantom{=\frac 1n\sum_{i=1}^n}+\bv_i'(\widehat{\boldsymbol{\beta}_1}-\boldsymbol{\beta}_1)\boldsymbol{\beta}_0'\bv_iy_i+\bv_i'(\widehat{\boldsymbol{\beta}_1}-\boldsymbol{\beta}_1)\boldsymbol{\beta}_1'\bv_iy_i^2+\bv_i'(\widehat{\boldsymbol{\beta}_1}-\boldsymbol{\beta}_1)y_i\epsilon_i\\
	&\phantom{=\frac 1n\sum_{i=1}^n}-\bv_i'\boldsymbol{\beta}_0\epsilon_i-\bv_i'\boldsymbol{\beta}_1y_i\epsilon_i-\epsilon_i^2\Big)-\frac 1n\sum_{i=1}^n(\hat{x_i}-x_i)\frac 1n\sum_{i=1}^nx_i
\end{align*}

We consider each term individually.

Terms 1-6: We can rewrite all six terms of the sum by defining a scalar $m_{i,j}$ for each one: 
\begin{align*}
	\bv_i'(\widehat{\boldsymbol{\beta}_0}-\boldsymbol{\beta}_0)\boldsymbol{\beta}_0'\bv_i&=\bv_i'(\widehat{\boldsymbol{\beta}_0}-\boldsymbol{\beta}_0)m_{i,1}\quad\text{letting }m_{i,1}=\boldsymbol{\beta}_0'\bv_i\\
	\bv_i'(\widehat{\boldsymbol{\beta}_0}-\boldsymbol{\beta}_0)\boldsymbol{\beta}_1'\bv_iy_i&=\bv_i'(\widehat{\boldsymbol{\beta}_0}-\boldsymbol{\beta}_0)m_{i,2}\quad\text{letting }m_{i,2}=\boldsymbol{\beta}_1'\bv_iy_i\\
	\bv_i'(\widehat{\boldsymbol{\beta}_0}-\boldsymbol{\beta}_0)\epsilon_i&=\bv_i'(\widehat{\boldsymbol{\beta}_0}-\boldsymbol{\beta}_0)m_{i,3}\quad\text{letting }m_{i,3}=\epsilon_i\\
	\bv_i'(\widehat{\boldsymbol{\beta}_1}-\boldsymbol{\beta}_1)\boldsymbol{\beta}_0'\bv_iy_i&=\bv_i'(\widehat{\boldsymbol{\beta}_1}-\boldsymbol{\beta}_1)m_{i,4}\quad\text{letting }m_{i,4}=\boldsymbol{\beta}_0'\bv_iy_i\\
	\bv_i'(\widehat{\boldsymbol{\beta}_1}-\boldsymbol{\beta}_1)\boldsymbol{\beta}_1'\bv_iy_i^2&=\bv_i'(\widehat{\boldsymbol{\beta}_1}-\boldsymbol{\beta}_1)m_{i,5}\quad\text{letting }m_{i,5}=\boldsymbol{\beta}_1'\bv_iy_i^2\\
	\bv_i'(\widehat{\boldsymbol{\beta}_1}-\boldsymbol{\beta}_1)y_i\epsilon_i&=\bv_i'(\widehat{\boldsymbol{\beta}_1}-\boldsymbol{\beta}_1)m_{i,6}\quad\text{letting }m_{i,6}=y_i\epsilon_i
\end{align*}

Then we note that for $j\in\{1,2,3,4,5,6\}$ and $k\in\{0,1\}$:
\begin{align*}
	&\left|\frac 1n \sum_{i=1}^n\bv_i'(\widehat{\boldsymbol{\beta}_k}-\boldsymbol{\beta}_k)m_{i,j}\right|\\
	&\leq \frac 1n \sum_{i=1}^n|\bv_i'(\widehat{\boldsymbol{\beta}_k}-\boldsymbol{\beta}_k)|\cdot |m_{i,j}|\quad\text{by the triangle inequality}\\
	&\leq \left(\frac 1n \sum_{i=1}^n||\bv_i||\cdot |m_{i,j}|\right)\cdot||\widehat{\boldsymbol{\beta}_k}-\boldsymbol{\beta}_k||\quad\text{by the Cauchy-Schwartz inequality}\\
	&\overset{P}{\to}E\left[||\bV||\cdot|M_j|\right]E\left[||\widehat{\boldsymbol{\beta}_k}-\boldsymbol{\beta}_k||\right]\quad\text{by the weak law of large numbers and Slutsky's}
\end{align*}
For $j=3,6$, $E\left[||\bV||\cdot|M_j|\right]=E\left[||\bV|| |Y^{\mathbb{I}\{j=6\}}\right]E[|\epsilon|]<\infty$ since all $\bV,Y$ are bounded and $Var(\epsilon)<\infty$ implies $E|\epsilon|<\infty$. For all other $j$, $E\left[||\bV||\cdot|M_j|\right]<\infty$ since all terms are bounded.

Therefore, for all $j$, $E\left[||\bV||\cdot|M_j|\right]E\left[||\widehat{\boldsymbol{\beta}_k}-\boldsymbol{\beta}_k||\right]=0$ since $\widehat{\boldsymbol{\beta}_k}$ is consistent. So all six terms converge in probability to 0. 

Terms 7-8: We also rewrite these two terms:
\begin{align*}
	\bv_i'\boldsymbol{\beta}_0\epsilon_i&=m_{i,7}\epsilon_i\quad\text{letting }m_{i,7}=\bv_i'\boldsymbol{\beta}_0\\
	\bv_i'\boldsymbol{\beta}_1y_i\epsilon_i&=m_{i,8}\epsilon_i\quad\text{letting }m_{i,8}=\bv_i'\boldsymbol{\beta}_1y_i
\end{align*}

And we see that for $j\in\{7,8\}$:
\begin{align*}
	&\frac 1n \sum_{i=1}^nm_{i,j}\epsilon_i\\
	&\overset{P}{\to}E\left[M_j\epsilon\right]\quad\text{by the weak law of large numbers}\\
	&=E\left[M_j\right]E\left[\epsilon\right]\quad\text{since the error }\epsilon\text{ is independent of the independent variables }\bV,Y\\
	&=0\quad\text{since }E\epsilon=0\text{ and }M_j \text{ are bounded} 
\end{align*}
So these two terms also converge in probability to 0.

Term 9: 
\begin{align*}
	\frac 1n\sum_{i=1}^n-\epsilon_i^2&\overset{P}{\to}E[-\epsilon^2]\quad\text{by the weak law of large numbers}\\
	&<0\quad\text{since }E[\epsilon^2]=Var(\epsilon)>0
\end{align*}

Term 10:
\begin{align*}
	&\frac 1n\sum_{i=1}^n(\hat{x_i}-x_i)\frac 1n\sum_{i=1}^nx_i = 0
\end{align*}
as $\sum \hat x_i = \sum x_i$ for a linear regression with an intercept. 

Combining the results of all 10 terms with Slutsky's theorem, we see that 
\begin{align*}
	(\ast)\overset{P}{\to}E[-\epsilon^2]<0
\end{align*}

Therefore, the covariance between the bias and the true pollutant concentration is asymptotically negative.

\end{proof}

\subsection{Proposition 3 Proof with Covariates}


We proved proposition 3 in the case where the inverse-regression model does not contain covariates in the appendix. Now, we consider the case with covariates in the model. 

\begin{proof} 
	
In the case of a model with covariates, the inverse model can be written as $y_i=\boldsymbol{\alpha_0}'\bz_{a,i}+\boldsymbol{\alpha_1}'\bz_{a,i} x_i +\epsilon_i$ where $\boldsymbol{\alpha_0}=(\beta_0,\boldsymbol{\beta_2}')'$, $\boldsymbol{\alpha_1}=(\beta_1,\boldsymbol{\beta_3}')'$, and $\bz_{a,i}\st=(1,\bz_i')'$. 

Assume that the true pollutant concentrations $x$ and covariates $z_{k}$ are bounded below and above and that $\epsilon$ is normally distributed with $Var(\epsilon)<\infty$. We also assume that 
$|{\boldsymbol{\alpha_1}}'\bz_{a}|>a$ for some $a>0$, so that division by this quantity does not result in overinflated predictions. 
Lastly, we assume that $\bX'\bX/n\overset{P}{\to}\bC$, for a positive definite $\bC$, where $\bX$ is the matrix of independent variables $\bz_{a,i}$ and $\bz_{a,i}x_i$. 

The final assumption means that 
$Var(\boldsymbol{\alpha})=Var(\epsilon)(\bX'\bX)^{-1}=Var(\epsilon)(\bX'\bX/n)^{-1}/n\overset{P}{\to}Var(\epsilon)\bC^{-1}/n$, where the inversion is well defined since $\bC$ is positive definite. Therefore for each term, $Var(\widehat{{\alpha_{1,d}}})=O(1/n)$ for a finite $C_d$. 

We first state two lemmas: 

\textbf{Lemma 1: } $\frac 1n\sum_i\frac{1}{|\widehat{\boldsymbol{\alpha_1}}'\bz_{a,i}|^2}$ is bounded in probability.
\begin{proof}
	Let $\epsilon>0$. We will show that there exist some $N$ such that for $n\geq N$, $P\left(\frac 1n\sum\frac{1}{|\widehat{\boldsymbol{\alpha_1}}'\bz_{a,i}|^2}>\frac{4}{a^2}\right)<\epsilon$. We have
	\begin{align*}
		&P\left(\frac 1n\sum\frac{1}{|\widehat{\boldsymbol{\alpha_1}}'\bz_{a,i}|^2}>\frac{4}{a^2}\right)\\
		&\leq P\left(\bigcup\left\{\frac{1}{|\widehat{\boldsymbol{\alpha_1}}'\bz_{a,i}|^2}>\frac{4}{a^2}\right\}\right)\quad\text{since at least one term must be larger than the average}\\
		&\leq\sum P\left(\frac{1}{|\widehat{\boldsymbol{\alpha_1}}'\bz_{a,i}|}>\frac{2}{a}\right)\quad\text{by sub-additivity}\\
		&=\sum P\left(|\widehat{\boldsymbol{\alpha_1}}'\bz_{a,i}|<\frac{a}{2}\right)\\
		&\leq \sum P\left(||\boldsymbol{\alpha_1}'\bz_{a,i}|-|(\widehat{\boldsymbol{\alpha_1}}-\boldsymbol{\alpha_1})'\bz_{a,i}||<\frac{a}{2}\right)\quad\text{by the inverse triangle inequality}\\
		&\leq \sum P\left(|(\widehat{\boldsymbol{\alpha_1}}-\boldsymbol{\alpha_1})'\bz_{a,i}|>\frac{a}{2}\right)\quad\text{since }|\boldsymbol{\alpha_1}'\bz_{a,i}|>a\\
		&\leq \sum P\left(||\widehat{\boldsymbol{\alpha_1}}-\boldsymbol{\alpha_1}||\cdot||\bz_{a,i}||>\frac{a}{2}\right)\quad\text{by the Cauchy-Schwartz inequality}\\
		&\leq \sum P\left(||\widehat{\boldsymbol{\alpha_1}}-\boldsymbol{\alpha_1}||>\frac{a}{2M}\right)\quad\text{since $||\bz_{a,i}||$ is bounded by } M\\
		&=nP\left(||\widehat{\boldsymbol{\alpha_1}}-\boldsymbol{\alpha_1}||>\frac{a}{2M}\right)\\
		&=nP\left({\sum_{d=1}^{D}(\widehat{{\alpha_{1,d}}}-{\alpha_{1,d}})^2}>\frac{a^2}{4M^2}\right)\quad\text{by the definition of the $L_2$ norm}\\
		&\leq nP\left(\bigcup\left\{(\widehat{{\alpha_{1,d}}}-{\alpha_{1,d}})^2>\frac{a^2}{4M^2D}\right\}\right)\quad\text{since at least one term must be larger than the average}\\
		&\leq n\sum_{d=1}^DP\left((\widehat{{\alpha_{1,d}}}-{\alpha_{1,d}})^2>\frac{a^2}{4M^2D}\right)\quad\text{by sub-additivity}\\
		&\leq n\sum_{d=1}^DP\left(|\widehat{{\alpha_{1,d}}}-{\alpha_{1,d}}|>\frac{a}{2M\sqrt{D}}\right)\\
		&\leq n\sum_{d=1}^D2\exp\left\{-\frac{a^2}{4M^2D}\Big/(2Var(\widehat{{\alpha_{1,d}}}))\right\}\quad\text{by the tail bound of a normal distribution}\\
		&=\sum_{d=1}^D2n\exp\left\{-\frac{a^2}{4M^2D}O(n)/2\right\}\quad\text{since $Var(\widehat{{\alpha_{1,d}}})=O(1/n)$  for all $d$}\\
		&<\epsilon \quad \text{for $n$ large enough since } 2n\exp\left\{-\frac{a^2}{4M^2D}O(n)/2\right\}\to 0\text{ as }n\to\infty \text{ for all }d
	\end{align*}
	So the quantity is bounded in probability.
\end{proof}

\textbf{Lemma 2: }If $A_n\overset{P}{\to}0$ and $B_n$ is bounded in probability, then $A_nB_n\overset{P}{\to}0$.

\begin{proof}
	Let $\epsilon,\delta>0$. Then $\exists M \ni P(|B_n|\geq M) \leq \epsilon/2$ for $n \geq n_B$. Since $A_n\overset{P}{\to}0$, there exists $N_A := N_A(M)$ such that $P(|A_n|\geq \delta/M)<\epsilon/2$.
	\begin{align*}
		P(|A_nB_n|\geq \delta)&=P(|A_nB_n|\geq \delta,|B_n|\geq M)+P(|A_nB_n|\geq \delta,|B_n|<M)\\
		&\leq P(|B_n\geq M)+P(|A_n|\geq \delta/M)
	\end{align*}
	
	 So for $n\geq\max\{N_A,N_B\}$,
	\begin{align*}
		P(|A_nB_n|\geq \delta)&<\epsilon/2+\epsilon/2=\epsilon
	\end{align*}
	So $|A_nB_n|\overset{P}{\to}0$. 
\end{proof}

Next, we see that
\begin{align*}
	\widehat{Cov}(\hat{\bx}-\bx,\bx)&=\frac 1n\sum_{i=1}^{n}(\hat{x}_i-x_i)x_i-\frac 1n\sum_{i=1}^{n}(\hat{x}_i-x_i)\frac 1n\sum_{i=1}^nx_i\\
	&=\frac 1n\sum_{i=1}^{n}\left(\frac{(\boldsymbol{\alpha_0}'\bz_{a,i}+\boldsymbol{\alpha_1}'\bz_{a,i}x_i+\epsilon_i)-\widehat{\boldsymbol{\alpha_0}}'\bz_{a,i}}{\widehat{\boldsymbol{\alpha_1}}'\bz_{a,i}}-x_i\right)x_i\\
	&\phantom{=}-\frac 1n\sum_{i=1}^{n}\left(\frac{(\boldsymbol{\alpha_0}'\bz_{a,i}+\boldsymbol{\alpha_1}'\bz_{a,i}x_i+\epsilon_i)-\widehat{\boldsymbol{\alpha_0}}'\bz_{a,i}}{\widehat{\boldsymbol{\alpha_1}}'\bz_{a,i}}-x_i\right)\frac 1n\sum_{i=1}^nx_i\\
	&=\frac 1n\sum_{i=1}^n\frac{1}{\widehat{\boldsymbol{\alpha_1}}'\bz_{a,i}}\left[(\boldsymbol{\alpha_0}'-\widehat{\boldsymbol{\alpha_0}}')\bz_{a,i}(x_i-\bar{x})\right.\\
	&\phantom{=\frac 1n\sum_{i=1}^n\frac{1}{\widehat{\boldsymbol{\alpha_1}}'\bz_{a,i}}}\left.+(\boldsymbol{\alpha_1}'-\widehat{\boldsymbol{\alpha_1}}')\bz_{a,i}(x_i^2-x_i\bar{x})+\epsilon_i(x_i-\bar{x})\right]
\end{align*}

Terms 1-2: we rewrite the first two terms as follows:
\begin{align*}
	\frac{1}{\widehat{\boldsymbol{\alpha_1}}'\bz_{a,i}}(\boldsymbol{\alpha_0}'-\widehat{\boldsymbol{\alpha_0}}')\bz_{a,i}(x_i-\bar{x})&=\frac{1}{\widehat{\boldsymbol{\alpha_1}}'\bz_{a,i}}(\boldsymbol{\alpha_0}'-\widehat{\boldsymbol{\alpha_0}}')\bz_{a,i}m_{i,0}\quad\text{where }m_{i,0}=x_i-\bar{x}\\
	\frac{1}{\widehat{\boldsymbol{\alpha_1}}'\bz_{a,i}}(\boldsymbol{\alpha_1}'-\widehat{\boldsymbol{\alpha_1}}')\bz_{a,i}(x_i^2-x_i\bar{x})&=\frac{1}{\widehat{\boldsymbol{\alpha_1}}'\bz_{a,i}}(\boldsymbol{\alpha_1}'-\widehat{\boldsymbol{\alpha_1}}')\bz_{a,i}m_{i,1}\quad\text{where }m_{i,1}=x_i^2-x_i\bar{x}
\end{align*}

Then for $j\in\{0,1\},k\in\{0,1\}$:
\begin{align*}
	&\left|\frac 1n\sum_{i=1}^n\frac{1}{\widehat{\boldsymbol{\alpha_1}}'\bz_{a,i}}(\boldsymbol{\alpha_k}'-\widehat{\boldsymbol{\alpha_k}}')\bz_{a,i}m_{i,j}\right|\\
	&\leq\frac 1n\sum_{i=1}^n\left|\frac{1}{\widehat{\boldsymbol{\alpha_1}}'\bz_{a,i}}\right|\left|(\boldsymbol{\alpha_k}'-\widehat{\boldsymbol{\alpha_k}}')\bz_{a,i}\right|\cdot \left|m_{i,j}\right|\quad \text{by the triangle inequality}\\
	&\leq\left(\frac 1n\sum_{i=1}^n\left|\frac{1}{\widehat{\boldsymbol{\alpha_1}}'\bz_{a,i}}\right|\cdot||\bz_{a,i}||\cdot \left|m_{i,j}\right|\right)\cdot||\boldsymbol{\alpha_k}-\widehat{\boldsymbol{\alpha_k}}||\quad\text{by the Cauchy-Schwartz inequality}\\
	&\leq\left(\frac 1n\sum_{i=1}^n\frac{1}{\left|\widehat{\boldsymbol{\alpha_1}}'\bz_{a,i}\right|^2}\right)^{1/2}\left(\frac 1n\sum_{i=1}^n||\bz_{a,i}||^2\cdot \left|m_{i,j}\right|^2\right)^{1/2}||\boldsymbol{\alpha_k}-\widehat{\boldsymbol{\alpha_k}}||\quad\text{by Cauchy-Schwartz}\\
\end{align*}

By Lemma 1, the first term is bounded in probability. Also, since the covariates are bounded, there exists a $B>0$ such that $||\bz_{a,i}||\leq B$, and $X$ being bounded means there exists $C>0$ such that $\left|m_{i,j}\right|\leq C$. Then $\frac 1n\sum_{i=1}^n||\bz_{a,i}||^2\cdot \left|m_{i,j}\right|^2\leq \frac 1n\sum_{i=1}^nB^2C^2$. Lastly, $||\boldsymbol{\alpha_k}-\widehat{\boldsymbol{\alpha_k}}||\overset{P}{\to}0$ by the consistency of $\widehat{\boldsymbol{\alpha_k}}$. 

By Lemma 2, the entire quantity converges to 0 in probability. So the first two terms converge to 0 in probability. 

For the third term, 
\begin{align*}
	\left|\frac 1n\sum_{i=1}^n\frac{1}{\widehat{\boldsymbol{\alpha_1}}'\bz_{a,i}}\epsilon_i(x_i-\bar{x})\right|
	&\leq \left|\frac 1n\sum_{i=1}^n\left(\frac{1}{\widehat{\boldsymbol{\alpha_1}}'\bz_{a,i}}\epsilon_i(x_i-\bar{x})-\frac{1}{{\boldsymbol{\alpha_1}}'\bz_{a,i}}\epsilon_i(x_i-\bar{x})\right)\right|\\
	&\phantom{=}+\left|\frac 1n\sum_{i=1}^n\frac{1}{{\boldsymbol{\alpha_1}}'\bz_{a,i}}\epsilon_i(x_i-\bar{x})\right|\quad\text{by the triangle inequality}
\end{align*}

Note that the second part of the expression tends to 0 asymptotically since
\begin{align*}
	\left|\frac 1n\sum_{i=1}^n\frac{1}{{\boldsymbol{\alpha_1}}'\bz_{a,i}}\epsilon_i(x_i-\bar{x})\right|
	&=\left|\frac 1n\sum_{i=1}^n\frac{1}{{\boldsymbol{\alpha_1}}'\bz_{a,i}}\epsilon_ix_i-\frac 1n\sum_{i=1}^n\frac{1}{{\boldsymbol{\alpha_1}}'\bz_{a,i}}\epsilon_i\bar{x}\right|
\end{align*}
We apply the weak law of large numbers directly to the first term to see $\frac 1n\sum_{i=1}^n\frac{1}{{\boldsymbol{\alpha_1}}'\bz_{a,i}}\epsilon_ix_i\overset{P}{\to}E\left[\frac{1}{{\boldsymbol{\alpha_1}}'\bZ_{a}}\epsilon X\right]=E[\epsilon]E\left[\frac{1}{{\boldsymbol{\alpha_1}}'\bZ_{a}} X\right]=0$ since $\epsilon$ is independent of $X,\bZ_{a}$ and $E[\epsilon]=0$. For the second term we write
\begin{align*}
	\frac 1n\sum_{i=1}^n\frac{1}{{\boldsymbol{\alpha_1}}'\bz_{a,i}}\epsilon_i\bar{x}&=\left(\frac 1n\sum_{i=1}^n\frac{1}{{\boldsymbol{\alpha_1}}'\bz_{a,i}}\epsilon_i\right)\left(\frac 1n\sum_{i=1}^nx_i\right)\\
	&\overset{P}{\to}E\left[\frac{1}{{\boldsymbol{\alpha_1}}'\bZ_{a}}\epsilon\right]E[X]\quad\text{by the weak law of large numbers and Slutsky's theorem}\\
	&=E[\epsilon]E\left[\frac{1}{{\boldsymbol{\alpha_1}}'\bZ_{a}}\right]E[X]\quad\text{since $\epsilon$ and $\bZ_{a}$ are independent}\\
	&=0\quad\text{since }E[\epsilon]=0
\end{align*}

For the first part of the expression,
\begin{align*}
	&\left|\frac 1n\sum_{i=1}^n\left(\frac{1}{\widehat{\boldsymbol{\alpha_1}}'\bz_{a,i}}\epsilon_i(x_i-\bar{x})-\frac{1}{{\boldsymbol{\alpha_1}}'\bz_{a,i}}\epsilon_i(x_i-\bar{x})\right)\right|\\
	&=\left|\frac 1n\sum_{i=1}^n\frac{({\boldsymbol{\alpha_1}}-\widehat{\boldsymbol{\alpha_1}})'\bz_{a,i}}{(\widehat{\boldsymbol{\alpha_1}}'\bz_{a,i})({\boldsymbol{\alpha_1}}'\bz_{a,i})}\epsilon_i(x_i-\bar{x})\right|\\
	&\leq\frac 1n\sum_{i=1}^n\frac{|({\boldsymbol{\alpha_1}}-\widehat{\boldsymbol{\alpha_1}})'\bz_{a,i}|}{|\widehat{\boldsymbol{\alpha_1}}'\bz_{a,i}|\cdot|{\boldsymbol{\alpha_1}}'\bz_{a,i}|}|\epsilon_i|\cdot|x_i-\bar{x}|\quad\text{by the triangle inequality}\\
	&\leq ||{\boldsymbol{\alpha_1}}-\widehat{\boldsymbol{\alpha_1}}||\frac 1n\sum_{i=1}^n\frac{||\bz_{a,i}||}{|\widehat{\boldsymbol{\alpha_1}}'\bz_{a,i}|\cdot|{\boldsymbol{\alpha_1}}'\bz_{a,i}|}|\epsilon_i|\cdot|x_i-\bar{x}|\quad\text{by the Cauchy-Schwartz inequality}\\
	&\leq ||{\boldsymbol{\alpha_1}}-\widehat{\boldsymbol{\alpha_1}}||\left(\frac 1n\sum_{i=1}^n\frac{||\bz_{a,i}||^2\epsilon_i^2(x_i-\bar{x})^2}{({\boldsymbol{\alpha_1}}'\bz_{a,i})^2}\right)^{\frac 12}\left(\frac 1n\sum_{i=1}^n\frac{1}{(\widehat{\boldsymbol{\alpha_1}}'\bz_{a,i})^2}\right)^{\frac 12}\quad\text{by Cauchy-Schwartz}
\end{align*}

In this expression, $||{\boldsymbol{\alpha_1}}-\widehat{\boldsymbol{\alpha_1}}||\overset{P}{\to}0$ by the consistency of the least squares estimator. 
$\left(\frac 1n\sum_{i=1}^n\frac{||\bz_{a,i}||^2\epsilon_i^2(x_i-\bar{x})^2}{({\boldsymbol{\alpha_1}}'\bz_{a,i})^2}\right)\leq\frac 1n\sum_{i=1}^n\frac{D^2F^2}{a^2}\epsilon_i^2\overset{P}{\to}\frac{D^2F^2}{a^2}Var(\epsilon) <\infty$ since $\bz_{a,i},x_i-\bar{x}$ are bounded (with the bounds denoted as $D,F$ respectively), and $Var(\epsilon)<\infty$. So this factor is bounded in probability. Finally, by Lemma 1, the last factor is bounded in probability. By Lemma 2, the entire third term converges to 0 in probability. 

%

Since all three terms tend to 0 in probability, we can use Slutsky's theorem to conclude that $\widehat{Cov}(\hat{\bx}-\bx,\bx)\overset{P}{\to}0$.

\end{proof}

\section{Predicting pollutant concentrations on a grid}\label{sec:supp_grid}
We provide the details of predicting pollutant concentrations on a grid of locations (Set D) with neither low-cost or reference data (Section \ref{sec:grid}). 
To write the conditional distributions of these random variables, we will drop indexing the time $t$ and use a subscript to notate the set (for example $\bx_{D}=\bx\sdt$). Also, let $\bar{D} = A \cup B \cup C$. 
Given $\bx_{\bar D}$ and $\by_B$ the conditional distributions of the true pollutant surface at the grid locations are: 
\begin{align}
\label{eq:jointkrig}
\begin{split} 
\bx_D|\bx_{\bar{D}}, \by_B =  \bx_D|\bx_{\bar{D}}&\sim N\left(\tilde{\bmu}_D,\bSigma_D\right)\\
	\text{where  }\quad \tilde \bmu_D &=\mu\bones+\bC_{D\bar{D}}\bC_{\bar{D}\bar{D}}^{-1}(\bx_{\bar{D}}-\mu\bones)\\
	\bSigma_D&=\bC_{DD}-\bC_{D\bar{D}}\bC_{\bar{D}\bar{D}}^{-1}\bC_{\bar{D}D}.
\end{split}
\end{align}

Given the true pollutant at the locations in $\bar D$, the conditional distribution at the grid locations $D$ does not depend on the low-cost data $\by_B$. For the Bayesian implementation, as $D$ will often be moderate or large, we will replace $\bSigma_D$ with $diag(\bSigma_D)$, i.e., conduct independent kriging for computational efficiency. For the frequentist implementation, given the predictions $\hat \bx_B =\bx_{update}\slt$ from (\ref{kalman}), the predictions on a grid of locations is simply:

\begin{equation}\label{eq:predgrid}
	\hat\bx_D = 	\mu\bones+\bA_1(\bx_A-\mu\bones)+\bA_3(\bx_C-\mu\bones)+\bA_2(\hat\bx_B-\mu\bones)
\end{equation}
where $\bA_1,\bA_2,\bA_3$ are blocks of columns of $\bC_{D\bar{D}}\bC_{\bar{D}\bar{D}}^{-1}$ 
corresponding to 
$\bx_A,\bx_B,\bx_C$ respectively.

\section{GP Filter without collocation}\label{sec:supplement_no_colocation}
We made the assumption that there is at least one site in Set A (the set of collocated sites) when developing the GP Filter. This is because most low-cost sensor networks use some form of regression-calibration for collocation and thereby need at-least one site with a reference device for collocation. However, in the event that a network does not have any collocated site, we explore a possible approach to perform the filtering. This approach can be explained in two steps:
\begin{enumerate}
    \item at each time point in the training set, we use kriging on the low-cost network data to predict the low-cost measurement at the reference site(s)
    \item train the observation model by treating the predicted low-cost measurements and the actual reference measurements as collocated data and run the GP Filter with this observation model.
\end{enumerate}

We apply this method and compare to GP Filter with collocation to demonstrate why not having collocation impacts the performance of the model. We also assess the regression-calibration (RegCal) approach for the same two settings, i.e., with or without collocation. 


		\begin{figure}[!htbp]
	\centering
\includegraphics[width=3.5in]{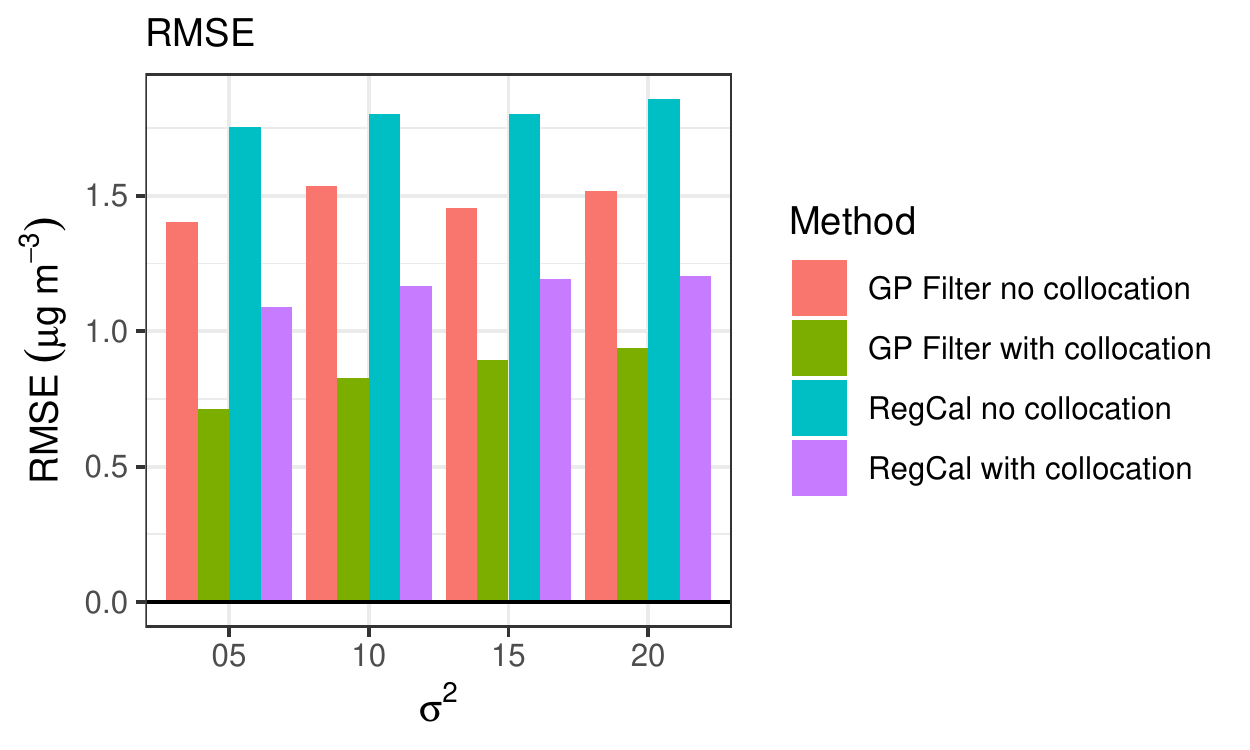}
	\caption{RMSE of GP Filter and regression-calibration with and without collocation}
	\label{supplement_no_colocation_performance}
\end{figure}

We see from Figure \ref{supplement_no_colocation_performance}  that empirically, the GP Filter still outperforms RegCal when both are used without collocation. Therefore, if a network does not have collocation, we still recommend using the GP Filter instead of RegCal, though we acknowledge this is not a typical setting. However, we also see that each method has considerably lower RMSE when there is collocation compared to when there is not, showing the benefit of having collocation. 
 
To explain this, note that kriging effectively corresponds to assuming a marginal model for the low-cost data $y(s) \sim GP(\mu_y, C_y(\cdot, \cdot))$. Typically, for kriging, the mean $\mu_y$ will either be assumed to be zero or a constant and a stationary model will be used for the covariance family. However, this leads to severe model specification. Our theoretical results on the under-estimation issue suggest that the model for the two data sources should be specified as $y | x \times x$ and not in the other direction. Hence, we specify a marginal GP model for the true pollution surface $x \sim GP(\mu_x, C_x(\cdot,\cdot))$ and assume that the low-cost measurements $y$ are a noisy function of the truth with the bias given by the observation model (\ref{obsmodel}). Omitting the indexing with time, we have the marginal model for the low-cost data as 
\begin{equation}\label{eq:margy}
    y(s) \sim GP(\beta_0 + \bbeta_2'\bz(s), (\beta_1 + \bbeta_3'\bz(s))^2 C_x(\cdot,\cdot) + \tau^2 \delta(\cdot,\cdot))
\end{equation}
where $\delta(s_1,s_2)=I(s_1=s_2)$ is the Kronecker delta specifying the nugget term. 

This implied marginal model for $y$ will be have both means and variances to be non-stationary in the covariates $\bz$.  
Thus kriging the low-cost measurements using a stationary model is not consistent with our  assumed data generation model and leads to severe misspecification in both the GP mean and variance. Hence, especially with smaller sample size (network size), there can be bias in predicting the low-cost measurement at the reference site. 

The second issue is that without exact collocation the observation model coefficients cannot be estimated in an unbiased manner even if the kriging model is correctly specified. To show this, we assume (unrealistically) that the meteorological covariates did not contribute to the bias of the low-cost data, i.e., $\bbeta_2=\bbeta_3=\bf{0}$ in  (\ref{eq:margy}) in which case the kriging model agrees with the true marginal model for $y$  as long as $C_y$ is chosen to be from the same family as $C_x$ plus a nugget. 

We show that, even in this favorable case, the observation model coefficients cannot be estimated in an unbiased manner without exact collocation. Consider a two-site network, with $s$ being the low-cost site and $s_0$ denote the reference site. Then the kriging prediction of $y$ at $s_0$ given $y(s)$ is
$$ \widehat{y(s_0)} = E(y(s_0) | y(s)) = \frac {C_y(s,s_0)}{C_y(s_0,s_0)} y(s)$$
Then the estimate of $\beta_1$ (slope of $y$ on $x$) based on this imputed data is given by \begin{equation}\label{eq:bias}
    \widehat {\beta}_{1, no\, colloc} = \frac {\widehat{Cov}(\widehat{y(s_0)},x(s_0))}{\widehat{Cov}(x(s_0),x(s_0))} = \frac {C_y(s,s_0)}{C_y(s_0,s_0)} \frac {\widehat{Cov}(y(s_0),x(s_0))}{\widehat{Cov}(x(s_0),x(s_0))} = \frac {C_y(s,s_0)}{C_y(s_0,s_0)}  \widehat {\beta}_{1, colloc}
\end{equation}
where $\widehat{Cov}$ denotes the sample covariance. It is easy to see that 
\begin{equation}\label{eq:biaslim}
    \lim \widehat {\beta}_{1, no\, colloc} = \frac {C_y(s,s_0)}{C_y(s_0,s_0)} \beta_1.
\end{equation}
Thus, without collocation, the estimation slope coefficient $\beta_1$ in the observation model will be biased with a multiplicative bias of $\frac {C_y(s,s_0)}{C_y(s_0,s_0)}$ that goes to zero as $s \rightarrow s_0$, i.e., as we approach exact collocation. 

		\begin{figure}[!hbp]
	\centering
	\includegraphics[width=5.5in]{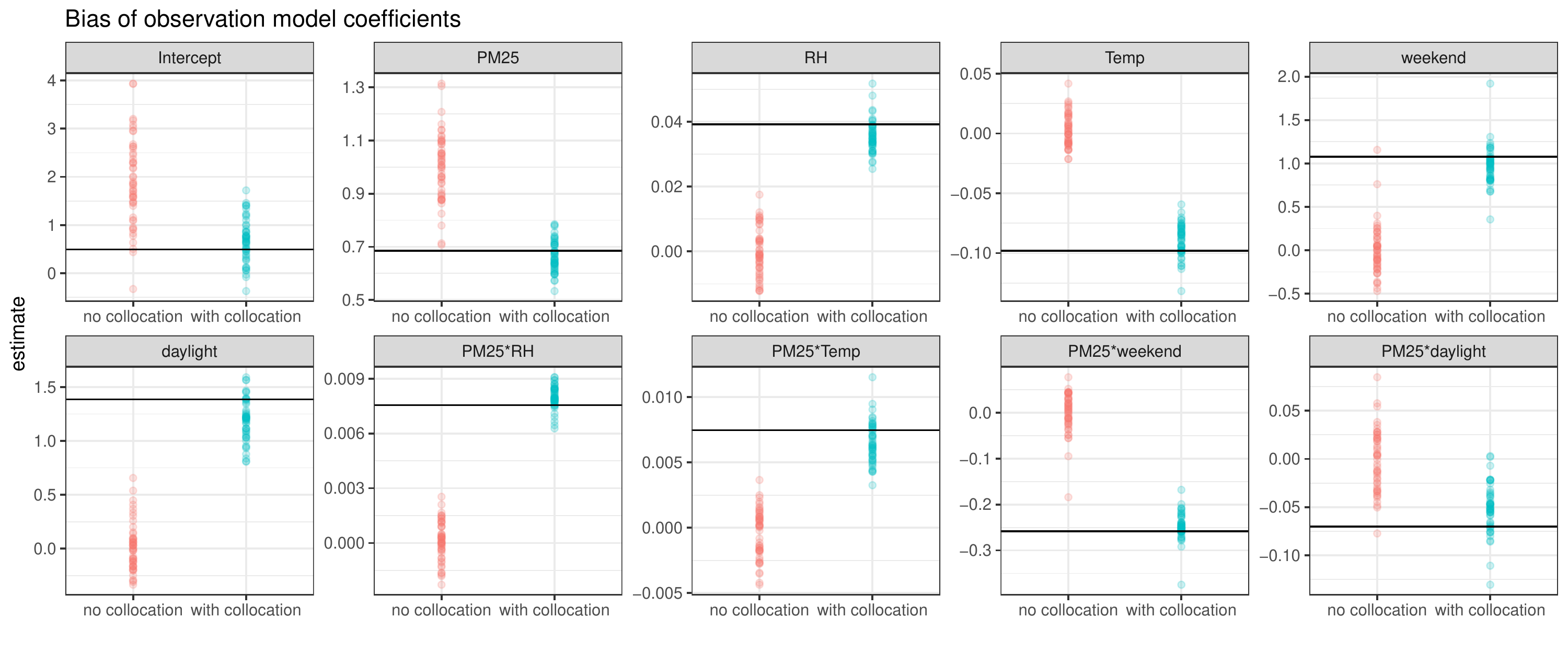}
	
	\caption{Bias of observation model coefficients when there is no collocated site and compared to when there is one collocated site, when $\sigma^2=10$. The true value of the coefficient is given by the black horizontal lines.}
	\label{supplement_no_colocation_bias}
\end{figure}

The more realistic case with covariates and more than two sites in the network is not analytically tractable. Hence, we study it using a simulation. In Figure \ref{supplement_no_colocation_bias} we plot the estimated observation model parameters  from the GP Filter with and without collocation based on the simulated datasets. We see the notable bias for the setting without collocation.


\section{Modeling time dependence}\label{sec:supst}

Our proposed spatial filtering is quite flexible and can easily be adapted to model temporal correlation both in the observation model and in the state-space model. Time structure in the observation model can be specified in two ways. Studies using longer-term deployment of low-cost sensors have often reported drift in the readings over time \citep{Miskell2018}. This would need to be modeled by including time-based bias terms (e.g. linear trend or non-linear splines) in the observation model. If there is no time-drift, there can also be temporal correlation in the biases (residuals) of the low-cost data. This can be modeled by including an adequate dependence structure in the error process of the observation model and using a generalized least squares for efficient estimation of the regression coefficients. 
However, only short-term time dependence in the residuals, without any longer-term time-drift, is of less concern. With abundant collocated training data the observation model coefficients can be estimated well with the working assumption of independent residuals. 

For modeling time in the state-space model for the true pollutants, we present one approach here, based on suggestions from one reviewer. We assume like before that the observation model has been trained with abundant collocated data. Let $t=1,\ldots,U$ denote the time points where we are interested in prediction. We now model the true pollutant levels as a spatio-temporal GP $x(\cdot,\cdot) = \{x(\bs,t) : \bs \in \calD, t \in [1,U]\} \sim GP(\mu,C)$ where $\mu(s,t)$ is the mean at time $t$ and location $\bs$ and $C$ denote any valid spatio-temporal covariance function specified by parameters $\theta$ which we temporarily assume to be known. Let $\calF_t(A \cup C) = \{X(s,u) : s \in A \cup C, u \leq t\}$ denote all reference data up to time $t$. Then we can modify the prediction equation (\ref{kriging}) to 
\begin{align*}
\label{kriging_st}
\begin{split}
\bx\slt|\calF_t(A \cup C) &\sim N(\tilde\bmu_t,\bSigma_t).
\end{split} 
\end{align*}
Here $\tilde \mu_t$ is now the spatio-temporal kriging (conditional) mean  of the true pollutant levels at time $t$ at the low-cost sites $B$ given all reference data at sites $A \cup C$ upto time $t$. The corresponding kriging variance is $\bSigma_t$. These quantities will have expressions similar to those provided in (7) but based on the larger spatio-temporal conditioning set $\calF_t(A \cup C)$. Given the predict step from this model the Kalman update step can proceed in the same way as (\ref{kalman}). Other spatio-temporal processes like dynamic space-time models that discretizes the time and uses an autoregressive temporal model can also be easily used. 

In addition to a temporal model potentially smoothing out localized peaks in time, a spatio-temporal filtering would also be computationally intensive. If one adopts a sequential strategy, the covariance parameters $\theta$ would need to be re-estimated for every new time $t$ using data from times upto $t-1$. Thus the process would need to be repeated $U$ times, each time with an increasing amount of training data. Alternatively, one can consider a joint approach but that will also need to estimate $\theta$ based on the likelihood for $\calF_U(A \cup C)$. As GP likelihoods scale cubically with respect to the total number of space-time points, for $n$ spatial locations, this would incur a cost of $O(n^3 U^3)$ and will be prohibitive for even moderate number of time points $U$ even if the number of reference locations $n$ is very small. In that case, one would need to resort to scalable approximations of spatio-temporal GP likelihoods \citep{dnngp}.

\section{Pareto calibration model}\label{sec:supplement_pareto}

The generalized Pareto distribution has the cumulative distribution function $$F(x)=\left\{\begin{array}{ll}
    1-\left(1+\frac{\xi(x-\mu)}{\sigma}\right)^{-1/\xi} & \text{  for }\xi\neq 0 \\
    1-e^{-\frac{(x-\mu)}{\sigma}} & \text{  for }\xi=0
\end{array}\right.$$ where $\mu$ is the threshold. The support is $x\geq \mu$ when $\xi\geq 0$ and $x\in[\mu,\mu-\sigma/\xi]$ when $\xi<0$. The mean of the distribution is $\mu+\frac{\sigma}{1-\xi}$ when $\xi<1$. We model 
\begin{align*}
    \sigma=&\exp(\gamma_1+\gamma_2\log (y)+\gamma_3\log (RH)+\gamma_4\log (T) +\gamma_5weekend+\gamma_6daylight\\
    &+\gamma_7\log (y)\log (RH)+\gamma_8\log (y)\log (T) +\gamma_9\log(y) weekend+\gamma_{10}\log (y) daylight)\\
    \xi=&\exp(\gamma_0)-0.5
\end{align*}

where we use the $\log$ transformation on the covariates -- the continuous low-cost measurements $y$ and the meteorological variables RH and T so that when they are exponentiated, large values of these covariates don't result in very inflated predictions. 
Since this distribution cannot take values less than $\mu$, we can set $\mu$ to be a value beyond which we wish to make predictions of concentrations. We apply this method to the data where the true concentration is $>12$ (the threshold for a ``moderate'' AQI classification according to the EPA). 

We can train it on the subset of the training window $\calW$ where reference sites measured moderate or unhealthy concentrations. Our simulations were made to resemble the PM$_{2.5}$ distributions in Baltimore, so there are not many observations in this range. Therefore, applying the Pareto model to the same training dataset would dramatically decrease the sample size and lead to huge uncertainty in estimates. In simulations, for any time points with ``good'' concentrations at the reference site, we therefore generated new values of the true concentrations that were ``moderate'' and corresponding low-cost measurements. This way, the training datasets were the same size in all models. 

To make predictions from the model, we would like to apply it to all location-time pairs where the true concentration $x(\bs,t)$ is greater than 12 (ie is moderate/unhealthy). 
However, we do not know which low-cost measurements at other sites correspond to true concentrations that are greater than 12. Instead, we run the model for all time points where the low-cost measurement is greater than 12, since we assume that the low-cost measurement is equal to the true concentration for true concentrations less than 12. 

We would also like to note that while the results presented here are for a threshold of 12, which has scientific meaning for air quality in US but may represent vastly different quantiles of different concentration distributions. We also looked at using the 75th percentile of the training dataset as the threshold. This produces relatively similar results but loses some interpretation of why we only apply the model to these values. 

As highlighted in the Discussion Section, the biggest drawback of threshold-based approaches is that they do not calibrate the noisy low-cost data at low concentrations and cannot be applied for calibrating air pollution data for a city like Baltimore with concentrations predominantly below threshold. We now discuss some other  drawbacks of the Pareto model. The threshold calibration methods also suffer from low sample size issue for this type of applications, as the majority of the air pollution data will be below the threshold in many US cities. In our simulations, we generated equal training data for the Pareto and other models to make sure the Pareto could be fairly compared. However, even in this case, the Pareto model was only at par with the GP Filter for higher concentrations (Figure \ref{sim_1a_rmse}, right). Since the GP Filter also calibrated the baseline concentrations, its overall performance across all concentrations was much better (Figure \ref{sim_1a_rmse}, left). 

On the other hand, in applications with more data above the threshold, the Pareto distribution would no longer be modelling the tail of the air quality distribution. Since the Pareto's usefulness is largely for tails of distributions, it is not as applicable for modelling large portions of the data. We see the disadvantage of using the Pareto for a larger proportion of the data to some extent in Figure \ref{sim_1a_rmse} (right), where the RMSE of the Pareto on moderate concentrations increases as $\sigma^2$ increases. Higher $\sigma^2$ results in a greater spread in the true concentrations, so more concentrations are above the moderate threshold. Therefore, at both lower and higher baseline concentrations, the GP Filter is a more natural choice compared to a Pareto regression.

Additionally, a threshold calibration approach has a risk of exposure misclassification around the threshold. The Pareto can be trained on all data where the true concentration $x$ is above the threshold $c$ ($x\geq c$), but when applying the trained model to predict the true concentration $x$, we do not know if $x$ is above or below the threshold. There is only knowledge of the low-cost data $y$, and one can use the same threshold on these low-cost measurements $y$, calibrating data where $y\geq c$. There can be two kinds of exposure misclassification when this is done. A datapoint with a true exposure above the threshold but low-cost measurement below the threshold ($x\geq c,y<c$) will not be calibrated despite having a high true concentration, hence $\hat x = y < c$. Second, a datapoint with a true concentration below the threshold but a low-cost measurement above the threshold ($x<c,y\geq c$) will be calibrated, but the prediction will have to be greater than the threshold by assumption, yielding $
\hat x > c$. Thus threshold-based approaches for calibration of low-cost air-pollution data will lead to exposure misclassification both above and below the threshold in opposite directions. These types of misclassification will likely occur in cities more often with true concentrations around the threshold. 

The Pareto model we presented was not spatial, as opposed to the GP filter which leverages spatial correlation with latest available regulatory data in the region, yielding dynamic calibrations. This is because while there are spatial/spatio-temporal Pareto models in the literature, they are often not suitable for the setting of calibration of low-cost air pollution sensor data. To illustrate, the majority of spatial or spatio-temporal Pareto models \citep{cooley2010spatial,fuentes2013nonparametric,sang2009hierarchical} use hierarchical models that include a spatial random effect on one or more of the parameters in the Pareto distribution. 
Generally, these models can be written as  
$$x(s) \sim Pareto(\mu=c, \sigma=\beta_\sigma^\top(y(s),Z(s))+w_\sigma(s), \xi= \beta_\xi^\top(y(s),Z(s))+w_\xi(s)),$$
where $c$ is the threshold, $x(s)$, $y(s)$  and $Z(s)$ are respectively the true PM$_{2.5}$, low-cost PM$_{2.5}$ and covariates at location $s$ and $w_\sigma(s)$ and $w_\xi(s)$ are spatial random effects. \cite{bacro2020hierarchical}, takes a similar but slightly different approach, writing the Pareto model as a conditional exponential with a latent gamma and uses gamma random fields in a second stage. The issue of these classes of models for our setting is that we may have collocated reference data $x(s)$ at only 1 location $s$ (as in our study in Baltimore, $s$ being the Oldtown location). So, there is no spatially resolved data to estimate the random effects for such a model. Even in most other cities there will be very few locations to properly estimate these random effects. 

Our approach, due to inverting the regression and writing $y(s) \sim N(\beta^\top(x(s), Z(s)), error)$ still works even if $x(s)$ is measured at only one location. This is because the GP prior for $x(s)$ induces a GP prior for $y(s)$ through this formulation, and $y(s)$ is observed at several locations, thereby allowing estimation of the spatial structure. 

Another spatial Pareto regression approach \citep{bortot2022model} uses a Gaussian Process along with the Pareto, but it reverses the order by first using a Pareto model and then applying a GP to the (transformed) predictions from the Pareto. This is similar to using the forward regression calibration to calibrate at each low-cost site and then using GP to spatially smooth the predictions, with the difference being that the Pareto regression is only done at the tails of the distribution. 
This essentially leads to two different generative models for the true low-cost pollution

\begin{align*}
&\mbox{Calibration model:}\qquad x(s) \sim Pareto(\mu=c, \sigma=\beta_\sigma^\top(y(s),Z(s)), \xi= \beta_\xi^\top(y(s),Z(s)))\\
&\mbox{Spatial smoothing model:} \qquad x\left(s\right) \sim GP
\end{align*}

Two generative models of the same quantity is generally undesirable from a modeling perspective. It leads to two different estimates of $x(s)$ where there is a reference site – the reference data itself and the prediction from the calibration model. Our approach avoids this by using a inverse regression which is a generative model for $y(s)$ and then the Gaussian Process as the generative model for $x(s)$. Thus, we do not get two estimates of $x(s)$, and at sites $s$ with reference data, our predictions exactly agree with the latest reference data $x(s)$. 

This coherent modeling of spatial correlation in the GP filter makes it naturally dynamic as we can use latest reference data to produce estimates of $x(s)$ at other sites $s$. This is one of the main reasons why it performs well both at high and low concentrations. When the concentrations within the city are high, the GP Filter uses the high true concentrations at reference sites to predict high concentrations across the network. Therefore, peaks in time are identifiable. Similarly, when reference concentrations are low, this information is being conditioned into the GP formulation. 

Also, using a forward regression (either linear or Pareto), as was done in the papers mentioned, essentially models the measurement error of collocated low-cost data as a Berkson error. We have argued that a classical measurement error is more appropriate in the setting of low-cost air pollution networks, and it is more organic to model the low-cost measurement as a function of the reference data at that location and not the other way round. 

Additionally, in the two generative model setting, the spatial smoothing model would generally just use point estimates from the calibration model, thereby not propagating uncertainty, whereas our formulation yields a coherent hierarchical Bayesian model with a natural propagation of uncertainty into the final predictions of $x(s)$.

\section{Extended analysis of the SEARCH low-cost network PM$_{2.5}$ data}\label{sec:supplement_search}

\subsection{Details of the SEARCH network}\label{sec:network}
The sensor locations were determined by a weighted random sampling approach with weights based on population, NO$_2$ concentrations derived from satellite, proximity to energy generating units and other point sources and roads. Monitors were sited to reduce bias from nearby sources or physical features, and data after installation was assessed for such interferences. This resulted in a well-distributed network of monitors across locations in the city of Baltimore that included a mix of neighborhood or environmental characteristics and degrees of urbanity (i.e., urban vs. suburban). For example, it included monitors spanning from industrial areas to parks.

The Plantower PM sensor used in SEARCH also used in many other networks \citep{feenstra2019performance,malings2020fine,magi2020evaluation}, including the PurpleAir network, a large nationwide network of low-cost air pollution sensors. The Plantower sensors have been shown to be moderately well correlated with collocated reference devices ($R^2$=0.9) however there is considerable upwards bias in absolute measurements. 
One reason for this is that many commercially available low-cost sensors are often manufactured and calibrated in another country and this manufacturing-phase calibration  may not work for field deployment in other ambient settings or countries due to differences in PM composition and size distribution. Additionally, measurements from these types of sensors are known to depend on meteorological conditions, such as relative humidity (RH) and temperature (T). Relative humidity affects measurements through hygroscopy. All PM sensors are optical sensors that measure pollution concentrations based on the scattering of light. Hygroscopy occurs when water molecules attach themselves to PM$_{2.5}$ and result in the optical sensor detecting more PM$_{2.5}$ than is present resulting in substantial bias as RH increases. There can also be a drift in measurements over time as the sensors remain deployed for years.

\subsection{Additional analysis}
The analysis presented in the main text trains on hourly data from November 2019 to apply the models daily or hourly in December 2019. Now, we show the results of the full data analysis on six month of data, from December 2019 to May 2020. 

\begin{figure}[!htbp]
	\centering
	\includegraphics[width=5in]{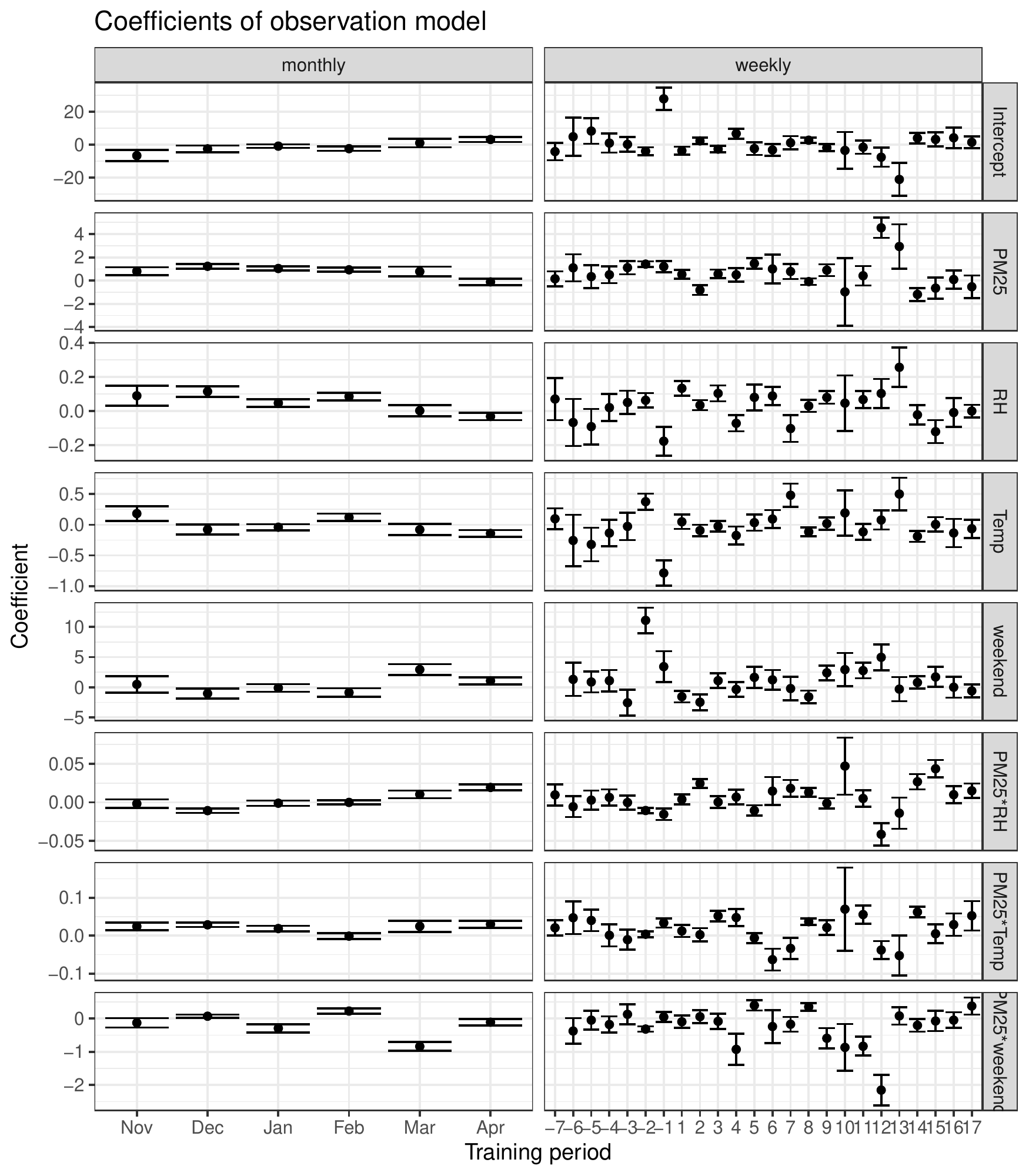}
	\caption{Coefficients of the observation model with different training windows. The left column has the coefficients trained over a month of data at a time. The right hand column is the coefficients trained over a week of data at a time, with the weeks with positive integer labels counting the week number in 2020, and the negative labels counting the weeks backwards into 2019. }
	\label{supp_search_obs_coefs}
\end{figure}

For the longer analysis, we begin by investigating best length of time window for training the observation model hold. We consider both monthly or weekly training windows. For the former, we train the observation model on each month of data between November 2019 and April 2020 to compare the coefficients. We also train on data at the weekly level within that six month training period to compare the performance. Figure \ref{supp_search_obs_coefs} shows the coefficients for each training window. We see in the left hand column that there is variation across months for most of the coefficients. Additionally, the confidence intervals for each month are narrow (owing to being estimated from a larger training data) and often  do not overlap, especially for months farther away from each other. This gives evidence that there is some change in the true observation model month to month. On the weekly scale, shown on the right column, we see variation between neighboring weeks but more overlapping confidence intervals for each coefficient due to lack of adequate sample size. There is more uncertainty when training on the weekly scale since there are at most 168 observations. We decide to train on the monthly scale since there are  variations across months and there is good certainty in training the coefficients. We will use each month's hourly data to train the observation model used to predict in the following month. We train the RegCal model on the same period as the observation model of the GP filter to be able to fairly compare the results. 

\begin{figure}[!htbp]
	\centering
	\includegraphics[width=3in]{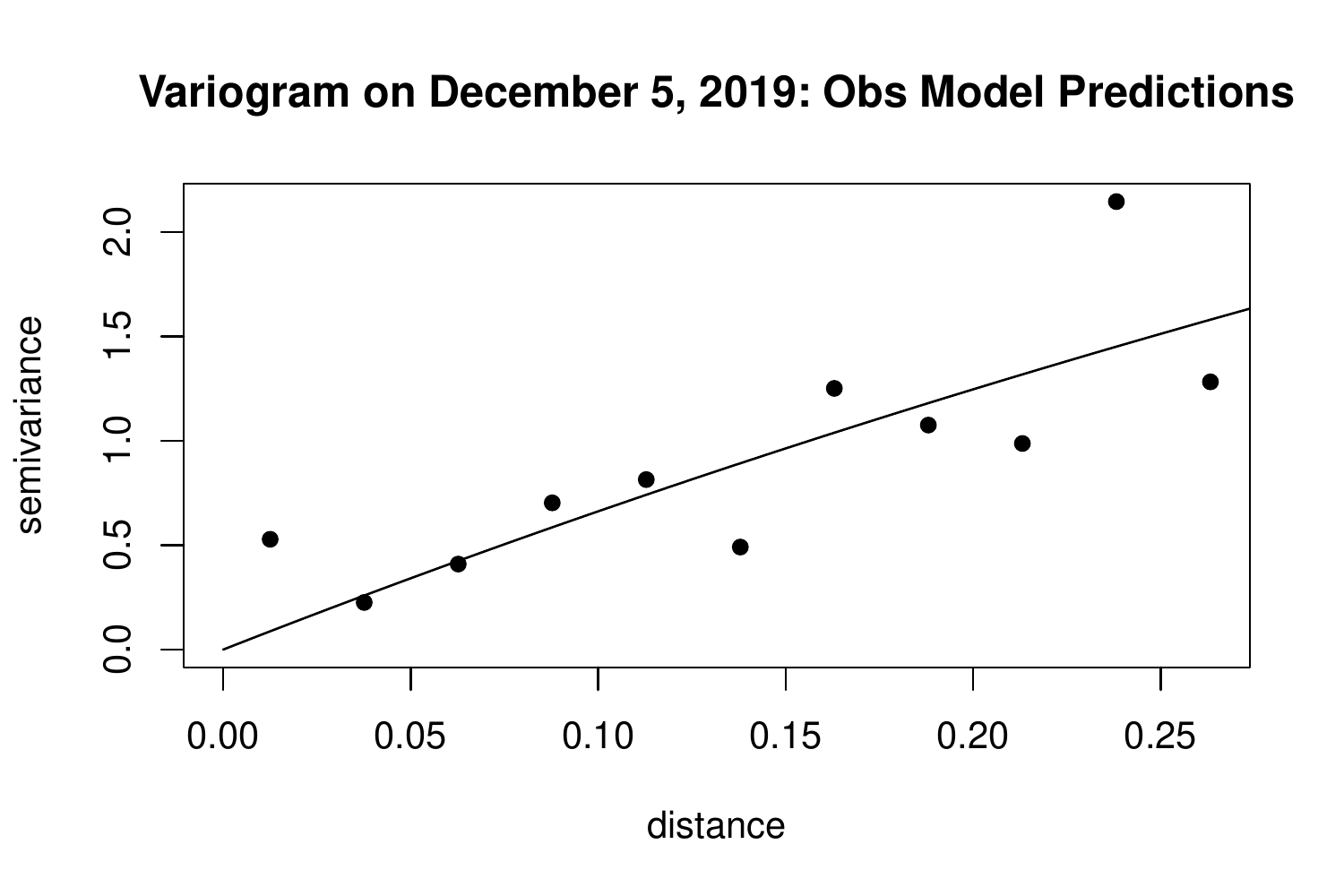}\includegraphics[width=3in]{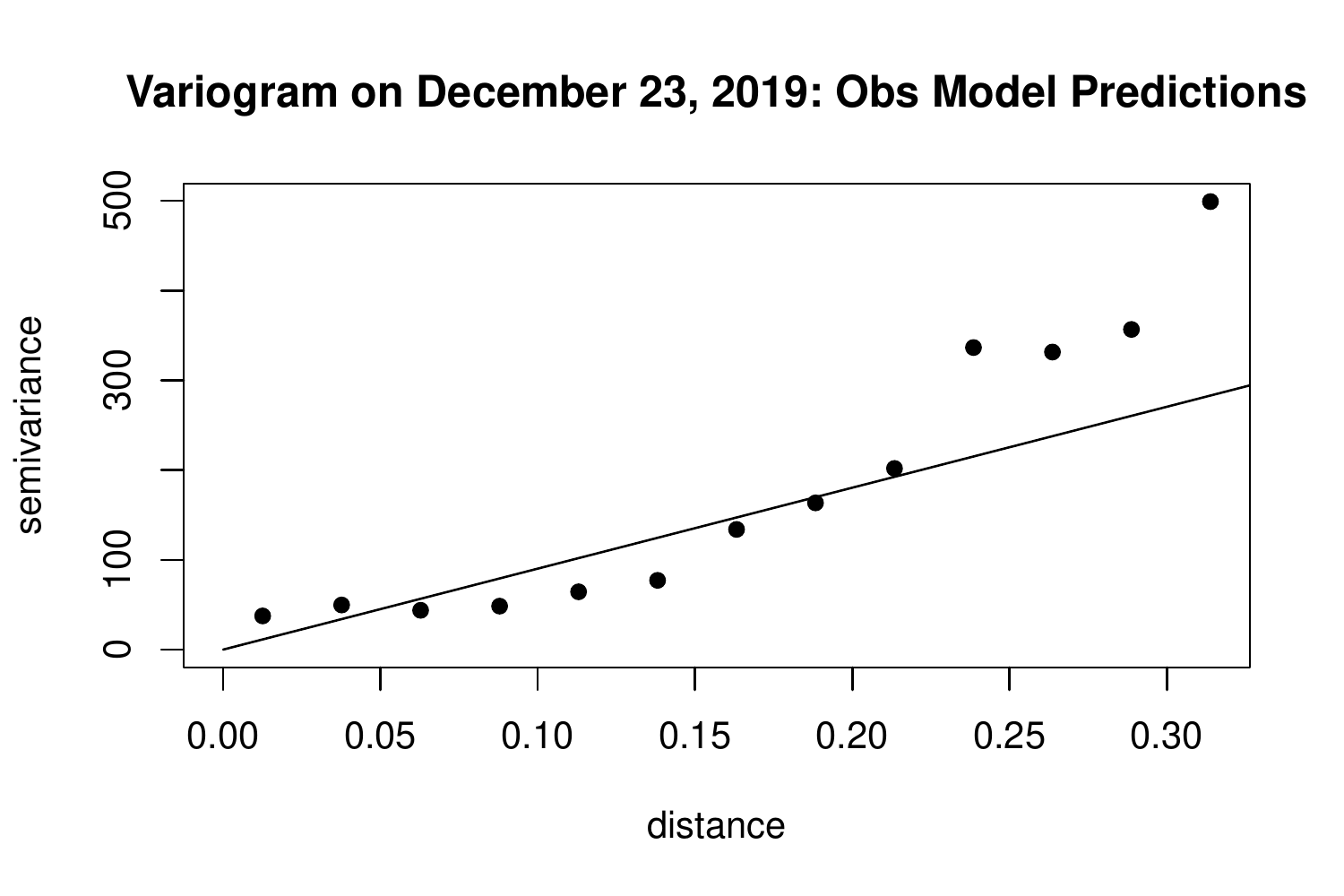}
	\caption{Variograms on selected days using the predicted true concentrations $\hat{x}$ from the observation model trained on the previous month, with the fitted exponential covariance function. }
	\label{supp_search_variog}
\end{figure}

Next we looked at variograms for a subset of time points to see if there is spatial correlation in the data. We use the predicted true concentrations $\hat{x}$ from the observation model given by Equation (\ref{eq:pred}) to make the variograms. Figure \ref{supp_search_variog} plots the variograms for two days -- December 5 and December 23. We see that the variograms are very noisy given the very small spatial sample size of the SEARCH network ($n=36$). Hence, the fitted exponential covariance function offers only a a moderately good fit to the noisy empirical variogram. However, we do see that there is clear spatial correlation in the predicted concentrations at the SEARCH network sites, with the variograms tending to increase with distance. This justifies the use of our second-stage spatial model for the true concentrations.  

\begin{figure}[!htbp]
	\centering
	\includegraphics[width=4in]{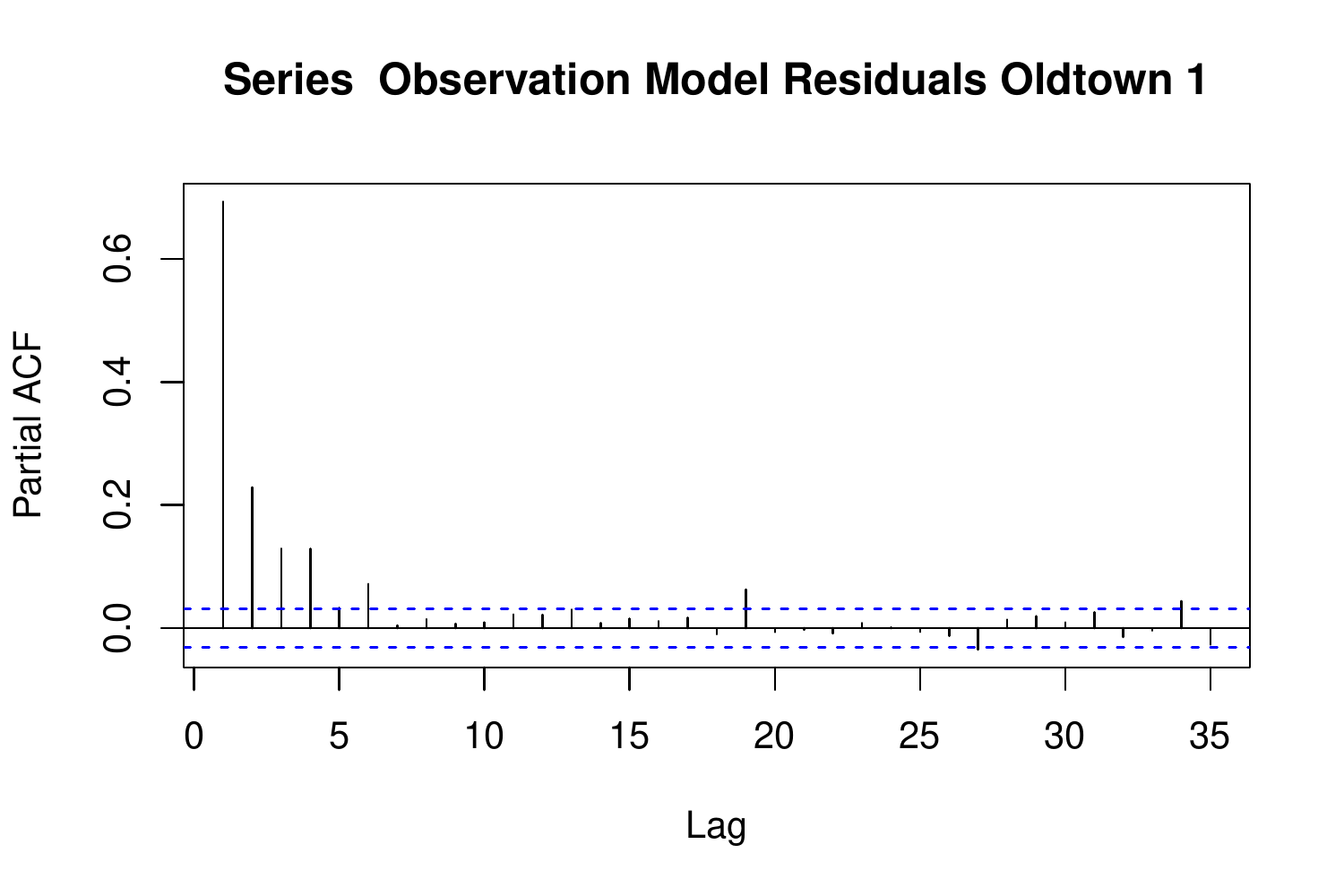}
	\caption{Partial autocorrelation function (PACF) for hourly residuals from the observation model at the first sensor placed at Oldtown. }
	\label{supp_search_acf}
\end{figure}

We also look at the partial autocorrelation function (PACF) for the hourly residuals from the observation model. We see that the residuals are highly correlated for a one hour lag, but beyond 5 or 6 hours there is very little remaining correlation in residuals. 
As we train the observation model on a monthly scale, this scale of autoregression is much smaller. Hence, while there is room to improve our estimation of the observation model by including these temporal correlations, the short-term nature of the dependence implies that even an independent working covariance model, like the one we deploy will provide reasonable estimates. 

\begin{figure}[!htbp]
	\centering
	\includegraphics[width=6in]{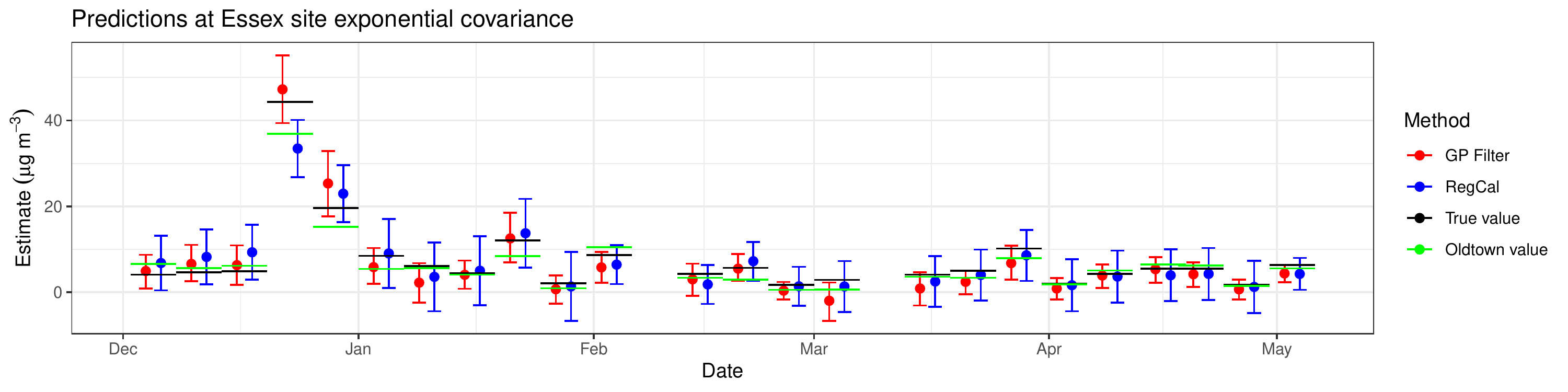}
	\includegraphics[width=6in]{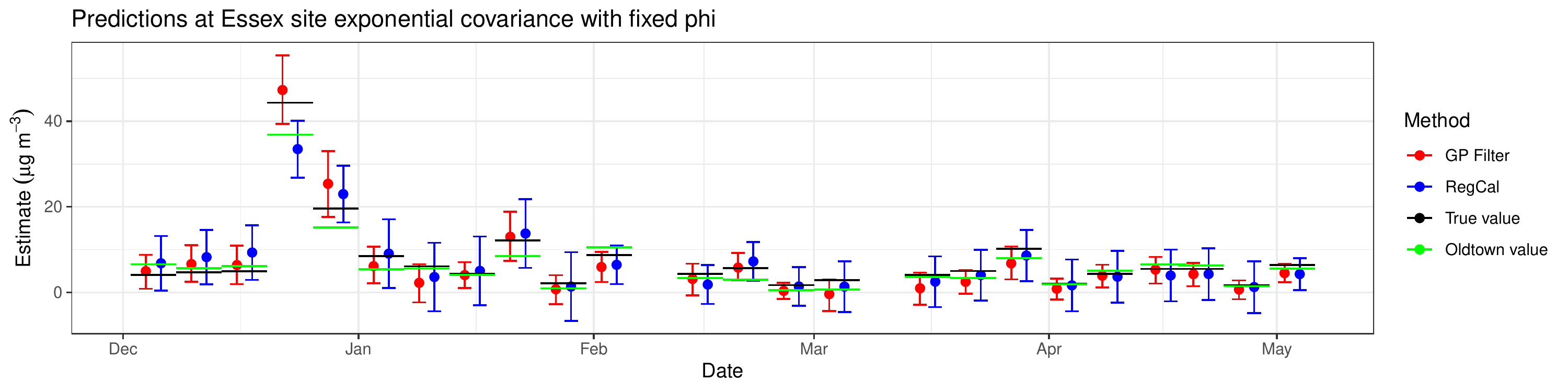}
	\includegraphics[width=6in]{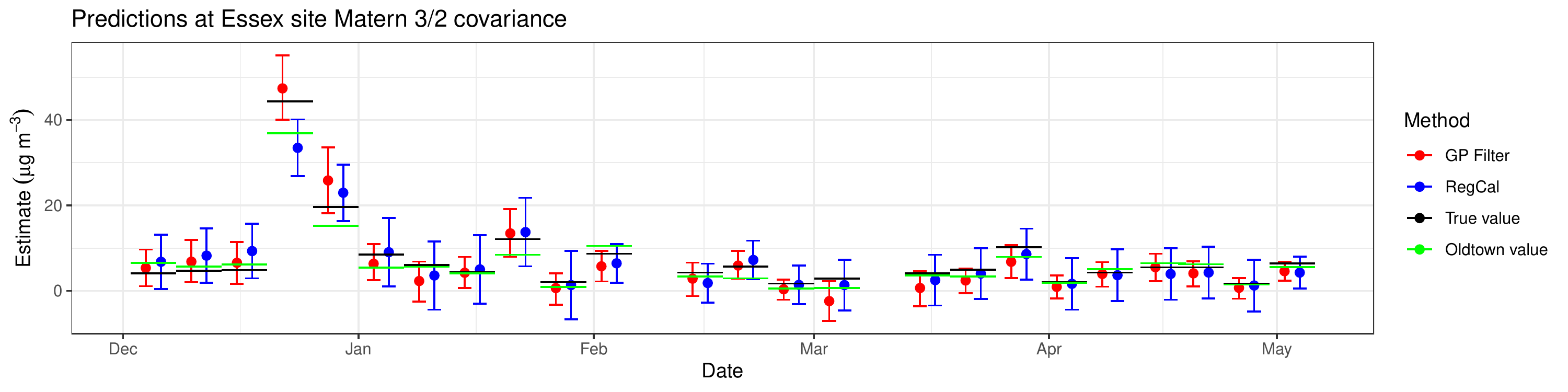}
	\includegraphics[width=6in]{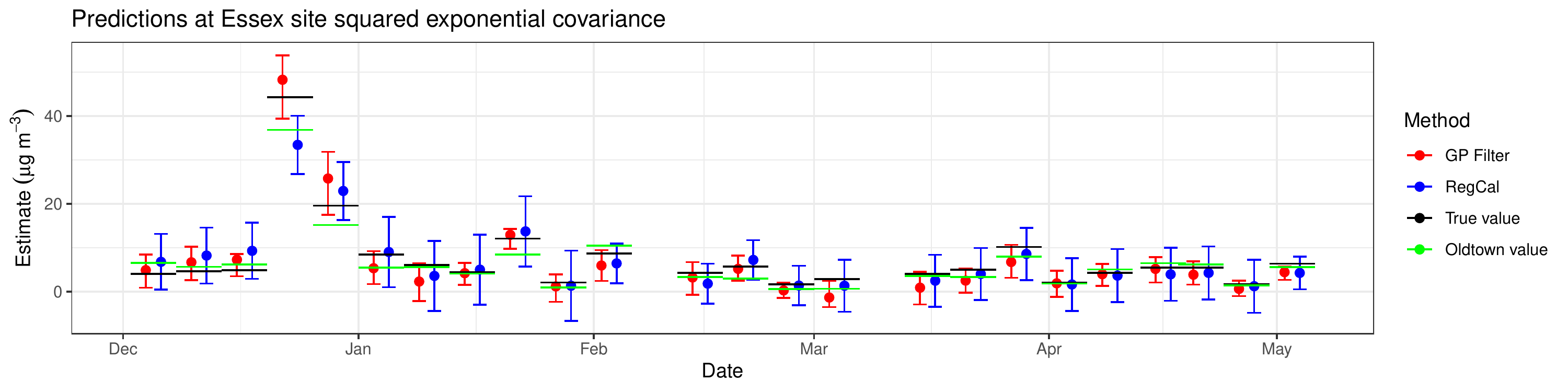}
	\caption{Time series of predictions at the Essex validation site with 95\% confidence intervals, using four different forms for the covariance. The true concentration at Essex is denoted by the black line, and the true concentration at Oldtown by the green line. }
	\label{supp_search_ts_4covariance}
\end{figure}

\begin{figure}[!htbp]
	\centering
	\includegraphics[width=5in]{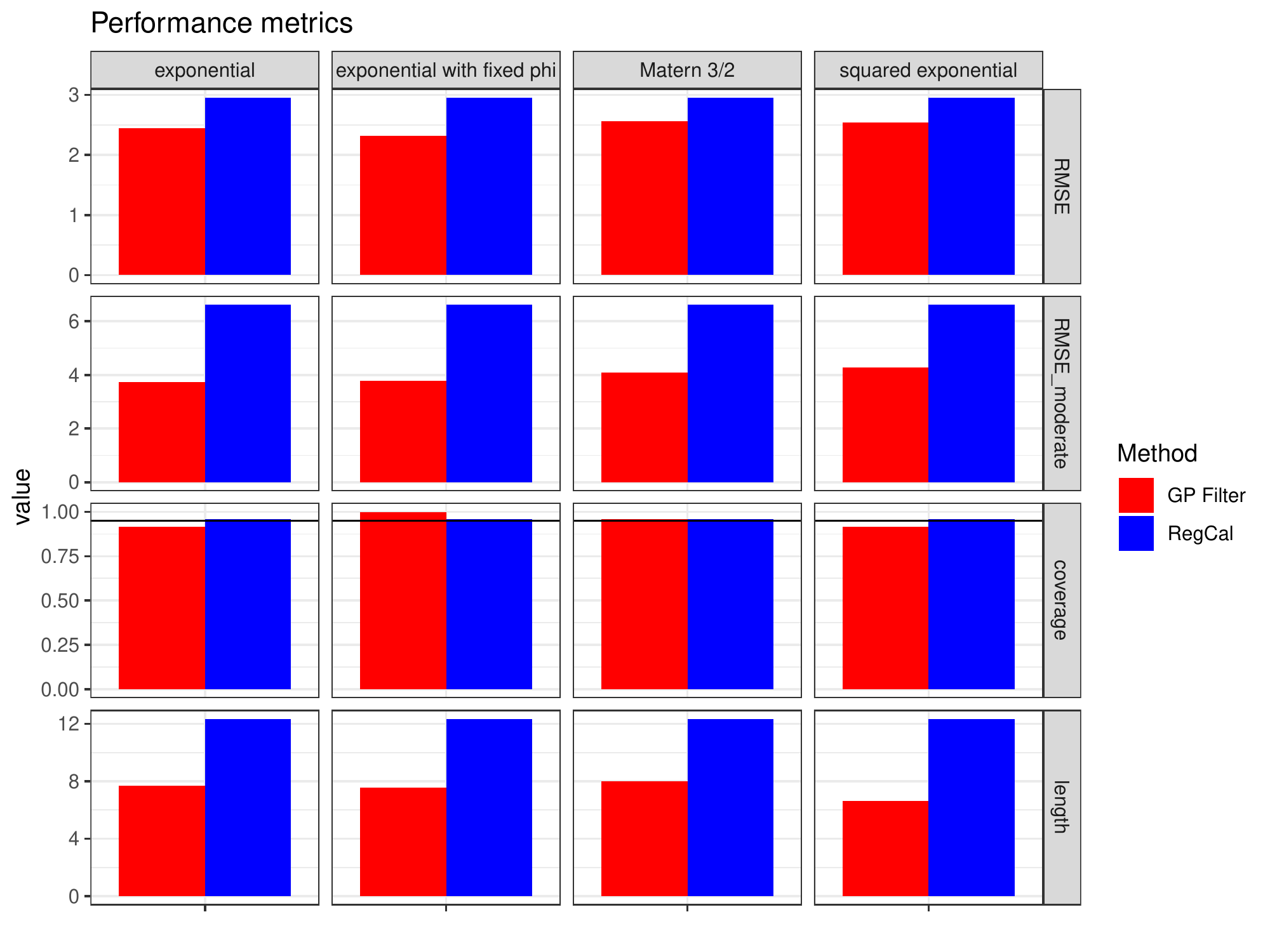}
	\caption{RMSE, RMSE on moderate time points, coverage probability, and width of confidence interval over the six month testing period with four different covariance functions. The black horizontal lines denotes a coverage of 95\%.}
	\label{supp_search_metrics}
\end{figure}

Now, we show the results of applying the GP Filter to six months of data. We use four different covariance functions: (1) the exponential covariance function, (2) the exponential where the value of the spatial decay parameter $\phi_t$ is fixed to its maximum likelihood estimate, since we have a small sample of SEARCH sensors with which to identify the spatial parameters, (3) the Mat\'ern  3/2 covariance function, (4) the squared exponential covariance function. A nugget variance was included in all of the covariance models. The time series using the four covariance functions are shown in Figure \ref{supp_search_ts_4covariance}. We note that the predictions do not change too dramatically with the choice of covariance function, so the model is robust to the choice of the covariance function. Figure \ref{supp_search_metrics} shows the RMSE over six months using all four covariance functions. We note that the exponential with fixed spatial decay has the lowest RMSE overall. The squared exponential has lower RMSE when restricted to ``moderate'' time points, but it performs less well on ``good'' time points. The FNR is exactly 0 for all methods and covariance functions, so it is not shown in the figure. All methods have coverage around 95\%. 
We also see that the length of the confidence interval is less for the GP Filter than for RegCal. From these results, we select the exponential with fixed $\phi_t$ as our preferred model as it produces the least overall RMSE. All subsequent results presented in this Section and the analysis in the main text use this covariance function. 

\begin{figure}[!htbp]
	\centering
	\includegraphics[width=5in]{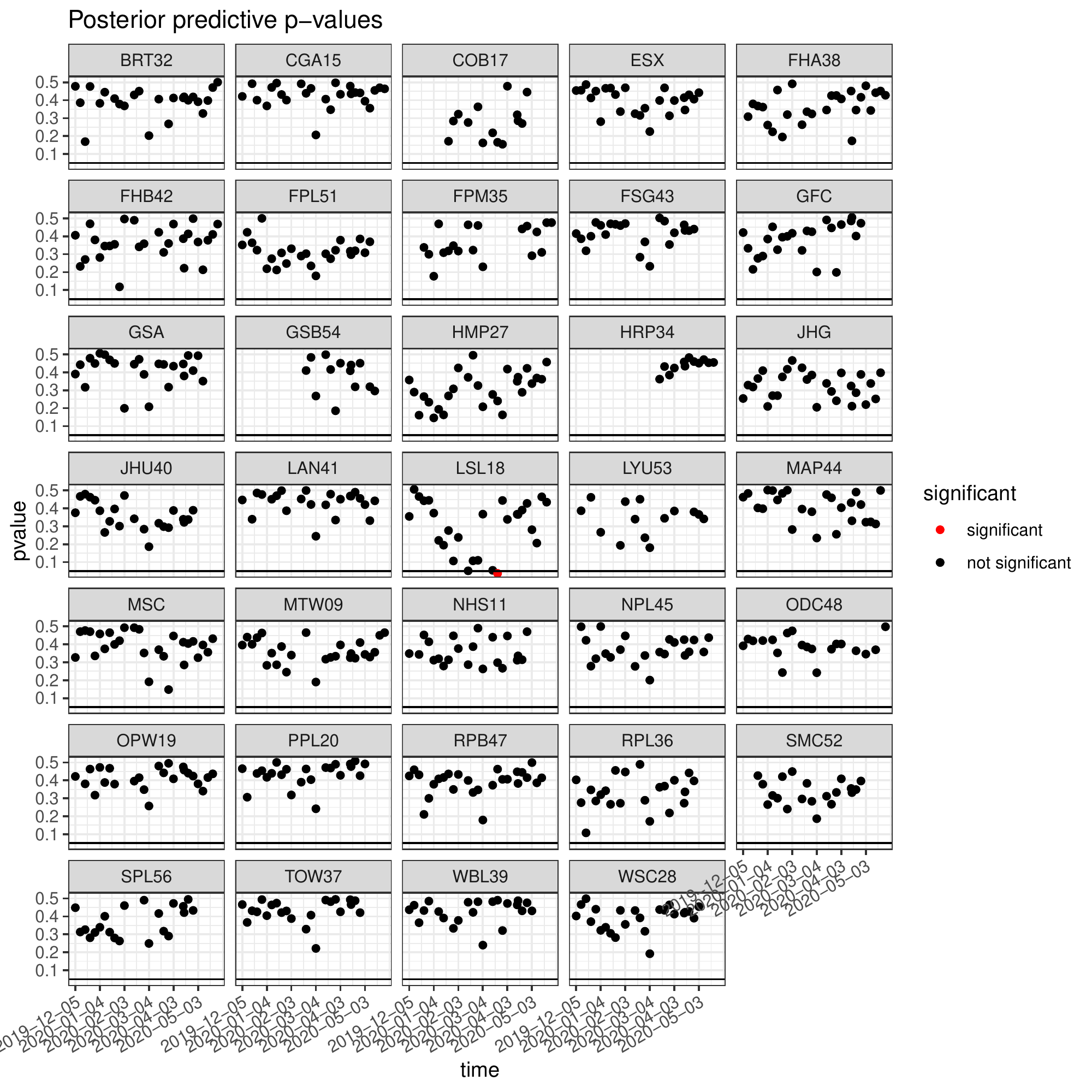}
	\caption{Posterior predictive $p$-values for the GP Filter with an exponential covariance with fixed $\phi_t$. The horizontal black line denotes a $p$-value of 0.05. The point in red has a significant $p$-value ($<0.05$) and black points are not significant. Each plot represents a different sensor in the SEARCH network.}
	\label{supp_search_ppp}
\end{figure}

To assess the impact of the different modeling choices, we conduct an extensive model adequacy analysis using posterior predictive $p$-values 
 \citep{gelman1995bayesian}. For each site $\bs$ in the network and each time point $t$, the posterior predictive $p$-value can be calculated by estimating the probability $$P\Big(|y_{sample}(\bs,t)-\mu(\bs,t)|>|y(\bs,t)-\mu(\bs,t)|\Big)$$ where $y(\bs,t)$ is the observed low-cost measurement at that site and time point. Therefore, we are calculating the probability that the model generates low-cost measurements more extreme than the observed measurements. 
To calculate this probability, for every time point $t$ in our testing window, we draw a sample $y_{sample}\slt$ from the observation model given the fitted $\boldsymbol{\beta}$s and $\tau^2$, as well as the final values of $x\slt$. Also, for every MCMC iteration, the sample $x^{(k)}\slt$ is plugged into the observation model in Equation \ref{obsmodel} to get the expected low-cost measurement $\mu^{(k)}\slt$ for that iteration. Then, the proportion $$\frac 1K\sum_{k=1}^K\mathbb{I}\left(|y_{sample}(\bs,t)-\mu^{(k)}(\bs,t)|>|y(\bs,t)-\mu^{(k)}(\bs,t)|\right)$$ estimates the posterior predictive probability. We calculate this $p$-value for every time point and sensor location.
The results are shown in Figure \ref{supp_search_ppp}. We see that the exponential model with a fixed $\phi_t$ has only one significant $p$-value out of 796 points. This validates the GP Filter, showing the fit is adequate and our model assumptions are reasonable. 



\newpage
\FloatBarrier
\section{Supplemental Figures and Tables from Simulations}

\phantom{text}

\begin{table}[!htbp]
\caption{Simulation 1a: Runtimes (seconds) for 100 time points and 50 sensors. Frequentist implementation and fully parallelized Bayesian implementation are compared.}
\begin{tabular}[t]{lcc}
\toprule
  & Frequentist & Bayesian\\
\midrule
Covariate model & 3.56 & 322.86\\
No covariate model & 2.37 & 296.92\\
\bottomrule
\end{tabular}
\label{sim_1a_runtime}
\end{table}

\begin{figure}[!htbp]
	\centering
	\includegraphics[width=5.5in]{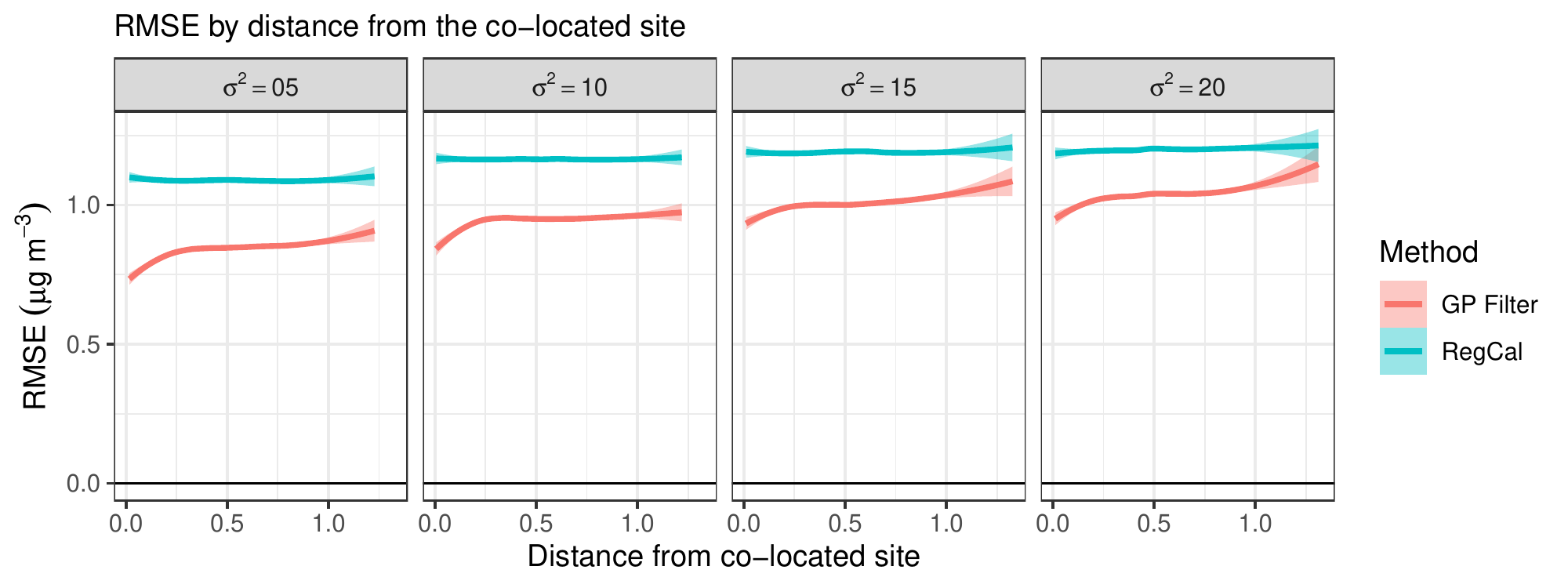}
	\caption{Simulation 1a: RMSE by distance of sensor from the collocated reference device, using correctly specified Gaussian process spatial model, across 50 datasets.}
	\label{1a_rmse_dist_all}
\end{figure}

\begin{figure}[!htbp]
	\centering
	\includegraphics[width=5.5in]{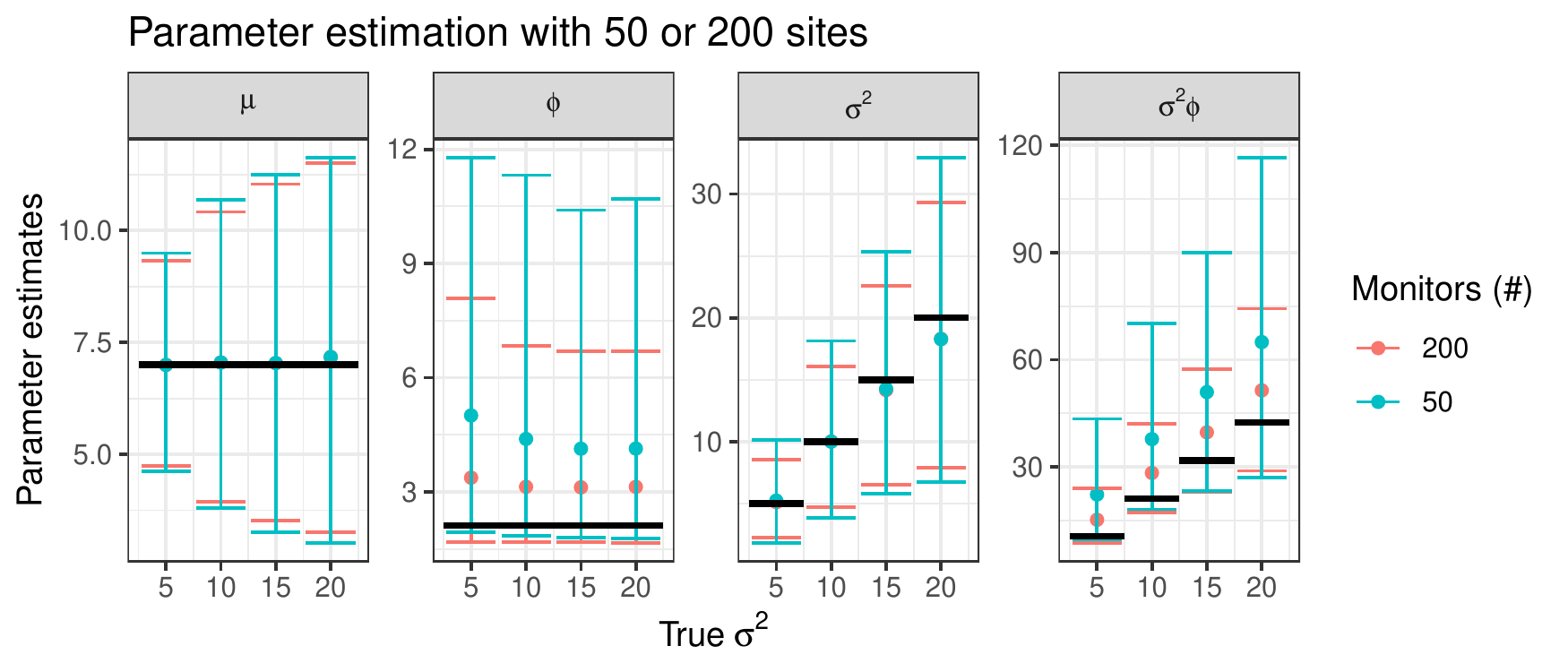}
	\caption{Simulation 1b: Average parameter estimates with 95\% interval from GP Filter, comparing 50 and 200 non-collocated sites, with 1 collocated site. True parameter values are shown in black. Averaged over 20 datasets.}
	\label{1b_parameters}
\end{figure}

\begin{figure}[!htbp]
	\centering
	\includegraphics[width=5.5in]{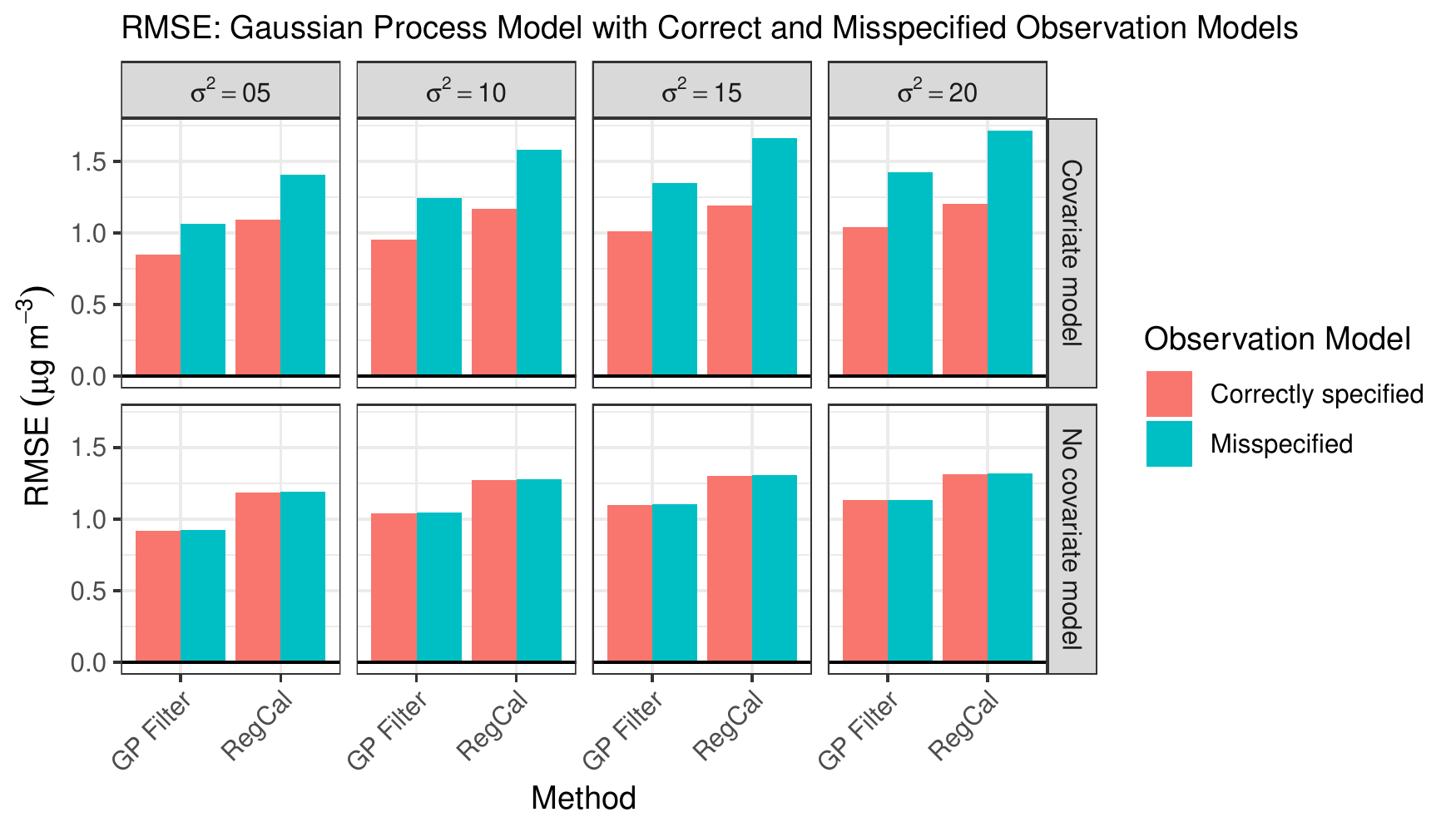}
	\caption{Simulation 2: RMSEs for correctly specified and misspecified covariate set in the observation model.}
	\label{2_rmse}
\end{figure}

\begin{figure}[!htbp]
	\centering
	\includegraphics[width=5.5in]{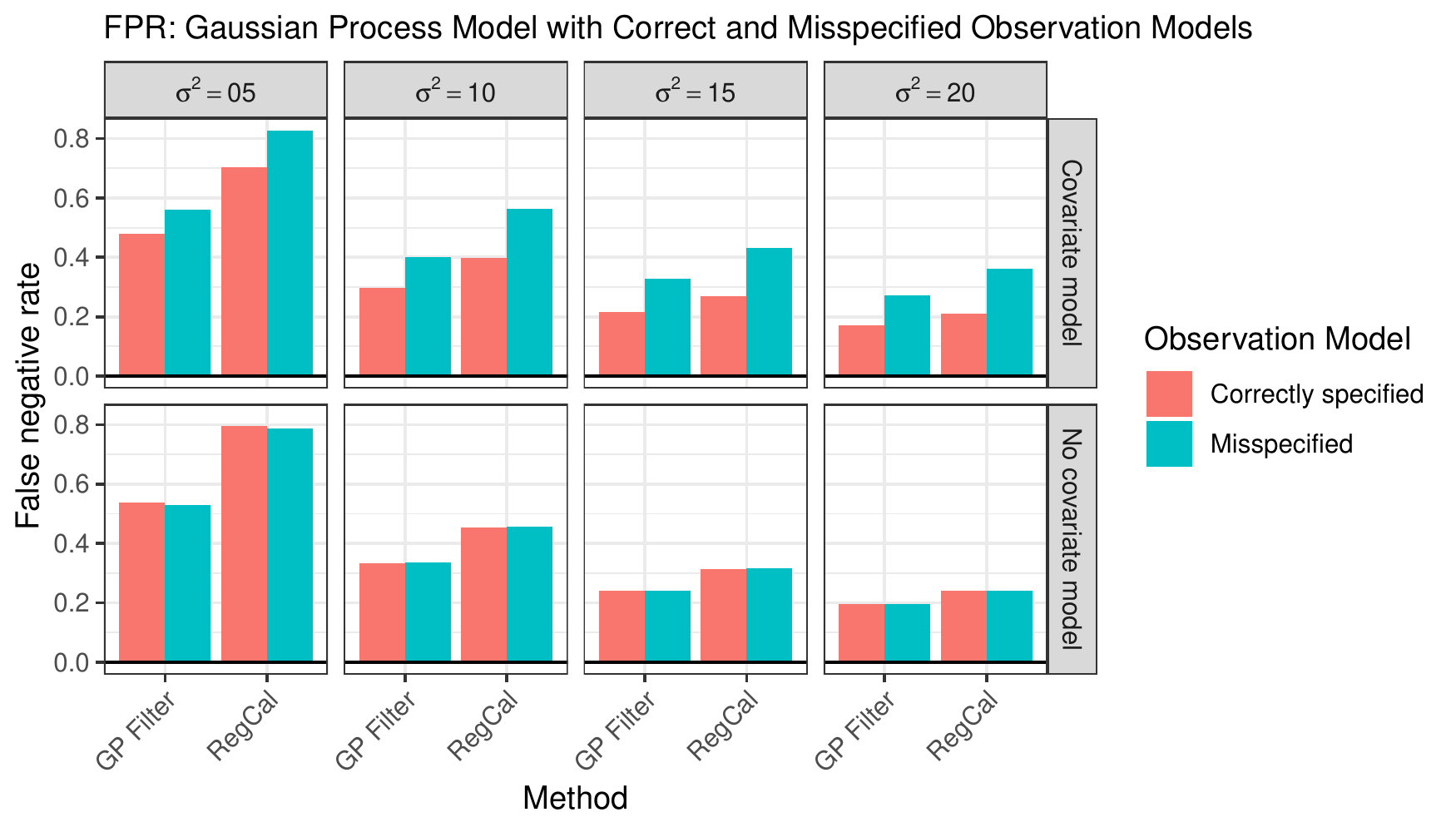}
	\caption{Simulation 2: False negative rates for correctly specified and misspecified covariate set in the observation model.}
	\label{2_fpr}
\end{figure}

\begin{figure}[!htbp]
	\centering
	\includegraphics[width=4.5in]{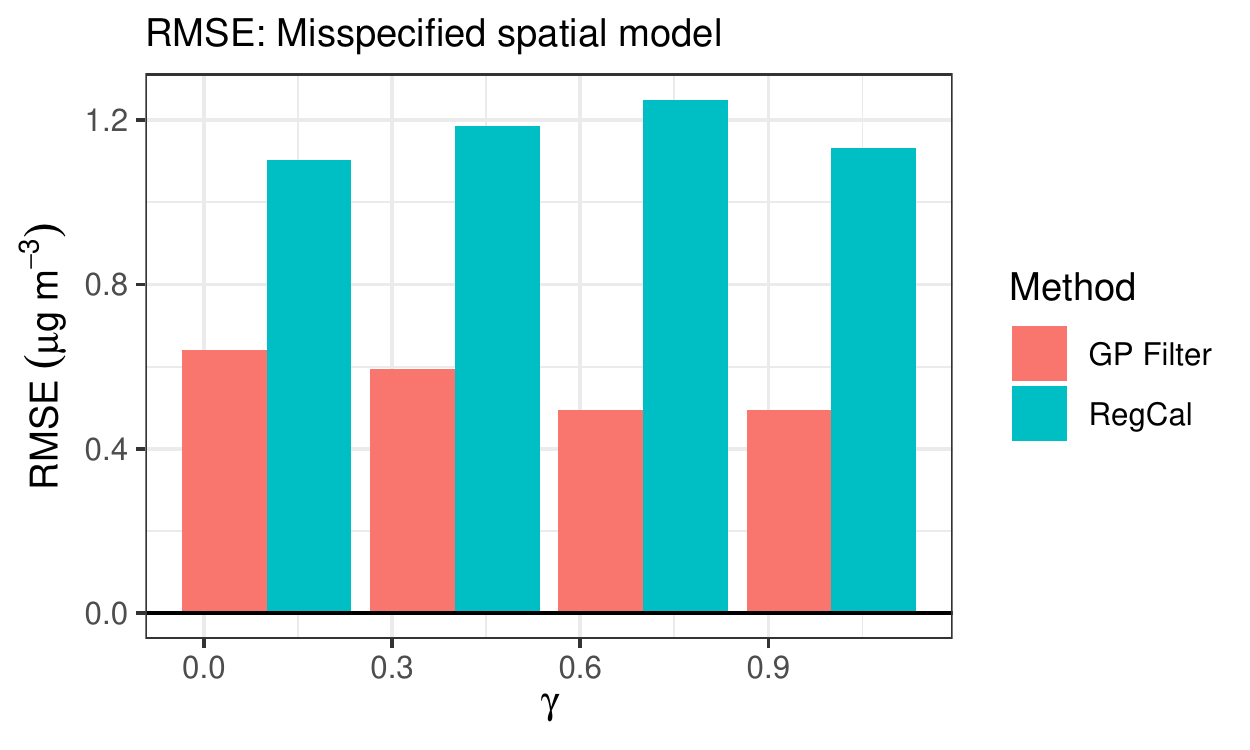}
	\caption{Simulation 3: RMSE when the underlying pollution surface is misspecified.}
	\label{3_rmse}
\end{figure}

\begin{figure}[!htbp]
	\centering
	\includegraphics[width=4.5in]{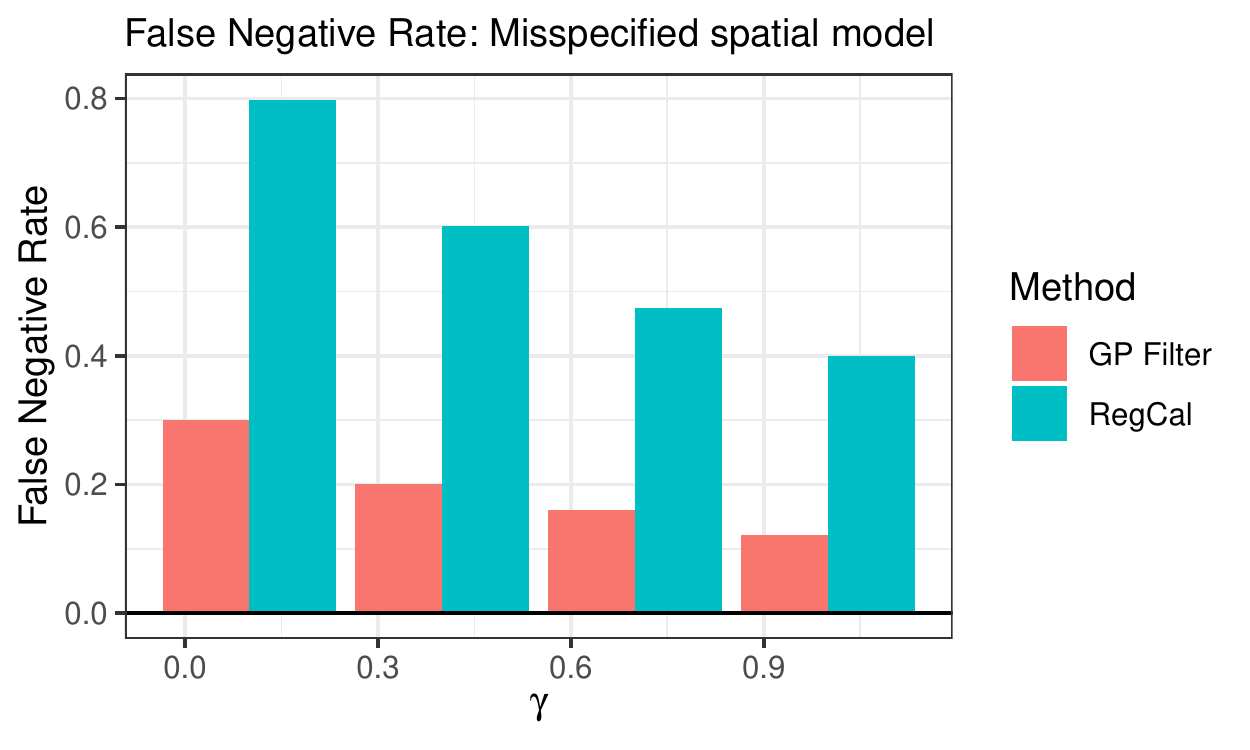}
	\caption{Simulation 3: FNR when the underlying pollution surface is misspecified.}
	\label{3_fpr}
\end{figure}

\end{document}